\documentclass[aps,twocolumn,pra,superscriptaddress,amsmath,showpacs,tightenlines,longbibliography]{revtex4-1}

\usepackage{graphicx,times}
\usepackage{amstext}
\usepackage{amsmath}            
\usepackage{amssymb}            
\usepackage{latexsym}
\usepackage{bm}
\usepackage{color}

\usepackage[dvipsnames]{xcolor}

\usepackage{etoolbox}

\makeatletter
\let\cat@comma@active\@empty
\makeatother

\def \mean#1{{\langle \hat #1 \rangle}}

\usepackage{url}
\usepackage[colorlinks]{hyperref}
\hypersetup{%
    plainpages=true,
    breaklinks=true,
    hypertexnames=false,
    pageanchor=true,
    colorlinks=true,
    linkcolor={blue},
    citecolor={red},
    urlcolor={blue},
    anchorcolor={black}
}

\usepackage{graphics}
\usepackage{epstopdf}

\usepackage[T1]{fontenc}
\usepackage[applemac]{inputenc}
\usepackage{lmodern}
\usepackage[english]{babel}

\usepackage{ae}
\usepackage{units}

\usepackage[americaninductors]{circuitikz}
\usepackage{tikz}
\usetikzlibrary{arrows,snakes,calc}

\usepackage{multirow}

\newcommand{\ket}[1]{|#1\rangle}

\newcommand{\ketbra}[2]{\left| #1 \rangle \langle #2 \right|}
\newcommand{\brakket}[3]{\langle #1 | #2 | #3 \rangle}

\newcommand{\figref}[1]{\mbox{Fig.~\ref{#1}}}
\newcommand{\tabref}[1]{\mbox{Table~\ref{#1}}}
\newcommand{\secref}[1]{\mbox{Sec.~\ref{#1}}}

\newcommand{\appref}[1]{\mbox{Appendix~\ref{#1}}}
\renewcommand{\eqref}[1]{\mbox{Eq.~(\ref{#1})}}

\newcommand{\sz}{\hat \sigma_z}
\newcommand{\sx}{\hat \sigma_x}
\newcommand{\sm}{\hat \sigma_-}
\renewcommand{\sp}{\hat \sigma_+}

\newcommand{\nn}{\nonumber}

\newcommand{\abs}[1]{\left|#1\right|}
\newcommand{\abssq}[1]{\left| #1 \right|^2}

\newcommand{\be}{\begin{equation}}
\newcommand{\ee}{\end{equation}}
\newcommand{\bea}{\begin{eqnarray}}
\newcommand{\eea}{\end{eqnarray}}

\newcommand{\rd}{\ensuremath{\mathrm{d}}}
\newcommand{\id}{\ensuremath{\,\rd}}

\begin{document}

\title{Deterministic quantum nonlinear optics with single atoms and virtual photons}

\author{Anton Frisk Kockum}
\email[e-mail:]{anton.frisk.kockum@gmail.com}
\affiliation{Center for Emergent Matter Science, RIKEN, Saitama 351-0198, Japan}

\author{Adam Miranowicz}
\affiliation{Faculty of Physics, Adam Mickiewicz University, PL-61-614 Poznan, Poland}
\affiliation{Center for Emergent Matter Science, RIKEN, Saitama 351-0198, Japan}

\author{Vincenzo Macrì}
\affiliation{Dipartimento di Fisica e di Scienze della Terra, Universit\`{a} di Messina, I-98166 Messina, Italy}
\affiliation{Center for Emergent Matter Science, RIKEN, Saitama 351-0198, Japan}

\author{Salvatore Savasta}
\affiliation{Dipartimento di Fisica e di Scienze della Terra, Universit\`{a} di Messina, I-98166 Messina, Italy}
\affiliation{Center for Emergent Matter Science, RIKEN, Saitama 351-0198, Japan}

\author{Franco Nori}
\affiliation{Center for Emergent Matter Science, RIKEN, Saitama 351-0198, Japan}
\affiliation{Physics Department, The University of Michigan, Ann Arbor, Michigan 48109-1040, USA}

\date{\today}

\begin{abstract}

We show how analogues of a large number of well-known nonlinear-optics phenomena can be realized with one or more two-level atoms coupled to one or more resonator modes. Through higher-order processes, where virtual photons are created and annihilated, an effective deterministic coupling between two states of such a system can be created. In this way, analogues of three-wave mixing, four-wave mixing, higher-harmonic and -subharmonic generation (i.e., up- and downconversion), multiphoton absorption, parametric amplification, Raman and hyper-Raman scattering, the Kerr effect, and other nonlinear processes can be realized. The effective coupling becomes weaker the more intermediate transition steps are needed. However, given the recent experimental progress in ultrastrong light-matter coupling, especially in the field of circuit quantum electrodynamics, we estimate that many of these nonlinear-optics analogues can be realized with currently available technology. 

\end{abstract}

\maketitle

\section{Introduction}

In nonlinear optics, a medium responds nonlinearly to incoming light of high intensity. This nonlinear response can give rise to a host of effects, including frequency conversion and amplification, many of which have important technological applications~\cite{Lindsay1975, Kielich1981, Shen1984, Boyd2008}. After the high-intensity light of a laser made possible the first experimental demonstration of second-harmonic generation (frequency upconversion) in 1961~\cite{Franken1961}, many more nonlinear-optics effects have been demonstrated using a variety of nonlinear media. The many applications and the fundamental nature of nonlinear optics have also inspired investigations of analogous effects in other types of waves. Prominent examples include nonlinear acoustics~\cite{Bunkin1986, Hamilton1998}, nonlinear spin waves~\cite{Cottam1994}, nonlinear atom optics~\cite{Lenz1993, Rolston2002}, nonlinear Josephson plasma waves~\cite{Savelev2006}, and nonlinear plasmonics~\cite{Kauranen2012}. Analogies of this kind can sometimes enable simulations or demonstrations of phenomena that are hard to realize in other systems~\cite{Buluta2009, Nation2012, Georgescu2014}.

In this article, we will show that analogues of many nonlinear-optics effects can also be realized by coupling one or more resonator modes to one or more two-level atoms. This stands in contrast to many other nonlinear-optics realizations, which require three or more atomic levels~\cite{Boyd2008, You2011}. The key to the analogues we propose lies in the full interaction between a two-level atom and an electromagnetic mode, which is given by the quantum Rabi Hamiltonian~\cite{Rabi1937}. This Hamiltonian includes terms that can change the number of excitations in the system, enabling higher-order processes via virtual photons. These photons are created and annihilated in a way that creates a deterministic coupling between two system states that otherwise do not have a direct coupling. In this way, we can realize analogues of various frequency-conversion processes, parametric amplification, Raman and hyper-Raman scattering, multiphoton absorption, the Kerr effect, and other nonlinear processes.

Just as nonlinear-optics effects usually require very high light intensity to manifest clearly, the higher-order processes we consider require a very strong light-matter coupling to become noticeable. Ultrastrong coupling (USC, where the coupling strength starts to become comparable to the resonance frequencies of the bare system components) between light and matter has only recently been reached in some solid-state experiments~\cite{Gunter2009, Forn-Diaz2010, Niemczyk2010, Todorov2010, Schwartz2011, Scalari2012, Geiser2012, Kena-Cohen2013, Gambino2014, Maissen2014, Goryachev2014, Baust2016, Forn-Diaz2017, Yoshihara2017, Chen2016, Langford2016, Braumuller2016, Yoshihara2016}. Among these systems, circuit quantum electrodynamics (QED)~\cite{Wallraff2004, Blais2004, You2011, Xiang2013} has provided some of the clearest examples~\cite{Forn-Diaz2010, Niemczyk2010, Baust2016, Forn-Diaz2017, Yoshihara2017, Chen2016, Langford2016, Braumuller2016, Yoshihara2016}, including the largest reported coupling strength~\cite{Yoshihara2017} and the first quantum simulations of the USC regime~\cite{Langford2016, Braumuller2016}.  

The experimental progress in USC physics has motivated many theoretical studies of the interesting new effects that occur in this regime~\cite{DeLiberato2007, Ashhab2010, Cao2010, Casanova2010, Beaudoin2011, Ridolfo2012, Stassi2013, DeLiberato2014, Sanchez-Burillo2014, Lolli2015, DiStefano2016, Cirio2016}. Some previous~\cite{Ma2015, Garziano2015, Garziano2016} and forthcoming~\cite{Kockum2017, Stassi2017} works explore processes in the USC regime where the number of excitations is not conserved, such as multiphoton Rabi oscillations~\cite{Garziano2015} and a single photon exciting multiple atoms~\cite{Garziano2016}. Several of these processes can be interpreted as analogues of nonlinear-optics phenomena. 

In this work, we present a unified picture of this type of processes and their relation to nonlinear optics. We also provide many more examples, not previously studied, which together allow us to make complete tables with translations between three- and four-wave mixing in nonlinear optics and analogous processes in USC systems. It should be noted that these analogues, many of which can be realized in one universal setup, do not use propagating waves, but instead mix excitations in resonators and atoms of different frequencies. Given the versatility and technological maturity of the circuit QED setups, we expect them to become the primary experimental platform for realising these deterministic nonlinear-optics analogues with single atoms and virtual photons. We believe that these deterministic analogues can find many important applications, including frequency conversion and the creation of superposition states for use in quantum information technology.

This article is organized as follows. In \secref{sec:Mechanisms}, we give a brief overview of how nonlinear processes in optics usually occur. We then describe how analogous deterministic processes become possible in the quantum Rabi model. In Secs.~\ref{sec:3WaveMixing} and \ref{sec:4WaveMixing}, we discuss three- and four-wave mixing, respectively, and give details of the analogous deterministic processes that can be realized with resonators ultrastrongly coupled to qubits. Other nonlinear processes, including higher-harmonic generation, parametric processes, and the Kerr effect, are discussed in \secref{sec:OtherProcesses}. We conclude in \secref{sec:SummaryOutlook}. Some details are left for the appendices: \appref{app:NonlinearSusceptibility} expands on the classical mechanisms for some nonlinear-optics phenomena, \appref{app:PerturbationTheory} gives a derivation of the perturbation-theory formula used to calculate the strength of the effective coupling between initial and final states in our three- and four-wave-mixing analogues, and \appref{app:4WaveMixing} contains details about a few four-wave-mixing processes not treated in the main text.

\section{Mechanisms for nonlinearity}
\label{sec:Mechanisms}

\subsection{Nonlinear optics}
\label{sec:MechanismsNonlinearOptics}

In conventional classical electro-optical processes, the polarization $\mathbf{P}$ of a given medium induced by the applied electric field ${\bf E}$ is linearly proportional to its strength, i.e., $\mathbf{P} = \epsilon_{0} \chi \mathbf{E}$, where $\epsilon_0$ is the vacuum permittivity and $\chi \equiv \chi^{(1)}$ is the linear susceptibility of the medium, which can be considered a scalar for linear, homogeneous, and isotropic dielectric media. Usually, the real and imaginary parts of $\chi$ describe, respectively, the refraction and damping of a light beam going through such medium.

For a strong electric field $\mathbf{E}$ and nonlinear media, the above linear relation for the induced polarization is generalized to
\be
{\bf P} = \epsilon_0 \left( \chi^{(1)} \mathbf{E} + \chi^{(2)} \mathbf{E}^{2} + \chi^{(3)} \mathbf{E}^{3} + \ldots \right),
\label{nonlinearP1}
\ee
which is considered a core principle of nonlinear optics~\cite{Lindsay1975, Kielich1981, Shen1984, Boyd2008}. In \eqref{nonlinearP1}, $\chi^{(2)}$ and $\chi^{(3)}$ are the second- and third-order nonlinear susceptibilities, respectively. In general, these susceptibilities are tensors $\chi_{kl}^{(1)}$, $\chi_{klm}^{(2)}$, and $\chi_{klmn}^{(3)}$. However, for simplicity, we consider them as scalars, which is usually valid for isotropic dielectric media.

Various nonlinear optical phenomena (including wave mixing) can be explained classically by recalling the nonlinear dependence of the induced polarization and electric-field strength, as given by \eqref{nonlinearP1}. Standard examples include Pockels and Kerr effects, which are, respectively, linear and quadratic electro-optical phenomena, in which the induced polarization (and, thus, also the refractive index) of a medium is proportional to the amplitude and its square of the applied constant electric field. 

For example, second-harmonic generation in a medium described by the second-order susceptibility $\chi^{(2)}$ can be described classically as follows. Assuming that a monochromatic scalar electric field $E(t) = E_0 \cos (\omega t)$ is applied to the medium, the second-order induced polarization $P^{(2)}$ of the medium is given by
\bea
P^{(2)} &=& \epsilon_{0} \chi^{(2)} E^2 = \epsilon_0 \chi^{(2)} E_0^2 \cos^2 (\omega t) \nn \\ 
&=& \epsilon_0 \chi^{(2)} E_0^2 \left( \frac{1 + \cos (2 \omega t)}{2} \right) \nn \\
&=& \frac{1}{2} \epsilon_0 \chi^{(2)} E_0^2 + \frac{1}{2} \epsilon_0 \chi^{(2)} E_0^2 \cos (2 \omega t),
\label{chi2a}
\eea
where the first term in the last line describes frequency-independent polarization, while the second term in the last line describes the polarization oscillating at twice the frequency of the input field. This doubling of the input frequency can be interpreted as second-harmonic generation.

In \appref{app:NonlinearSusceptibility}, we present a few additional pedagogical classical explanations, based on \eqref{nonlinearP1}, of phenomena arising due to the $\chi^{(2)}$ and $\chi^{(3)}$ nonlinearities.

\subsection{The quantum Rabi model}
\label{sec:QuantumRabiModel}

The Hamiltonian for a single two-level atom (a qubit) coupled to a single resonator mode can be written as ($\hbar = 1$ here and in the rest of the article)
\be
\hat H = \omega_a \hat a^\dag \hat a + \omega_q \frac{\sz}{2} + \hat H_{\rm int}.
\label{eq:H1res1qb}
\ee
In the quantum Rabi model~\cite{Rabi1937}, the interaction is given by
\be
\hat H^{\rm Rabi}_{\rm int} = g \left( \hat a + \hat a^\dag \right) \sx = g \left( \hat a + \hat a^\dag \right) \left( \sm + \sp \right),
\label{eq:HintRabi}
\ee
where $g$ is the coupling strength. Here, and in the following parts of the paper that discuss deterministic realisations of nonlinear optics, we use the notation that $\hat a$, $\hat b$, $\hat c$, and $\hat d$ are the annihilation operators of resonator modes with frequencies $\omega_a$, $\omega_b$, $\omega_c$, and $\omega_d$, respectively. In setups with a single qubit, or with several identical qubits, the qubit transition frequency is denoted $\omega_q$. In setups with qubits having different transition frequencies, the frequencies are denoted $\omega_{q1}$, $\omega_{q2}$, \textit{etc.} The qubit operators $\sz$ and $\sx = \sm + \sp$ are Pauli matrices; $\sm$ and $\sp$ are the qubit lowering and raising operators, respectively.

In the limit $g \ll \omega_a, \omega_q$, the terms $\hat a^\dag \sp$ and $\hat a \sm$ in $H^{\rm Rabi}_{\rm int}$ can be neglected in the rotating-wave approximation (RWA), leading to the Jaynes--Cummings (JC) model~\cite{Jaynes1963}
\be
\hat H^{\rm JC}_{\rm int} = g \left( \hat a \sp + \hat a^\dag \sm \right).
\label{eq:HintJC}
\ee
Note that in the JC model, the number of excitations is conserved. In the quantum Rabi model, the number of excitations can change, but their parity is conserved. However, the quantum Rabi model can be generalized to
\be
\hat H^{\rm gen}_{\rm int} = g \left( \hat a + \hat a^\dag \right) \left(\sx \cos \theta + \sz \sin \theta \right),
\label{eq:HintGenRabi} 
\ee
where $\theta$ parameterizes the amount of longitudinal and transversal couplings as, for example, in experiments with USC of flux qubits to resonators~\cite{Niemczyk2010, Forn-Diaz2010, Baust2016, Yoshihara2017}. This generalized quantum Rabi model does not impose any restrictions on the number of excitations.

All these models can be straightforwardly extended to include additional resonators and qubits. The presence of one or more qubits provides the necessary nonlinearity to realize various deterministic analogues of nonlinear-optics processes that we will discuss in this article. For some of these processes, such as three-wave mixing (see \secref{sec:3WaveMixing}), the number of excitations changes by one; this requires setups with the generalized quantum Rabi model and its extensions. For other processes, e.g., four-wave mixing (see \secref{sec:4WaveMixing}), the number of excitations changes by an even number or not at all; these processes can be realized with extensions of the standard quantum Rabi model or of the JC model, respectively.

In a majority of the nonlinear-optics analogues considered in this article, higher-order processes, mediated by the interaction Hamiltonians in Eqs.~(\ref{eq:HintRabi})-(\ref{eq:HintGenRabi}) (and their extensions to additional resonators and qubits), connect an initial state $\ket{i}$ with a final state $\ket{f}$ of the same energy through an effective interaction Hamiltonian
\be
\hat H_{\rm int}^{\rm eff} = g_{\rm eff} \ketbra{f}{i} +  {\rm H.c.},
\label{eq:Heffint}
\ee
where $g_{\rm eff}$ is the strength of the effective coupling and H.c.~denotes the Hermitian conjugate of the preceding terms. In many of the intermediate transitions that contribute to this effective coupling, virtual photons are created and destroyed. We provide a multitude of examples of this in the following sections. 

To calculate the effective coupling strength $g_{\rm eff}$ analytically, three different techniques have been employed in previous works. In Ref.~\cite{Moon2005}, an effective Hamiltonian with explicit up- and downconversion terms was derived through a series of unitary transformations combined with approximations that only retained terms of lowest order in $g_{j}/\abs{\omega_n - \omega_m}$, where $g_{j}$ are the relevant coupling strengths in the setup and $\abs{\omega_n - \omega_m}$ are the energy differences of the relevant intermediate transitions. In Refs.~\cite{Ma2015, Garziano2015, Kockum2017}, the intermediate virtual transitions were adiabatically eliminated, relying on the approximation that the population of the intermediate levels will not change significantly if $g_{j} \ll \abs{\omega_n - \omega_m}$. In this article, we follow the approach of Ref.~\cite{Garziano2016}, which calculated $g_{\rm eff}$ using standard perturbation theory. Specifically, if the shortest path between $\ket{i}$ and $\ket{f}$ is an $n$th-order process, the effective coupling is given to lowest order by
\be
g_{\rm eff} = \sum_{j_1, j_2, \ldots, j_{n-1}} \frac{V_{f j_{n-1}} \ldots V_{j_2 j_1} V_{j_1 i}}{\left( E_i - E_{j_1} \right) \left( E_i - E_{j_2} \right) \ldots \left( E_i - E_{j_{n-1}} \right) },
\label{eq:PerturbationFormula}
\ee
where the sum goes over all virtual transitions forming $n$-step paths between $\ket{i}$ and $\ket{f}$, $E_k$ denotes the energy of state $\ket{k}$, and $V_{km} = \brakket{k}{\hat H_{\rm int}}{m}$. A derivation of this formula is given in \appref{app:PerturbationTheory}.

In general, the perturbation-theory method of \eqref{eq:PerturbationFormula} appears to be the simplest way to calculate $g_{\rm eff}$, especially for higher-order processes involving many virtual transitions. The other methods mentioned above can be more cumbersome, but provide more information such as energy-level shifts and higher-order corrections to the effective coupling.

\section{Three-wave mixing}
\label{sec:3WaveMixing}

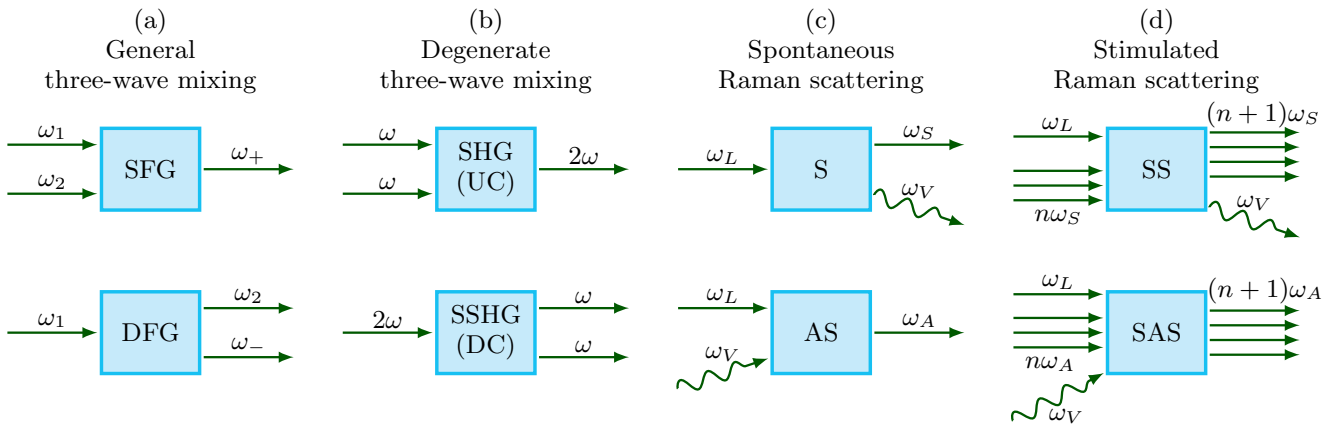
\begin{figure*}
\centerline{
  \resizebox{2.23\columnwidth}{!}{
    \begin{tikzpicture}[
      scale=1,
      level/.style={thick},
      virtual/.style={thick,dashed},
      ztrans/.style={thick,->,shorten >=0.1cm,shorten <=0.1cm,>=stealth,densely dashed,color=red},
      nrtrans/.style={thick,->,shorten >=0.1cm,shorten <=0.1cm,>=stealth,densely dashed,color=blue},
      rtrans/.style={thick,->,shorten >=0.1cm,shorten <=0.1cm,>=stealth,color=blue},
      strans/.style={rtrans, snake=snake, line after snake=1.5mm},
      classical/.style={thin,double,<->,shorten >=4pt,shorten <=4pt,>=stealth},
      mybox/.style={color=cyan!70, fill=cyan!20, very thick},
      myarrow/.style={color=green!35!black, thick, ->, shorten >=0.05cm, shorten <=0.05cm, >=latex},
      mysquigglyarrow/.style={color=green!35!black, thick, ->, shorten >=0.05cm, shorten <=0.05cm, >=latex, snake=snake, line after snake=2mm}
    ]
    \coordinate (corner1) at (0cm, 0cm); 
    \coordinate (rh) at (0, 1); 
    \coordinate (rl) at (1.2, 0); 
    \coordinate (vcol) at (0, 1); 
    \coordinate (hrow) at (0.5, 0); 
    \coordinate (al) at ($(rl) + (0,0)$); 
    \coordinate (corner2) at ($(corner1) + (rl) + (al) + (hrow) + (al)$); 
    \coordinate (corner3) at ($(corner2) + (rl) + (al) + (hrow) + (al)$); 
    \coordinate (corner4) at ($(corner3) + (rl) + (al) + (hrow) + (al)$); 
    \newcommand\labelyshift{-0.06}; 
    \newcommand\labelxshift{0}; 
    \newcommand\labelyshifttwo{0.8}; 
    \newcommand\labelxshifttwo{0}; 
    \filldraw[mybox](corner1) rectangle ($(corner1) + (rh) + (rl)$) node[color=black, midway, above, shift={(\labelxshifttwo,\labelyshifttwo)}] {\parbox{5cm}{\centering (a) \\ General \\ three-wave mixing}} node[color=black, midway] {SFG};
    \draw[myarrow] ($(corner1) + 0.8*(rh) - (al)$) -- ($(corner1) + 0.8*(rh)$) node[color=black, midway, above, shift={(\labelxshift,\labelyshift)}] {$\omega_1$};
    \draw[myarrow] ($(corner1) + 0.2*(rh) - (al)$) -- ($(corner1) + 0.2*(rh)$) node[color=black, midway, above, shift={(\labelxshift,\labelyshift)}] {$\omega_2$};
    \draw[myarrow] ($(corner1) + 0.5*(rh) + (rl)$) -- ($(corner1) + 0.5*(rh) + (rl) + (al)$) node[color=black, midway, above, shift={(\labelxshift,\labelyshift)}] {$\omega_+$};
    \filldraw[mybox] ($(corner1) - (vcol) - (rh)$) rectangle ($(corner1) - (vcol) - (rh) + (rh) + (rl)$) node[color=black, midway] {DFG};
    \draw[myarrow] ($(corner1) - (vcol) - (rh) + 0.5*(rh) - (al)$) -- ($(corner1) - (vcol) - (rh) + 0.5*(rh)$) node[color=black, midway, above, shift={(\labelxshift,\labelyshift)}] {$\omega_1$};
    \draw[myarrow] ($(corner1) - (vcol) - (rh) + 0.8*(rh) + (rl)$) -- ($(corner1) - (vcol) - (rh) + 0.8*(rh) + (rl) + (al)$) node[color=black, midway, above, shift={(\labelxshift,\labelyshift)}] {$\omega_2$};
    \draw[myarrow] ($(corner1) - (vcol) - (rh) + 0.2*(rh) + (rl)$) -- ($(corner1) - (vcol) - (rh) + 0.2*(rh) + (rl) + (al)$) node[color=black, midway, above, shift={(\labelxshift,-0.09)}] {$\omega_-$};
    \filldraw[mybox](corner2) rectangle ($(corner2) + (rh) + (rl)$) node[color=black, midway, above, shift={(\labelxshifttwo,\labelyshifttwo)}] {\parbox{5cm}{\centering (b) \\ Degenerate \\ three-wave mixing}} node[color=black, midway] {\parbox{5cm}{\centering SHG \\ (UC)}};
    \draw[myarrow] ($(corner2) + 0.8*(rh) - (al)$) -- ($(corner2) + 0.8*(rh)$) node[color=black, midway, above, shift={(\labelxshift,\labelyshift)}] {$\omega$};
    \draw[myarrow] ($(corner2) + 0.2*(rh) - (al)$) -- ($(corner2) + 0.2*(rh)$) node[color=black, midway, above, shift={(\labelxshift,\labelyshift)}] {$\omega$};
    \draw[myarrow] ($(corner2) + 0.5*(rh) + (rl)$) -- ($(corner2) + 0.5*(rh) + (rl) + (al)$) node[color=black, midway, above, shift={(\labelxshift,\labelyshift)}] {$2\omega$};
    \filldraw[mybox] ($(corner2) - (vcol) - (rh)$) rectangle ($(corner2) - (vcol) - (rh) + (rh) + (rl)$) node[color=black, midway] {\parbox{5cm}{\centering SSHG \\ (DC)}};
    \draw[myarrow] ($(corner2) - (vcol) - (rh) + 0.5*(rh) - (al)$) -- ($(corner2) - (vcol) - (rh) + 0.5*(rh)$) node[color=black, midway, above, shift={(\labelxshift,\labelyshift)}] {$2\omega$};
    \draw[myarrow] ($(corner2) - (vcol) - (rh) + 0.8*(rh) + (rl)$) -- ($(corner2) - (vcol) - (rh) + 0.8*(rh) + (rl) + (al)$) node[color=black, midway, above, shift={(\labelxshift,\labelyshift)}] {$\omega$};
    \draw[myarrow] ($(corner2) - (vcol) - (rh) + 0.2*(rh) + (rl)$) -- ($(corner2) - (vcol) - (rh) + 0.2*(rh) + (rl) + (al)$) node[color=black, midway, above, shift={(\labelxshift,\labelyshift)}] {$\omega$};
     \filldraw[mybox](corner3) rectangle ($(corner3) + (rh) + (rl)$) node[color=black, midway, above, shift={(\labelxshifttwo,\labelyshifttwo)}] {\parbox{5cm}{\centering (c) \\ Spontaneous \\ Raman scattering}} node[color=black, midway] {S};
    \draw[myarrow] ($(corner3) + 0.5*(rh) - (al)$) -- ($(corner3) + 0.5*(rh)$) node[color=black, midway, above, shift={(\labelxshift,\labelyshift)}] {$\omega_L$};
    \draw[myarrow] ($(corner3) + 0.8*(rh) + (rl)$) -- ($(corner3) + 0.8*(rh) + (rl) + (al)$) node[color=black, midway, above, shift={(\labelxshift,\labelyshift)}] {$\omega_S$};
    \draw[mysquigglyarrow] ($(corner3) + 0.2*(rh) + (rl)$) -- ($(corner3) - 0.2*(rh) + (rl) + (al)$) node[color=black, midway, above, shift={(0,0.03)}] {$\omega_V$};
    \filldraw[mybox] ($(corner3) - (vcol) - (rh)$) rectangle ($(corner3) - (vcol) - (rh) + (rh) + (rl)$) node[color=black, midway] {AS};
    \draw[myarrow] ($(corner3) - (vcol) - (rh) + 0.8*(rh) - (al)$) -- ($(corner3) - (vcol) - (rh) + 0.8*(rh)$) node[color=black, midway, above, shift={(\labelxshift,\labelyshift)}] {$\omega_L$};
    \draw[mysquigglyarrow] ($(corner3) - (vcol) - (rh) - 0.2*(rh) - (al)$) -- ($(corner3) - (vcol) - (rh) + 0.2*(rh)$) node[color=black, midway, above, shift={(0,0.03)}] {$\omega_V$};
    \draw[myarrow] ($(corner3) - (vcol) - (rh) + 0.5*(rh) + (rl)$) -- ($(corner3) - (vcol) - (rh) + 0.5*(rh) + (rl) + (al)$) node[color=black, midway, above, shift={(\labelxshift,\labelyshift)}] {$\omega_A$};    
    \filldraw[mybox](corner4) rectangle ($(corner4) + (rh) + (rl)$) node[color=black, midway, above, shift={(\labelxshifttwo,\labelyshifttwo)}] {\parbox{5cm}{\centering (d) \\ Stimulated \\ Raman scattering}} node[color=black, midway] {SS};
    \draw[myarrow] ($(corner4) + 0.9*(rh) - (al)$) -- ($(corner4) + 0.9*(rh)$) node[color=black, midway, above, shift={(\labelxshift,\labelyshift)}] {$\omega_L$};
    \draw[myarrow] ($(corner4) + 0.48*(rh) - (al)$) -- ($(corner4) + 0.48*(rh)$);
    \draw[myarrow] ($(corner4) + 0.3*(rh) - (al)$) -- ($(corner4) + 0.3*(rh)$);
    \draw[myarrow] ($(corner4) + 0.12*(rh) - (al)$) -- ($(corner4) + 0.12*(rh)$) node[color=black, midway, below, shift={(0,0)}] {$n\omega_S$};            
    \draw[myarrow] ($(corner4) + 0.95*(rh) + (rl)$) -- ($(corner4) + 0.95*(rh) + (rl) + (al)$) node[color=black, midway, above, shift={(0.1,\labelyshift)}] {$(n+1)\omega_S$};
    \draw[myarrow] ($(corner4) + 0.77*(rh) + (rl)$) -- ($(corner4) + 0.77*(rh) + (rl) + (al)$);
    \draw[myarrow] ($(corner4) + 0.59*(rh) + (rl)$) -- ($(corner4) + 0.59*(rh) + (rl) + (al)$);
    \draw[myarrow] ($(corner4) + 0.41*(rh) + (rl)$) -- ($(corner4) + 0.41*(rh) + (rl) + (al)$);        
    \draw[mysquigglyarrow] ($(corner4) + 0.05*(rh) + (rl)$) -- ($(corner4) - 0.35*(rh) + (rl) + (al)$) node[color=black, midway, above, shift={(0,0.03)}] {$\omega_V$};
    \filldraw[mybox]($(corner4) - (vcol) - (rh)$) rectangle ($(corner4) - (vcol) - (rh) + (rh) + (rl)$) node[color=black, midway] {SAS};
    \draw[myarrow] ($(corner4) - (vcol) - (rh) + 0.97*(rh) - (al)$) -- ($(corner4) - (vcol) - (rh) + 0.97*(rh)$) node[color=black, midway, above, shift={(\labelxshift,\labelyshift)}] {$\omega_L$};
    \draw[myarrow] ($(corner4) - (vcol) - (rh) + 0.68*(rh) - (al)$) -- ($(corner4) - (vcol) - (rh) + 0.68*(rh)$);
    \draw[myarrow] ($(corner4) - (vcol) - (rh) + 0.5*(rh) - (al)$) -- ($(corner4) - (vcol) - (rh) + 0.5*(rh)$);
    \draw[myarrow] ($(corner4) - (vcol) - (rh) + 0.32*(rh) - (al)$) -- ($(corner4) - (vcol) - (rh) + 0.32*(rh)$) node[color=black, midway, below, shift={(-0.1,0)}] {$n\omega_A$};
    \draw[mysquigglyarrow] ($(corner4) - (vcol) - (rh) - 0.6*(rh) - (al)$) -- ($(corner4)  - (vcol) - (rh) + 0.03*(rh)$) node[color=black, midway, below, shift={(0.1,-0.03)}] {$\omega_V$};                
    \draw[myarrow] ($(corner4) - (vcol) - (rh) + 0.77*(rh) + (rl)$) -- ($(corner4) - (vcol) - (rh) + 0.77*(rh) + (rl) + (al)$) node[color=black, midway, above, shift={(0.13,-0.03)}] {$(n+1)\omega_A$};
    \draw[myarrow] ($(corner4) - (vcol) - (rh) + 0.59*(rh) + (rl)$) -- ($(corner4) - (vcol) - (rh) + 0.59*(rh) + (rl) + (al)$);
    \draw[myarrow] ($(corner4) - (vcol) - (rh) + 0.41*(rh) + (rl)$) -- ($(corner4) - (vcol) - (rh) + 0.41*(rh) + (rl) + (al)$);
    \draw[myarrow] ($(corner4) - (vcol) - (rh) + 0.23*(rh) + (rl)$) -- ($(corner4) - (vcol) - (rh) + 0.23*(rh) + (rl) + (al)$);        
    \end{tikzpicture}
  }
}
\caption{Schematic representations (Feynman-like diagrams) of three-wave-mixing processes. (a) The two general types of three-wave mixing are sum-frequency generation (SFG, above) and difference-frequency generation (DFG, below). (b) When two of the involved frequencies are degenerate, we have either second-harmonic generation [SHG, or upconversion (UC), above] or second-subharmonic generation [SSHG, or downconversion (DC), below]. (c) Another special case is spontaneous Raman scattering, where a small part of the total energy is carried by a phonon (pictured as a wavy arrow), which is either outgoing [Stokes Raman scattering (S), above] or incoming [anti-Stokes Raman scattering (AS), also known as sideband cooling, below]. (d) In stimulated Raman scattering, the rate is increased due to the presence of $n$ additional photons of the same frequency as the outgoing one. Stimulated Stokes Raman scattering (SS) is shown above and stimulated anti-Stokes Raman scattering (SAS) is shown below. \label{fig:3WaveMixing}}
\end{figure*}

In this section, we look at three-wave mixing, starting with a general description of sum- and difference-frequency generation and then treating special cases such as upconversion, downconversion, and Raman scattering; see \figref{fig:3WaveMixing} for an overview. We provide deterministic analogues based on the generalized quantum Rabi model for each case.

\subsection{General description: Generation of sum- and difference-frequency fields}
\label{sec:3WaveMixingGeneral}

\subsubsection{Nonlinear optics}
\label{sec:3WaveMixingGeneralNonlinear}

The creation and annihilation of a photon with sum frequency $\omega_+$ can be described in the Fock representation as $\ket{n_1,n_2,n_+}\rightarrow \ket{n_1-1,n_2-1,n_+ +1},$ and $\ket{n_1,n_2,n_+}\rightarrow \ket{n_1+1,n_2+1,n_+-1}$, respectively; see also \figref{fig:3WaveMixing}(a). The interaction Hamiltonian $\hat H^{(+)}_{\rm int}$ for this sum-frequency generation can be given by
\be
\hat H^{(+)}_{\rm int} = g \hat a_1 \hat a_2 \hat a_+^\dag + g^* \hat a_1^\dag \hat a_2^\dag \hat a_+,
\label{TWMsum}
\ee
in terms of the annihilation ($\hat a_j$) and creation ($\hat a_j^\dag$) operators for the input modes (for $j=1,2$) and the output sum-frequency mode (for $j=+$), and the three-mode complex coupling constant $g$.

Analogously, the creation and annihilation of a photon with difference frequency $\omega_-$ can be described in the Fock representation as $\ket{n_1, n_2, n_-} \rightarrow \ket{n_1-1, n_2+1, n_- +1}$ and $\ket{n_1, n_2, n_-} \rightarrow \ket{n_1+1,n_2-1,n_--1}$; see also \figref{fig:3WaveMixing}(a). The interaction Hamiltonian $\hat H^{(-)}_{\rm int}$ describing this process can be given by
\be
\hat H^{(-)}_{\rm int} = g \hat a_1 \hat a_2^\dag \hat a_-^\dag + g^* \hat a_1^\dag \hat a_2 \hat a_-,
\label{TWMdif}
\ee
using the same notation as in \eqref{TWMsum} except that the subscript `$+$', corresponding to the sum-frequency mode, is
replaced by `$-$' for the difference-frequency mode.

The energy conservation principle implies that $\omega_+ = \omega_1 + \omega_2$ and $\omega_- = \omega_1 - \omega_2$, and the momentum conservation principle implies that ${\bf k}_+ = {\bf k}_1 + {\bf k}_2$ and ${\bf k}_- = {\bf k}_1 - {\bf k}_2$ for the wave vectors ${\bf k}_j$.

\subsubsection{Analogous processes}
\label{sec:3WaveMixingGeneralDeterministic}

There are several possible setups that can realize analogues of sum- and difference-frequency generation deterministically. One such setup would be three resonators coupled to a single qubit using the generalized Rabi interaction in \eqref{eq:HintGenRabi}. If the resonator frequencies satisfy $\omega_a + \omega_b \approx \omega_c$, the two states $\ket{1,1,0,g}$ and $\ket{0,0,1,g}$ become resonant. Here, and in all the following discussions of deterministic processes, kets list photon excitation numbers, starting from the resonator with frequency $\omega_a$, followed by qubit state(s) ($g$ for ground state, $e$ for excited state). The transition $\ket{1,1,0,g} \to \ket{0,0,1,g}$ then corresponds to sum-frequency generation [$a=1$, $b=2$, $c=+$ makes the connection explicit with \secref{sec:3WaveMixingGeneralNonlinear} and \figref{fig:3WaveMixing}(a)] and the transition $\ket{0,0,1,g} \to \ket{1,1,0,g}$ corresponds to difference-frequency generation ($a=2$, $b=-$, $c=1$).

The transition $\ket{1,1,0,g} \to \ket{0,0,1,g}$ is enabled by paths with several intermediate virtual transitions. One example of such a path is $\ket{1,1,0,g} \xrightarrow{\hat b \sp} \ket{1,0,0,e} \xrightarrow{\hat c^\dag \sm} \ket{1,0,1,g} \xrightarrow{\hat a \sz} \ket{0,0,1,g}$, where the terms from \eqref{eq:HintGenRabi} that generate the transitions are shown above the arrows. Note that the last transition changes the number of excitations in the system by one, which is only possible when the interaction is given by the generalized quantum Rabi Hamiltonian of \eqref{eq:HintGenRabi}. The last transition is also an example of how a virtual photon is annihilated in the process. For the transition in the opposite direction (difference-frequency generation), a virtual photon is created instead.

By adiabatic elimination, or suitable unitary transformations combined with perturbation expansion in $g$ over some frequency, it can be shown that these virtual transitions combine to give an effective interaction Hamiltonian
\be
\hat H_{\rm int}^{\rm eff} = g_{\rm eff} \ketbra{0,0,1,g}{1,1,0,g} +  {\rm H.c.},
\ee
where the effective coupling $g_{\rm eff}$, in general, becomes weaker the more intermediate steps are needed. Later in this section, we will provide examples with detailed diagrams of the virtual transitions and calculations of the effective coupling.

We note that, if at least one of the excitations in the three-wave mixing can be hosted in a qubit, other setups become possible. Both a single resonator coupled to two qubits and two resonators coupled to a single qubit could be used to implement the processes in \figref{fig:3WaveMixing}(a). In particular, Ref.~\cite{Garziano2016} analyzed the former case with $\omega_a \approx \omega_{q1} + \omega_{q2}$, showing an effective coupling between $\ket{1,g,g}$ and $\ket{0,e,e}$. In the latter case, the effective coupling of interest would be that between the states $\ket{1,1,g}$ and $\ket{0,0,e}$ when $\omega_a + \omega_b \approx \omega_q$.

\subsection{Degenerate three-wave mixing: Second-harmonic and second-subharmonic generation}
\label{sec:Degenerate3WaveMixing}

\subsubsection{Nonlinear optics}
\label{sec:Degenerate3WaveMixingNonlinear}

Let us assume the degenerate process of three-wave mixing for which $\hat a_1=\hat a_2 \equiv \hat a$ and $\omega_1 = \omega_2 \equiv \omega$. The energy conservation principle implies $\omega_+ = 2 \omega$. The processes of the creation and annihilation of photons can be written as $\ket{n,n_+} \to \ket{n-2, n_+ +1}$ and $\ket{n, n_+} \to \ket{n+2,n_+ -1}$; see also \figref{fig:3WaveMixing}(b). The interaction Hamiltonian reads as
\be
\hat H_{\rm int} = g \hat a^2 \hat a_+^\dag + g^* \hat a^{\dag 2} \hat a_+.
\label{eq:UpDownConversionHint}
\ee
For second-harmonic generation (also referred to as upconversion), one can assume that the initial pure state is $\ket{\psi(t_0)}=\sum_{n=0}^{\infty}c_n \ket{n,0}$. For second-subharmonic generation (also called downconversion), one can assume that the initial pure state is $\ket{\psi(t_0)}=\sum_{n_+=0}^{\infty}c_{n_+} \ket{0,n_+}$. Here, $c_n$ and $c_{n_+}$ denote arbitrary complex superposition amplitudes satisfying the normalization conditions. It is seen that our description of second-subharmonic generation can be the same as that for second-harmonic generation except with a different initial state.

\subsubsection{Analogous processes}
\label{sec:Degenerate3WaveMixingDeterministic}

\paragraph{Two resonators}
\label{sec:Degenerate3WaveMixingDeterministicTwoResonators}

\begin{figure*}
\centerline{
  \resizebox{\linewidth}{!}{
    \begin{tikzpicture}[
      scale=1,
      level/.style={thick},
      virtual/.style={thick,dashed},
      ztrans/.style={thick,->,shorten >=0.1cm,shorten <=0.1cm,>=stealth,densely dashed,color=red},
      nrtrans/.style={thick,->,shorten >=0.1cm,shorten <=0.1cm,>=stealth,densely dashed,color=blue},
      rtrans/.style={thick,->,shorten >=0.1cm,shorten <=0.1cm,>=stealth,color=blue},
      strans/.style={rtrans, snake=snake, line after snake=1.5mm},
      rtrans2/.style={thick,->,shorten >=0.1cm,shorten <=0.1cm,>=stealth,color=green!35!black},
      strans2/.style={rtrans2, snake=snake, line after snake=1.5mm},
      classical/.style={thin,double,<->,shorten >=4pt,shorten <=4pt,>=stealth},
      mybox/.style={color=cyan!70, fill=cyan!20, very thick},
      myarrow/.style={color=green!35!black, thick, ->, shorten >=0.05cm, shorten <=0.05cm, >=latex},
      mysquigglyarrow/.style={color=green!35!black, thick, ->, shorten >=0.05cm, shorten <=0.05cm, >=latex, snake=snake, line after snake=2mm}
    ]
    \newcommand\labelyshift{-0.05}; 
    \newcommand\labelxshift{0}; 
    \newcommand\labelyshifttwo{0.8}; 
    \newcommand\labelxshifttwo{0}; 
    \coordinate (corner) at (0,0); 
    \coordinate (rh) at (0,1); 
    \coordinate (rl) at (5cm, 0cm); 
    \coordinate (ccenter) at ($(corner) + 0.5*(rl) + (0cm, 3cm) + 1.5*(rh)$); 
    \coordinate (ccenter2) at ($(corner) + 0.5*(rl) - (0cm, 2cm) - 0.5*(rh)$); 
    \coordinate (al) at (1.5cm, 0cm); 
    \filldraw[mybox](corner) rectangle ($(corner) + (rh) + (rl)$) node[color=black, midway] {\parbox{5cm}{\centering SSHG \\ (DC)}};
    \draw[color=cyan!70, thick, dashed] (ccenter) circle (3cm);
    \draw[color=cyan!70, thick, densely dashed] ($(corner) + (rh) + 0.3*(rl)$) --  ($(corner) + (rh) + (0.5,1.2)$);
    \draw[color=cyan!70, thick, densely dashed] ($(corner) + (rh) + 0.7*(rl)$) --  ($(corner) + (rh) + (rl) + (-0.5,1.2)$);
    \draw[color=cyan!70, thick, dashed] ($(ccenter) - (6.5,0)$) circle (3cm);
    \draw[color=cyan!70, thick, densely dashed] ($(corner) + (rh)$) --  ($(corner) + (rh) + (-4.3,0.5)$);
    \draw[color=cyan!70, thick, densely dashed] ($(corner) + (rh) + 0.05*(rl)$) --  ($(corner) + (rh) + (-0.95,3.6)$);
    \draw[color=cyan!70, thick, dashed] ($(ccenter) + (6.5,0)$) circle (3cm);
    \draw[color=cyan!70, thick, densely dashed] ($(corner) + (rh) + (rl)$) --  ($(corner) + (rh) + (rl) + (4.3,0.5)$);
    \draw[color=cyan!70, thick, densely dashed] ($(corner) + (rh) + 0.95*(rl)$) --  ($(corner) + (rh) + (rl) + (0.95,3.6)$);
    \draw[color=cyan!70, thick, dashed] (ccenter2) circle (2cm);
    \draw[color=cyan!70, thick, densely dashed] ($(corner) + 0.3*(rl)$) --  ($(corner) + (0.85,-1.3)$);
    \draw[color=cyan!70, thick, densely dashed] ($(corner) + 0.7*(rl)$) --  ($(corner) + (rl) + (-0.85,-1.3)$);
    \draw[myarrow] ($(corner) + 0.5*(rh) - (al)$) -- ($(corner) + 0.5*(rh)$) node[color=black, midway, above, shift={(\labelxshift,\labelyshift)}] {$2\omega$};
    \draw[myarrow] ($(corner) + 0.75*(rh) + (rl)$) -- ($(corner) + 0.75*(rh) + (rl) + (al)$) node[color=black, midway, above, shift={(\labelxshift,\labelyshift)}] {$\omega$};
    \draw[myarrow] ($(corner) + 0.25*(rh) + (rl)$) -- ($(corner) + 0.25*(rh) + (rl) + (al)$) node[color=black, midway, above, shift={(\labelxshift,\labelyshift)}] {$\omega$};
    \coordinate (v) at (0cm, 0.4cm); 
    \coordinate (l) at (1.2cm, 0cm); 
    \coordinate (h) at (0.2cm, 0cm); 
    \coordinate (10g) at ($(ccenter) + (-9.2,0) - 1.4*(v) $); 
    \coordinate (11e) at ($(10g) + (l) + (h) + 5*(v)$);
    \coordinate (11g) at ($(10g) + (l) + (h) + 2*(v)$);
    \coordinate (00e) at ($(10g) + (l) + (h) - 1*(v)$);
    \coordinate (00g) at ($(10g) + (l) + (h) - 4*(v)$);
    \coordinate (12e) at ($(10g) + 2*(l) + 2*(h) + 7*(v)$);
    \coordinate (12g) at ($(10g) + 2*(l) + 2*(h) + 4*(v)$);
    \coordinate (01e) at ($(10g) + 2*(l) + 2*(h) + 1*(v)$);
    \coordinate (01g) at ($(10g) + 2*(l) + 2*(h) - 2*(v)$);
    \coordinate (02g) at ($(10g) + 3*(l) + 3*(h)$);
    \draw[level] (10g) -- ($(10g) + (l)$) node[midway,below,xshift=-0.1cm] {\footnotesize{$\ket{1,0,g}$}};
    \draw[level] (11e) -- ($(11e) + (l)$) node[midway,above] {\footnotesize{$\ket{1,1,e}$}};
    \draw[level] (11g) -- ($(11g) + (l)$) node[midway,above] {\footnotesize{$\ket{1,1,g}$}};
    \draw[level] (00e) -- ($(00e) + (l)$) node[midway,below] {\footnotesize{$\ket{0,0,e}$}};
    \draw[level] (00g) -- ($(00g) + (l)$) node[midway,below] {\footnotesize{$\ket{0,0,g}$}};
    \draw[level] (12e) -- ($(12e) + (l)$) node[midway,above] {\footnotesize{$\ket{1,2,e}$}};
    \draw[level] (12g) -- ($(12g) + (l)$) node[midway,above] {\footnotesize{$\ket{1,2,g}$}};
    \draw[level] (01e) -- ($(01e) + (l)$) node[midway,above] {\footnotesize{$\ket{0,1,e}$}};
    \draw[level] (01g) -- ($(01g) + (l)$) node[midway,below] {\footnotesize{$\ket{0,1,g}$}};
    \draw[level] (02g) -- ($(02g) + (l)$) node[midway,below,xshift=-0.1cm] {\footnotesize{$\quad\:\:\ket{0,2,g}$}};
    \draw[nrtrans] ($(10g) + (0.9cm,0cm)$) -- ($(11e) + (0.3cm,0cm)$);
    \draw[ztrans] ($(10g) + (1.0cm,0cm)$) -- ($(11g) + (0.3cm,0cm)$);
    \draw[rtrans] ($(10g) + (1.0cm,0cm)$) -- ($(00e) + (0.3cm,0cm)$);
    \draw[ztrans] ($(10g) + (0.9cm,0cm)$) -- ($(00g) + (0.3cm,0cm)$);
    \draw[ztrans] ($(11e) + (1.0cm,0cm)$) -- ($(12e) + (0.2cm,0cm)$);
    \draw[rtrans] ($(11e) + (1.0cm,0cm)$) -- ($(12g) + (0.2cm,0cm)$);
    \draw[ztrans] ($(11e) + (1.05cm,0cm)$) -- ($(01e) + (0.15cm,0cm)$);
    \draw[nrtrans] ($(11e) + (1.0cm,0cm)$) -- ($(01g) + (0.25cm,0cm)$);
    \draw[nrtrans] ($(11g) + (1.0cm,0cm)$) -- ($(12e) + (0.3cm,0cm)$);
    \draw[ztrans] ($(11g) + (1.0cm,0cm)$) -- ($(12g) + (0.2cm,0cm)$);
    \draw[rtrans] ($(11g) + (1.0cm,0cm)$) -- ($(01e) + (0.1cm,0cm)$);
    \draw[ztrans] ($(11g) + (1.0cm,0cm)$) -- ($(01g) + (0.1cm,0cm)$);
    \draw[ztrans] ($(00e) + (1.0cm,0cm)$) -- ($(01e) + (0.3cm,0cm)$);
    \draw[rtrans] ($(00e) + (1.0cm,0cm)$) -- ($(01g) + (0.05cm,0cm)$);
    \draw[nrtrans] ($(00g) + (1.0cm,0cm)$) -- ($(01e) + (0.4cm,0cm)$);
    \draw[ztrans] ($(00g) + (1.0cm,0cm)$) -- ($(01g) + (0.2cm,0cm)$);
    \draw[nrtrans] ($(12e) + (1.0cm,0cm)$) -- ($(02g) + (0.6cm,0cm)$);
    \draw[ztrans] ($(12g) + (1.0cm,0cm)$) -- ($(02g) + (0.4cm,0cm)$);
    \draw[rtrans] ($(01e) + (1.0cm,0cm)$) -- ($(02g) + (0.2cm,0cm)$);
    \draw[ztrans] ($(01g) + (1.0cm,0cm)$) -- ($(02g) + (0.2cm,0cm)$);
    %
    \coordinate (0e) at ($(ccenter) + (-2.0cm,0cm)$); 
    \coordinate (1e) at ($(0e) + (l) + (h) + 3*(v)$);
    \coordinate (1g) at ($(0e) + (l) + (h) - 3*(v)$);
    \coordinate (2g) at ($(0e) + 2*(l) + 2*(h)$);
    \draw[level] (0e) -- ($(0e) + (l)$) node[midway,below,xshift=-0.1cm] {\footnotesize{$\ket{0,e}$}};
    \draw[level] (1e) -- ($(1e) + (l)$) node[midway,above] {\footnotesize{$\ket{1,e}$}};
    \draw[level] (1g) -- ($(1g) + (l)$) node[midway,below] {\footnotesize{$\ket{1,g}$}};
    \draw[level] (2g) -- ($(2g) + (l)$) node[midway,below,xshift=0.1cm] {\footnotesize{$\ket{2,g}$}};
    \draw[ztrans] ($(0e) + (0.9cm,0cm)$) -- ($(1e) + (0.3cm,0cm)$);
    \draw[rtrans] ($(0e) + (0.9cm,0cm)$) -- ($(1g) + (0.3cm,0cm)$);
    \draw[rtrans] ($(1e) + (0.9cm,0cm)$) -- ($(2g) + (0.3cm,0cm)$);
    \draw[ztrans] ($(1g) + (0.9cm,0cm)$) -- ($(2g) + (0.3cm,0cm)$);
    \coordinate (1gg) at ($(ccenter) + (3.8,-0.15) - 1*(v)$); 
    \coordinate (2eg) at ($(1gg) + (l) + (h) + 6*(v)$);
    \coordinate (2gg) at ($(1gg) + (l) + (h) + 4*(v)$);
    \coordinate (0eg) at ($(1gg) + (l) + (h) - 2*(v)$);
    \coordinate (0gg) at ($(1gg) + (l) + (h) - 4*(v)$);
    \coordinate (1ee) at ($(1gg) + 2*(l) + 2*(h) + 4*(v)$);
    \coordinate (1eg) at ($(1gg) + 2*(l) + 2*(h) + 2*(v)$);
    \coordinate (0ee) at ($(1gg) + 3*(l) + 3*(h)$);
    \draw[level] (1gg) -- ($(1gg) + (l)$) node[midway,below,xshift=-0.15cm] {\footnotesize{$\ket{1,g,g}$}};
    \draw[level] (2eg) -- ($(2eg) + (l)$) node[midway,above,xshift=0.2cm,yshift=-0.1cm] {\footnotesize{$\frac{1}{\sqrt{2}}\left(\ket{2,e,g} + \ket{2,g,e} \right)$}};
    \draw[level] (2gg) -- ($(2gg) + (l)$) node[midway,above] {\footnotesize{$\ket{2,g,g}$}};
    \draw[level] (0eg) -- ($(0eg) + (l)$) node[midway,below,xshift=1.6cm,yshift=0.05cm] {\footnotesize{$\frac{1}{\sqrt{2}}\left(\ket{0,e,g} + \ket{0,g,e} \right)$}};
    \draw[level] (0gg) -- ($(0gg) + (l)$) node[midway,below] {\footnotesize{$\ket{0,g,g}$}};
    \draw[level] (1ee) -- ($(1ee) + (l)$) node[midway,above] {\footnotesize{$\ket{1,e,e}$}};
    \draw[level] (1eg) -- ($(1eg) + (l)$) node[midway,below,xshift=1.45cm,yshift=0.5cm] {\parbox{1.5cm}{\centering \footnotesize{$\frac{1}{\sqrt{2}}(\ket{1,e,g}$ \\ $\quad + \ket{1,g,e})$}}};
    \draw[level] (0ee) -- ($(0ee) + (l)$) node[midway,below,xshift=0.1cm] {\footnotesize{$\ket{0,e,e}$}};
    \draw[nrtrans] ($(1gg) + (0.85cm,0cm)$) -- ($(2eg) + (0.15cm,0cm)$);
    \draw[ztrans] ($(1gg) + (0.9cm,0cm)$) -- ($(2gg) + (0.3cm,0cm)$);
    \draw[rtrans] ($(1gg) + (0.9cm,0cm)$) -- ($(0eg) + (0.3cm,0cm)$);
    \draw[ztrans] ($(1gg) + (0.85cm,0cm)$) -- ($(0gg) + (0.25cm,0cm)$);
    \draw[rtrans] ($(2eg) + (1.0cm,0cm)$) -- ($(1ee) + (0.15cm,0cm)$);
    \draw[rtrans] ($(2gg) + (0.9cm,0cm)$) -- ($(1eg) + (0.3cm,0cm)$);
    \draw[nrtrans] ($(0eg) + (0.45cm,0cm)$) -- ($(1ee) + (0.15cm,0cm)$);
    \draw[nrtrans] ($(0gg) + (0.4cm,0cm)$) -- ($(1eg) + (0.1cm,0cm)$);
    \draw[ztrans] ($(1ee) + (0.9cm,0cm)$) -- ($(0ee) + (0.2cm,0cm)$);
    \draw[rtrans] ($(1eg) + (0.8cm,0cm)$) -- ($(0ee) + (0.05cm,0cm)$);
    %
    \coordinate (ground) at ($(ccenter2) + (-1cm,-1.5cm)$); 
    \coordinate (htot) at (0cm, 3cm); 
    \coordinate (ltot) at (2cm, 0cm); 
    \coordinate (lphon) at (1cm, 0cm); 
    \draw[level] (ground) -- ($(ground) + (ltot)$);
    \draw[virtual] ($(ground) + (htot)$) -- ($(ground) + (htot) + (ltot)$);
    \draw[virtual] ($(ground) + 0.5*(htot) + (ltot) - (lphon)$) -- ($(ground) + 0.5*(htot) + (ltot)$);
    \draw[rtrans2] ($(ground) + (0.5cm,0cm)$) -- ($(ground) + (htot) + (0.5cm,0cm)$) node[color=black, midway, left] {$2\omega$};
    \draw[rtrans2] ($(ground) + (htot) + (1.5cm,0cm)$) -- ($(ground) + 0.5*(htot) + (1.5cm,0cm)$) node[color=black, midway, right] {$\omega$};
    \draw[rtrans2] ($(ground) + 0.5*(htot) + (1.5cm,0cm)$) -- ($(ground) + (1.5cm,0cm)$) node[color=black, midway, right] {$\omega$};
    \end{tikzpicture}
  }
}
\caption{Realizations of down- and upconversion. The upper left panel shows all virtual transitions that contribute to the downconversion process (second-subharmonic generation) $\ket{1,0,g} \to \ket{0,2,g}$ to lowest order. Blue solid arrows mark transitions that do \textit{not} change the number of excitations [these transitions are mediated by the terms in the JC model, \eqref{eq:HintJC}], blue dashed arrows correspond to transitions that change the number of excitations by \textit{two} [these transitions are mediated by the non-JC terms in the quantum Rabi model, \eqref{eq:HintRabi}], and red dashed arrows show transitions that change the number of excitations by \textit{one} [these transitions are mediated by the additional terms in the generalized Rabi model, \eqref{eq:HintGenRabi}]. We have set $\omega_a = 2\omega_b$ and $\omega_q = 1.5\omega_b$. Similarly, the upper middle panel shows all virtual transitions that contribute to the downconversion process $\ket{0,e} \to \ket{2,g}$ to lowest order. Here, we have set $\omega_a = 2\omega_b$ and $\omega_q = 1.5\omega_b$. The upper right panel shows all virtual transitions that contribute to the downconversion process $\ket{1,g,g} \to \ket{0,e,e}$ to lowest order. In this case, we have set $\omega_a = 2\omega_q$. The lower panel shows the generic level diagram for the process in nonlinear optics. Dashed horizontal lines denote virtual levels. If the directions of all arrows in the entire figure are reversed, upconversion (second-harmonic generation) is shown instead. \label{fig:SecondSubharmonicDetailed}}
\end{figure*}
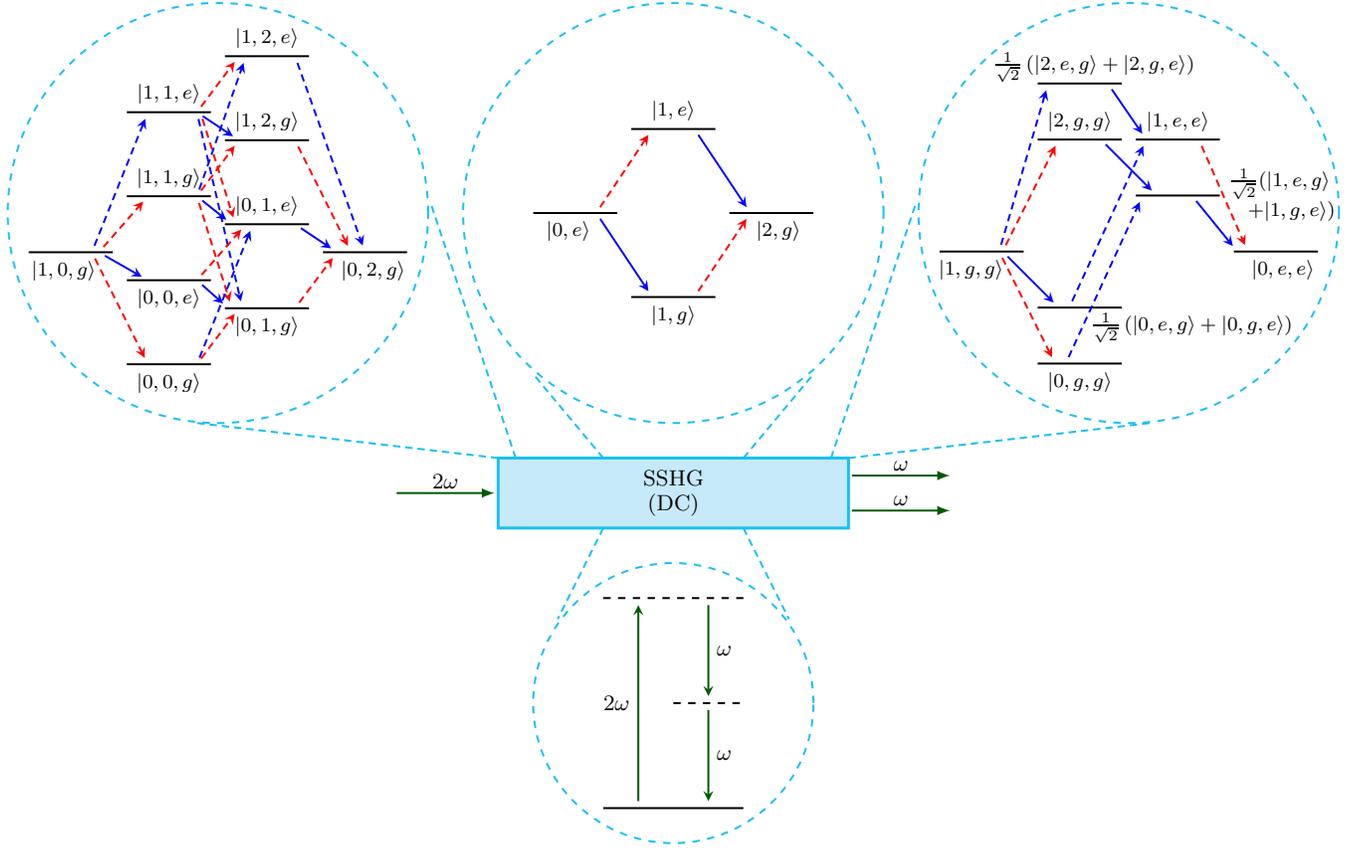

There are, again, several possible setups to realize analogues of up- and downconversion deterministically. In Ref.~\cite{Moon2005}, it was shown that an effective Hamiltonian like that of \eqref{eq:UpDownConversionHint} can be achieved with two resonator modes coupled to a qubit with the interaction given by \eqref{eq:HintGenRabi}. However, in that work some additional assumptions were made, since ultrastrong coupling had not yet been experimentally demonstrated at the time. With strong enough coupling, the virtual transitions shown in the upper left panel of \figref{fig:SecondSubharmonicDetailed} combine to achieve a robust effective coupling between the states $\ket{1,0,g}$ and $\ket{0,2,g}$, which results in both up- and downconversion. Note how virtual photons and qubit excitations are created or annihilated in all transitions marked with dashed arrows. 

To be precise, the full Hamiltonian of the system is here given by
\bea
\hat H &=& \omega_a \hat a^\dag \hat a + \omega_b \hat b^\dag \hat b + \omega_q \frac{\sz}{2} + \hat H_{\rm int},
\label{eq:H2res1qb} \\
\hat H_{\rm int} &=& \left[g_a \left( \hat a + \hat a^\dag \right) + g_b \left( \hat b + \hat b^\dag \right) \right]\nn\\
&&\times \left( \sx \cos\theta + \sz \sin\theta \right),
\label{eq:Hint2res1qbRabiGen}
\eea
and the effective interaction due to the virtual transitions shown in \figref{fig:SecondSubharmonicDetailed} becomes
\be
\hat H_{\rm int}^{\rm eff} = g_{\rm eff} \ketbra{1,0,g}{0,2,g} +  {\rm H.c.}
\label{eq:Heffint10g02g}
\ee
The effective coupling $g_{\rm eff}$ can be calculated with third-order perturbation theory. From \eqref{eq:PerturbationFormula}, we have
\be
g_{\rm eff} = \sum_{n,m} \frac{\brakket{f}{\hat H_{\rm int}}{n} \brakket{n}{\hat H_{\rm int}}{m} \brakket{m}{\hat H_{\rm int}}{i}}{\left(E_i - E_n \right) \left(E_i - E_m \right)}.
\ee
Looking at the upper left panel of \figref{fig:SecondSubharmonicDetailed}, we see that there are 12 paths contributing to the effective coupling between $\ket{i} = \ket{0,2,g}$ and $\ket{f} = \ket{1,0,g}$. Three of these paths consist solely of $\sz$-mediated transitions (dashed red arrows in the figure). Their contribution is
\be
\sqrt{2}g_a g_b^2 \sin^3 \theta \left( \frac{1}{\omega_a \Delta_{ba}} - \frac{1}{\omega_b \Delta_{ba}} - \frac{1}{2\omega_b^2} \right),
\ee
where we introduced the notation $\Delta_{nm} = \omega_n - \omega_m$. This contribution sums to zero on resonance ($\omega_a = 2\omega_b$). The contribution from the remaining 9 paths is (introducing the notation $\Omega_{nm} = \omega_n + \omega_m$)
\bea
&&\sqrt{2} g_a g_b^2 \sin \theta \cos^2 \theta \Big( \frac{1}{\left(\Delta_{ab} + \omega_q \right) \Omega_{aq}} - \frac{1}{\omega_a \left(\Delta_{ab} + \omega_q \right)} \nn\\
&&- \frac{1}{\left(\Delta_{ab} + \omega_q \right) \Delta_{bq}} + \frac{1}{\omega_b \left(\Delta_{ab} + \omega_q \right)} - \frac{1}{\Delta_{ab} \Omega_{aq}} + \frac{1}{\Delta_{ab}\Delta_{bq}} \nn\\
&& + \frac{1}{\left(2\omega_b - \omega_q \right) \Omega_{aq}} - \frac{1}{\omega_b \left(2\omega_b - \omega_q \right)} - \frac{1}{2\omega_b \Delta_{bq}} \Big),
\eea
which on resonance reduces to
\be
g_{\rm eff} = \frac{3\sqrt{2} g_a g_b^2 \omega_q^2 \sin (2\theta) \cos \theta}{4\omega_b^4 - 5\omega_b^2 \omega_q^2 + \omega_q^4}.
\ee
Since the transition paths in the upper left panel of \figref{fig:SecondSubharmonicDetailed} go via two intermediate levels, $g_{\rm eff}$ becomes on the order of $(g_j/\omega)^2$ weaker than $g_j$ ($j=a,b$). This expression is slightly more complicated than that derived in Ref.~\cite{Moon2005}, where unitary transformations were combined with perturbation expansions using the additional simplifying assumptions that $g_b \ll \abs{\omega_q - \omega_b} \ll \omega_a$.

\begin{figure}
\centering
\includegraphics[width=\linewidth]{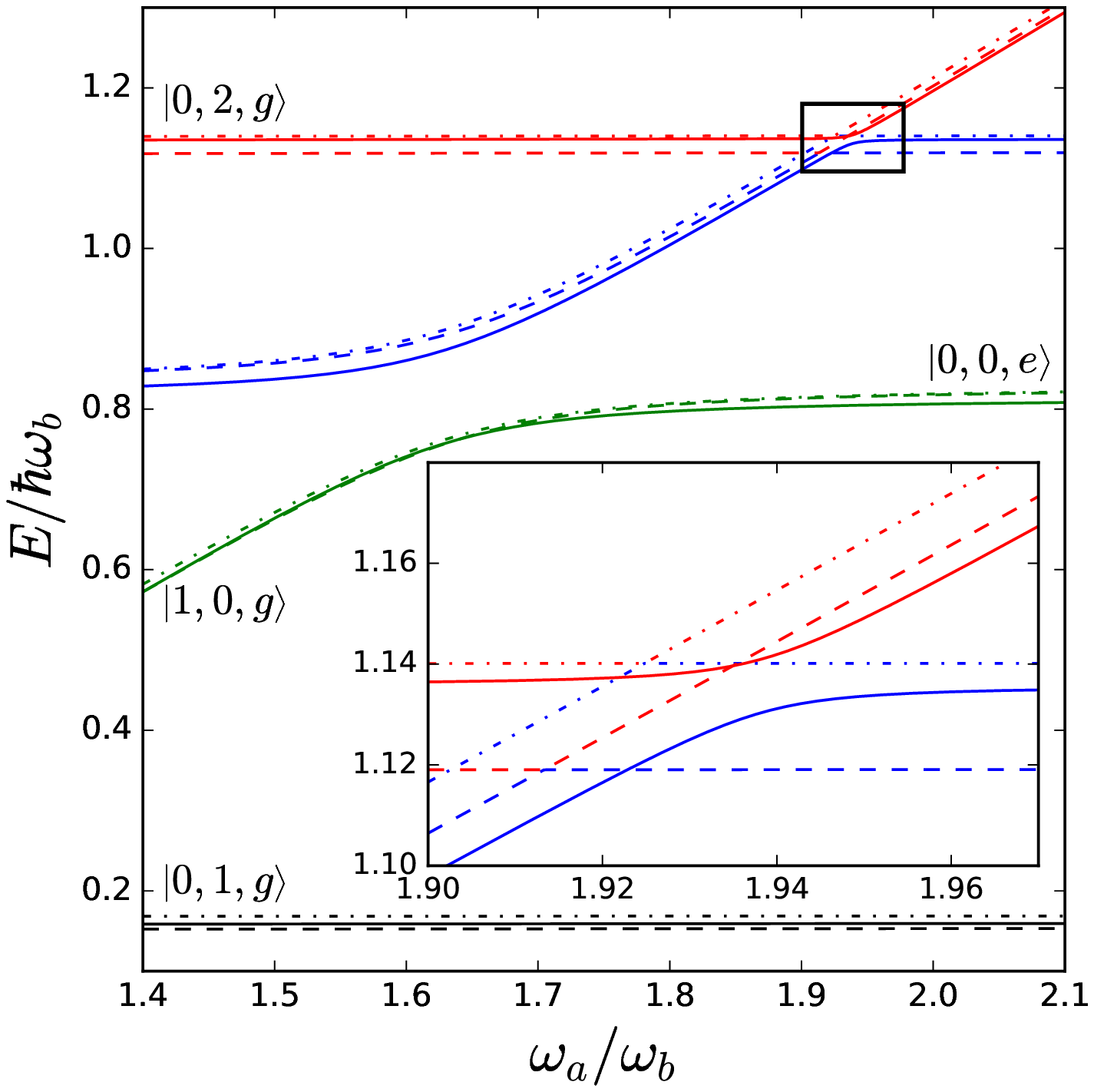}
\caption{Energy levels for two resonator modes coupled to a qubit via the JC [\eqref{eq:HintJC}, dashed-dotted lines], quantum Rabi [\eqref{eq:HintRabi}, dashed lines], and generalized quantum Rabi [\eqref{eq:HintGenRabi}, solid lines] interactions, as a function of the resonance frequency $\omega_a$ of the first resonator mode. The inset shows a zoom-in of the area marked by the black rectangle in the upper right corner. Parameters: $\omega_q = 1.6 \omega_b$, $g_a = 0.07 \omega_b$, $g_b = 2 g_a$, and $\theta = \pi/6$. \label{fig:10g02gEnergyLevels}}
\end{figure}

A further demonstration of the effective coupling in \eqref{eq:Heffint10g02g} is given in \figref{fig:10g02gEnergyLevels}, where we plot some of the energy levels in the system as a function of $\omega_a$ for the JC (dashed-dotted lines), Rabi (dashed lines), and generalized Rabi (solid lines) interactions. The inset shows a clear avoided crossing between $\ket{1,0,g}$ and $\ket{0,2,g}$; the splitting is set by $g_{\rm eff}$. The JC and Rabi interactions do not give rise to such an avoided crossing since they cannot change the excitation number by one. However, all three interactions give rise to an avoided crossing between $\ket{1,0,g}$ and $\ket{0,0,e}$ to the left in the figure, since those two states have the same number of excitations. 

\paragraph{Multiphoton Rabi oscillations}

An alternative implementation of up- and downconversion is multiphoton Rabi oscillations, illustrated in the upper middle panel of \figref{fig:SecondSubharmonicDetailed} and discussed in Ref.~\cite{Garziano2015}. In this case, virtual transitions induce an effective coupling (and, thus, Rabi oscillations) between the states $\ket{0,e}$ and $\ket{2,g}$ for a single resonator coupled to a single qubit with $\omega_q \approx 2\omega_a$. The transitions are mediated by the generalized Rabi Hamiltonian \eqref{eq:HintGenRabi} and give rise to an effective interaction
\be
\hat H_{\rm int}^{\rm eff} = g_{\rm eff} \ketbra{0,e}{2,g} +  {\rm H.c.}
\ee
The effective coupling is easily calculated with second-order perturbation theory. With $\ket{i} = \ket{2,g}$ and $\ket{f} = \ket{0,e}$, \eqref{eq:PerturbationFormula} gives
\be
g_{\rm eff} = \sqrt{2} g^2 \sin\theta \cos\theta \left( \frac{1}{\Delta_{aq}} - \frac{1}{\omega_{a}} \right).
\ee
Using that on resonance $\omega_a = \omega_q/2$, this reduces to
\be
g_{\rm eff} = - 2 \sqrt{2} \sin (2\theta) \frac{g^2}{\omega_q},
\ee
which was also derived in Ref.~\cite{Garziano2015} using adiabatic elimination. We note that the effective coupling acquires a factor $g/\omega_q$ due to the fact that each path between $\ket{i}$ and $\ket{f}$ contains one intermediate level.

\paragraph{Two identical qubits}

Yet another option, illustrated in the upper right panel of \figref{fig:SecondSubharmonicDetailed} and discussed in Ref.~\cite{Garziano2016}, is to couple a single resonator to two identical qubits such that the process $\ket{1,g,g} \leftrightarrow \ket{0,e,e}$ is realized. The Hamiltonian for this setup is
\bea
\hat H &=& \omega_a \hat a^\dag \hat a + \sum_{j=1}^2 \omega_{q}\frac{\sz^{(j)}}{2} + \hat H_{\rm int}, \\
\hat H_{\rm int} &=& g \left( \hat a + \hat a^\dag \right) \sum_{j=1}^2 \left( \sx^{(j)} \cos\theta + \sz^{(j)} \sin\theta \right). \quad
\label{eq:Hint1res2qb}
\eea
and the effective interaction becomes
\be
\hat H_{\rm int}^{\rm eff} = g_{\rm eff} \ketbra{1,g,g}{0,e,e} +  {\rm H.c.}
\ee
The third-order-perturbation-theory calculations for this process following \eqref{eq:PerturbationFormula} were already performed in the appendix of Ref.~\cite{Garziano2016}. Here, we merely restate their result
\be
g_{\rm eff} = -\frac{8}{3}\sin\theta \cos^2\theta \frac{g^3}{\omega_q^2},
\ee
which is valid on resonance, when $\omega_a = 2\omega_q$. Again, we see that the effective coupling has a factor $\left(g/\omega \right)^2$, since each path contributing to the coupling contains two intermediate states.

In conclusion, we note that the multiphoton Rabi oscillation only requires two intermediate transitions, while the other two proposals require three. This means that the multiphoton Rabi oscillation has a larger effective coupling than the other two setups and is easier to implement.

\subsection{Raman scattering}
\label{sec:3WaveMixingRaman}

\subsubsection{Nonlinear optics}

In nonlinear optics, Raman scattering is a special case of nondegenerate three-wave mixing, mixing photons and optical phonons of the scattering nonlinear medium. Usually Raman scattering refers to the scattering of a light beam on optical phonons, which results in changing the frequency of the light beam~\cite{Miranowicz1994}. We note that analogous scattering of photons on acoustic phonons is referred to as Brillouin scattering.

We consider the following fields: a driving laser ($L$) mode of frequency $\omega_L$, a Stokes ($S$) mode of frequency $\omega_S$, an anti-Stokes ($A$) mode of frequency $\omega_A$, and optical vibrational phonon ($V$) modes of frequencies $\omega_{Vj}$ ($j = 1, 2, \dots$) as described by the corresponding creation ($\hat a_k^\dag$) and annihilation ($\hat a_k$) operators for $k = L, A, S, Vj$.

\paragraph{Stokes Raman scattering}
\label{sec:3WaveMixingStokesRamanScattering}

Raman scattering with Stokes frequency $\omega_S < \omega_L$ is shortly referred to as Stokes (Raman) scattering. The process is illustrated in \figref{fig:3WaveMixing}(c) and the interaction Hamiltonian can be written as
\be
\hat{H}^{(S)}_{\rm int} = \sum_{j} g_{Sj} \hat{a}_L \hat{a}_S^\dag \hat{a}_{Vj}^\dag + {\rm H.c.},
\label{eq:StokesRamanSum}
\ee
or its simpler single-phonon version
\be
\hat{H}^{(S)}_{\rm int} = g_{S} \hat{a}_L \hat{a}_S^\dag \hat{a}_{V}^\dag + {\rm H.c.}
\ee

\paragraph{Anti-Stokes Raman scattering (sideband cooling)}
\label{sec:3WaveMixingAntiStokesRamanScattering}

One can also analyze the Raman scattering with anti-Stokes frequency $\omega_A > \omega_L$, referred to as anti-Stokes (Raman) scattering and illustrated in \figref{fig:3WaveMixing}(c). The interaction Hamiltonian for the anti-Stokes Raman scattering can be written as
\be
\hat{H}^{(A)}_{\rm int} = \sum_{j} g_{Aj}^* \hat{a}_L \hat{a}^\dag_A \hat{a}_{Vj} + {\rm H.c.},
\ee
or its simpler single-phonon version
\be
\hat{H}^{(A)}_{\rm int} = g_{A}^* \hat{a}_L \hat{a}^\dag_A \hat{a}_{V} + {\rm H.c.}
\ee
Since a phonon is absorbed in this process, it can also be referred to as sideband cooling of the phononic mode.

\paragraph{Stimulated Raman scattering}
\label{sec:3WaveMixingStimulatedRamanScattering}

The presence of additional photons in the $S$ or $A$ modes, as shown in \figref{fig:3WaveMixing}(d), can increase the rate of Raman scattering. This is called stimulated Raman scattering. To further distinguish the processes in \figref{fig:3WaveMixing}(c) from those in \figref{fig:3WaveMixing}(d), the former can be referred to as spontaneous Raman scattering.

\subsubsection{Analogous processes}
\label{sec:3WaveMixingRamanDeterministic}

We can achieve close analogues of Raman scattering in our deterministic setups by letting a qubit play the role of a phonon. The qubit is coupled to two resonators, one representing the laser mode and the other representing the Stokes or anti-Stokes mode. The Hamiltonian of the system is given by Eqs.~(\ref{eq:H2res1qb}) and (\ref{eq:Hint2res1qbRabiGen}).

\paragraph{Stokes Raman scattering}

\begin{figure}[ht!]
\centerline{
  \resizebox{0.9\columnwidth}{!}{
    \begin{tikzpicture}[
      scale=1,
      level/.style={thick},
      virtual/.style={thick,dashed},
      ztrans/.style={thick,->,shorten >=0.1cm,shorten <=0.1cm,>=stealth,densely dashed,color=red},
      nrtrans/.style={thick,->,shorten >=0.1cm,shorten <=0.1cm,>=stealth,densely dashed,color=blue},
      rtrans/.style={thick,->,shorten >=0.1cm,shorten <=0.1cm,>=stealth,color=blue},
      strans/.style={rtrans, snake=snake, line after snake=1.5mm},
      rtrans2/.style={thick,->,shorten >=0.1cm,shorten <=0.1cm,>=stealth,color=green!35!black},
      strans2/.style={rtrans2, snake=snake, line after snake=1.5mm},
      classical/.style={thin,double,<->,shorten >=4pt,shorten <=4pt,>=stealth},
    ]
    \coordinate (corner) at (0cm, 0cm); 
    \coordinate (rh) at (0cm, 1cm); 
    \coordinate (rl) at (2cm, 0cm); 
    \coordinate (ccenter) at ($(corner) + 0.5*(rl) + (0cm, 3cm) + 1.5*(rh)$); 
    \coordinate (ccenter2) at ($(corner) + 0.5*(rl) - (0cm, 2cm) - 0.5*(rh)$); 
    \coordinate (al) at (1.5cm, 0cm); 
    \filldraw[color=cyan!70, fill=cyan!20, very thick](corner) rectangle ($(corner) + (rh) + (rl)$) node[color=black, midway] {S};
    \draw[color=cyan!70, thick, dashed] (ccenter) circle (3cm);
    \draw[color=cyan!70, thick, densely dashed] ($(corner) + (rh)$) --  ($(corner) + (rh) + (-1.2cm,1.4cm)$);
    \draw[color=cyan!70, thick, densely dashed] ($(corner) + (rh) + (rl)$) --  ($(corner) + (rh) + (rl) + (1.2cm,1.4cm)$);
    \draw[color=cyan!70, thick, dashed] (ccenter2) circle (2cm);
    \draw[color=cyan!70, thick, densely dashed] (corner) --  ($(corner) + (-0.5cm,-1.2cm)$);
    \draw[color=cyan!70, thick, densely dashed] ($(corner) + (rl)$) --  ($(corner) + (rl) + (0.5cm,-1.2cm)$);
    \draw[color=green!35!black, thick, ->, shorten >=0.05cm, shorten <=0.05cm, >=latex] ($(corner) + 0.5*(rh) - (al)$) -- ($(corner) + 0.5*(rh)$) node[color=black, midway, above] {$\omega_L$};
    \draw[color=green!35!black, thick, ->, shorten >=0.05cm, shorten <=0.05cm, >=latex] ($(corner) + 0.75*(rh) + (rl)$) -- ($(corner) + 0.75*(rh) + (rl) + (al)$) node[color=black, midway, above] {$\omega_S$};
    \draw[color=green!35!black, thick, ->, shorten >=0.05cm, shorten <=0.05cm, >=latex, snake=snake, line after snake=1.5mm] ($(corner) + 0.25*(rh) + (rl)$) -- ($(corner) - 0.25*(rh) + (rl) + (al)$) node[color=black, midway, above, shift={(0,0.05)}] {$\omega_V$};
    \coordinate (10g) at ($(ccenter) + (-1.9cm,0cm) $); 
    \coordinate (v) at (0cm, 0.8cm); 
    \coordinate (l) at (1.2cm, 0cm); 
    \coordinate (h) at (0.1cm, 0cm); 
    \coordinate (00g) at ($(10g) + (l) + (h) - 3*(v)$);
    \coordinate (00e) at ($(10g) + (l) + (h) - 2*(v)$);
    \coordinate (11g) at ($(10g) + (l) + (h) + 2*(v)$);
    \coordinate (11e) at ($(10g) + (l) + (h) + 3*(v)$);
    \coordinate (01e) at ($(10g) + 2*(l) + 2*(h)$);
    \draw[level] (00g) -- ($(00g) + (l)$) node[midway,below] {\footnotesize{$\ket{0,0,g}$}};
    \draw[level] (00e) -- ($(00e) + (l)$) node[midway,below] {\footnotesize{$\ket{0,0,e}$}};
    \draw[level] (10g) -- ($(10g) + (l)$) node[left,xshift=-1.1cm] {\footnotesize{$\ket{1,0,g}$}};
    \draw[level] (01e) -- ($(01e) + (l)$) node[right,xshift=-0.5cm] {\footnotesize{$\quad\:\:\ket{0,1,e}$}};
    \draw[level] (11g) -- ($(11g) + (l)$) node[midway,above] {\footnotesize{$\ket{1,1,g}$}};
    \draw[level] (11e) -- ($(11e) + (l)$) node[midway,above] {\footnotesize{$\ket{1,1,e}$}};
    \draw[nrtrans] ($(10g) + (0.2cm,0cm)$) -- ($(11e) + (0.2cm,0cm)$);
    \draw[ztrans] ($(10g) + (0.6cm,0cm)$) -- ($(11g) + (0.4cm,0cm)$);
    \draw[rtrans] ($(10g) + (0.6cm,0cm)$) -- ($(00e) + (0.4cm,0cm)$);
    \draw[ztrans] ($(10g) + (0.2cm,0cm)$) -- ($(00g) + (0.2cm,0cm)$);
    \draw[ztrans] ($(11e) + (l) - (0.2cm,0cm)$) -- ($(01e) + (l) - (0.2cm,0cm)$);
    \draw[rtrans] ($(11g) + (l) - (0.4cm,0cm)$) -- ($(01e) + (l) - (0.6cm,0cm)$);
    \draw[ztrans] ($(00e) + (l) - (0.4cm,0cm)$) -- ($(01e) + (l) - (0.6cm,0cm)$);
    \draw[nrtrans] ($(00g) + (l) - (0.2cm,0cm)$) -- ($(01e) + (l) - (0.2cm,0cm)$);
    %
    \coordinate (ground) at ($(ccenter2) + (-1cm,-1.5cm)$); 
    \coordinate (htot) at (0cm, 3cm); 
    \coordinate (hphon) at (0cm, 1cm); 
    \coordinate (ltot) at (2cm, 0cm); 
    \coordinate (lphon) at (1cm, 0cm); 
    \draw[level] (ground) -- ($(ground) + (ltot)$);
    \draw[virtual] ($(ground) + (htot)$) -- ($(ground) + (htot) + (ltot)$);
    \draw[level] ($(ground) + (hphon) + (ltot) - (lphon)$) -- ($(ground) + (hphon) + (ltot)$);
    \draw[rtrans2] ($(ground) + (0.5cm,0cm)$) -- ($(ground) + (htot) + (0.5cm,0cm)$) node[color=black, midway, left] {$\omega_L$};
    \draw[rtrans2] ($(ground) + (htot) + (1.5cm,0cm)$) -- ($(ground) + (hphon) + (1.5cm,0cm)$) node[color=black, midway, right, shift={(0.03,0)}] {$\omega_S$};
    \draw[strans2] ($(ground) + (hphon) + (1.5cm,0cm)$) -- ($(ground) + (1.5cm,0cm)$) node[color=black, midway, right, shift={(0.03,0)}] {$\omega_V$};
    \end{tikzpicture}
  }
}
\caption{Raman scattering and its deterministic analogue. The upper panel shows all virtual transitions that contribute, to lowest order, to the process $\ket{1,0,g} \to \ket{0,1,e}$, which corresponds to Stokes Raman scattering. The lower panel shows the generic level diagram for the process in nonlinear optics. The same arrow and level styles as in \figref{fig:SecondSubharmonicDetailed} are used; we have set $\omega_a = 3\omega_q$ and $\omega_b = 2\omega_q$. If the directions of all arrows in the entire figure are reversed, and the labels are changed such that $S \to L$ and $L \to A$, anti-Stokes scattering is shown instead. Stimulated Stokes (anti-Stokes) Raman scattering is given by adding $n$ to the photon number in the second (first) resonator mode in the upper panel and adding $n$ incoming and outgoing photons to the $S$ ($A$) mode in the rest of the figure. \label{fig:RamanDetails}}
\end{figure}
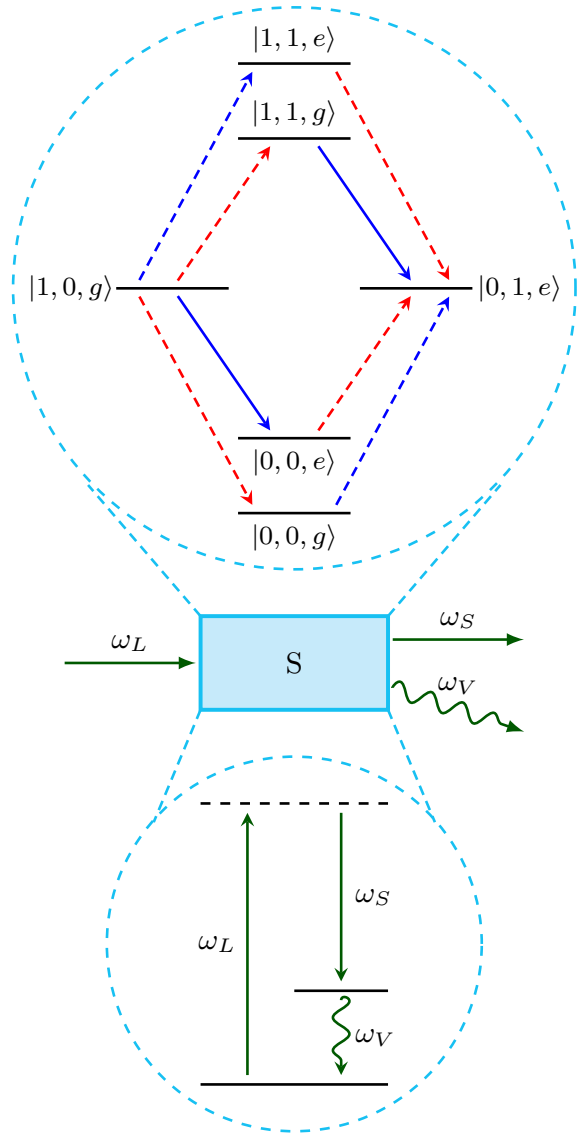

Setting $\omega_a = \omega_b + \omega_q$, and making the connections $a = L$, $b = S$, and $q = V$, the transition $\ket{1,0,g} \to \ket{0,1,e}$ emulates Stokes Raman scattering. The virtual transitions involved are shown in \figref{fig:RamanDetails}. This process is further discussed in the forthcoming work of Ref.~\cite{Kockum2017} as means to achieve single-photon frequency conversion controlled by the qubit. The effective interaction due to the virtual transitions when $\omega_q \approx \omega_a - \omega_b$ becomes
\be
\hat H_{\rm int}^{\rm eff} = g_{\rm eff} \ketbra{1,0,g}{0,1,e} +  {\rm H.c.}
\ee
Second-order perturbation theory using \eqref{eq:PerturbationFormula} and \figref{fig:RamanDetails} gives
\be
g_{\rm eff} = g_a g_b \sin\theta \cos\theta \left( \frac{1}{-\omega_a} - \frac{1}{\Delta_{qa}} + \frac{1}{\omega_b} - \frac{1}{\Omega_{bq}} \right),
\ee
which reduces to
\be
g_{\rm eff} = g_a g_b \left( \frac{1}{\omega_b} - \frac{1}{\omega_a} \right) \sin (2\theta)
\ee
on resonance ($\omega_q = \omega_a - \omega_b$). This agrees with the result obtained using adiabatic elimination in Ref.~\cite{Kockum2017}.

We also note that it has been shown that a photon scattering off a qubit ultrastrongly coupled to an open transmission line can be downconverted in frequency, leaving some of its energy with the qubit~\cite{Sanchez-Burillo2014}. However, this downconversion process is not deterministic.

\paragraph{Anti-Stokes Raman scattering} 

The same setup as for Stokes Raman scattering, but considering the reverse transition $\ket{0,1,e} \to \ket{1,0,g}$, implements anti-Stokes Raman scattering. In this case, we need to make the identifications $a = A$, $b = L$, and $q = V$.

\paragraph{Stimulated Raman scattering}

We can again consider the same setup, but instead look at the transitions $\ket{1,n,g} \to \ket{0,n+1,e}$ and $\ket{n,1,e} \to \ket{n+1,0,g}$ to obtain stimulated Stokes Raman scattering and stimulated anti-Stokes Raman scattering, respectively. In calculating the effective coupling $g_{\rm eff}$ between the initial and final states, as done above and in Ref.~\cite{Kockum2017} for the case $n=0$, we will, for each possible path between them, multiply the corresponding transition matrix elements. As can be seen from \figref{fig:RamanDetails}, each path contains exactly one transition that changes the number of excitations in one of the modes from $n$ to $n+1$. This contributes a factor $\sqrt{n+1}$ to the effective coupling, showing that the presence of the additional photons stimulates the transition.

\section{Four-wave mixing}
\label{sec:4WaveMixing}

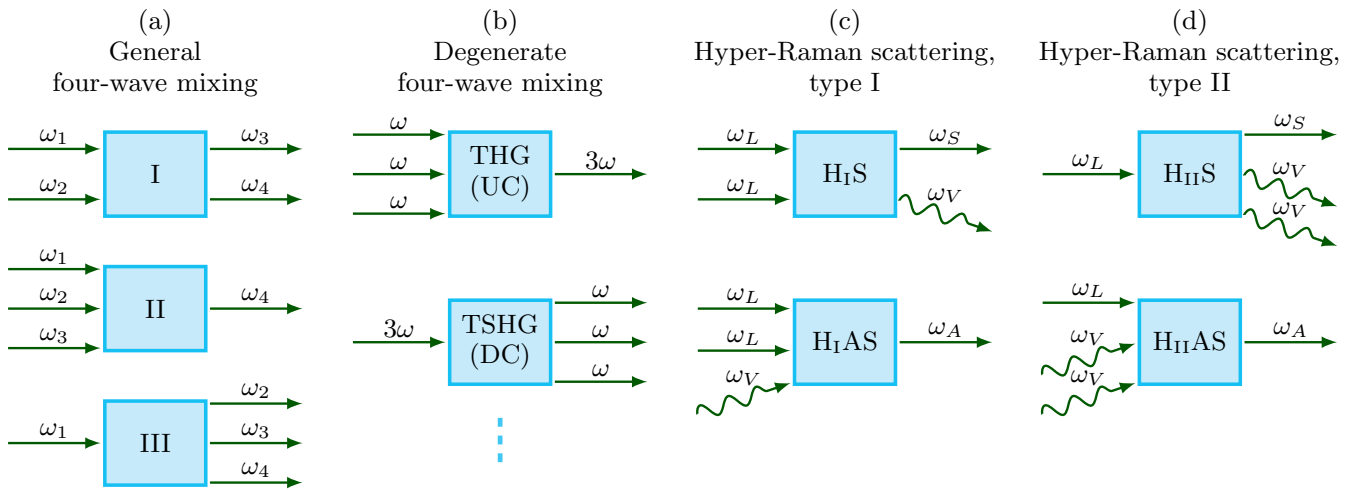
\begin{figure*}
\centerline{
  \resizebox{2.1\columnwidth}{!}{
    \begin{tikzpicture}[
      scale=1,
      level/.style={thick},
      virtual/.style={thick,dashed},
      ztrans/.style={thick,->,shorten >=0.1cm,shorten <=0.1cm,>=stealth,densely dashed,color=red},
      nrtrans/.style={thick,->,shorten >=0.1cm,shorten <=0.1cm,>=stealth,densely dashed,color=blue},
      rtrans/.style={thick,->,shorten >=0.1cm,shorten <=0.1cm,>=stealth,color=blue},
      strans/.style={rtrans, snake=snake, line after snake=1.5mm},
      classical/.style={thin,double,<->,shorten >=4pt,shorten <=4pt,>=stealth},
      mybox/.style={color=cyan!70, fill=cyan!20, very thick},
      myarrow/.style={color=green!35!black, thick, ->, shorten >=0.05cm, shorten <=0.05cm, >=latex},
      mysquigglyarrow/.style={color=green!35!black, thick, ->, shorten >=0.05cm, shorten <=0.05cm, >=latex, snake=snake, line after snake=2mm}
    ]
    \coordinate (corner1) at (0cm, 0cm); 
    \coordinate (rh) at (0, 1); 
    \coordinate (rl) at (1.2, 0); 
    \coordinate (vcol) at (0, 1); 
    \coordinate (hrow) at (0.5, 0); 
    \coordinate (al) at ($(rl) + (0,0)$); 
    \coordinate (corner2) at ($(corner1) + (rl) + (al) + (hrow) + (al)$); 
    \coordinate (corner3) at ($(corner2) + (rl) + (al) + (hrow) + (al)$); 
    \coordinate (corner4) at ($(corner3) + (rl) + (al) + (hrow) + (al)$); 
    \newcommand\labelyshift{-0.06}; 
    \newcommand\labelxshift{0}; 
    \newcommand\labelyshifttwo{0.8}; 
    \newcommand\labelxshifttwo{0}; 
    \filldraw[mybox](corner1) rectangle ($(corner1) + (rh) + (rl)$) node[color=black, midway, above, shift={(\labelxshifttwo,\labelyshifttwo)}] {\parbox{3cm}{\centering (a) \\ General \\ four-wave mixing}} node[color=black, midway] {I};
    \draw[myarrow] ($(corner1) + 0.8*(rh) - (al)$) -- ($(corner1) + 0.8*(rh)$) node[color=black, midway, above, shift={(\labelxshift,\labelyshift)}] {$\omega_1$};
    \draw[myarrow] ($(corner1) + 0.2*(rh) - (al)$) -- ($(corner1) + 0.2*(rh)$) node[color=black, midway, above, shift={(\labelxshift,\labelyshift)}] {$\omega_2$};
    \draw[myarrow] ($(corner1) + 0.8*(rh) + (rl)$) -- ($(corner1) + 0.8*(rh) + (rl) + (al)$) node[color=black, midway, above, shift={(\labelxshift,\labelyshift)}] {$\omega_3$};
    \draw[myarrow] ($(corner1) + 0.2*(rh) + (rl)$) -- ($(corner1) + 0.2*(rh) + (rl) + (al)$) node[color=black, midway, above, shift={(\labelxshift,\labelyshift)}] {$\omega_4$};    
    \filldraw[mybox] ($(corner1) - 0.6*(vcol) - (rh)$) rectangle ($(corner1) - 0.6*(vcol) - (rh) + (rh) + (rl)$) node[color=black, midway] {II};
    \draw[myarrow] ($(corner1) - 0.6*(vcol) - (rh) + 0.97*(rh) - (al)$) -- ($(corner1) - 0.6*(vcol) - (rh) + 0.97*(rh)$) node[color=black, midway, above, shift={(\labelxshift,\labelyshift)}] {$\omega_1$};
    \draw[myarrow] ($(corner1) - 0.6*(vcol) - (rh) + 0.5*(rh) - (al)$) -- ($(corner1) - 0.6*(vcol) - (rh) + 0.5*(rh)$) node[color=black, midway, above, shift={(\labelxshift,\labelyshift)}] {$\omega_2$};
    \draw[myarrow] ($(corner1) - 0.6*(vcol) - (rh) + 0.03*(rh) - (al)$) -- ($(corner1) - 0.6*(vcol) - (rh) + 0.03*(rh)$) node[color=black, midway, above, shift={(\labelxshift,\labelyshift)}] {$\omega_3$};
    \draw[myarrow] ($(corner1) - 0.6*(vcol) - (rh) + 0.5*(rh) + (rl)$) -- ($(corner1) - 0.6*(vcol) - (rh) + 0.5*(rh) + (rl) + (al)$) node[color=black, midway, above, shift={(\labelxshift,\labelyshift)}] {$\omega_4$};
    \filldraw[mybox] ($(corner1) - 1.2*(vcol) - 2*(rh)$) rectangle ($(corner1) - 1.2*(vcol) - 2*(rh) + (rh) + (rl)$) node[color=black, midway] {III};
    \draw[myarrow] ($(corner1) - 1.2*(vcol) - 2*(rh) + 0.5*(rh) - (al)$) -- ($(corner1) - 1.2*(vcol) - 2*(rh) + 0.5*(rh)$) node[color=black, midway, above, shift={(\labelxshift,\labelyshift)}] {$\omega_1$};
    \draw[myarrow] ($(corner1) - 1.2*(vcol) - 2*(rh) + 0.97*(rh) + (rl)$) -- ($(corner1) - 1.2*(vcol) - 2*(rh) + 0.97*(rh) + (rl) + (al)$) node[color=black, midway, above, shift={(\labelxshift,\labelyshift)}] {$\omega_2$};
    \draw[myarrow] ($(corner1) - 1.2*(vcol) - 2*(rh) + 0.5*(rh) + (rl)$) -- ($(corner1) - 1.2*(vcol) - 2*(rh) + 0.5*(rh) + (rl) + (al)$) node[color=black, midway, above, shift={(\labelxshift,\labelyshift)}] {$\omega_3$};        
    \draw[myarrow] ($(corner1) - 1.2*(vcol) - 2*(rh) + 0.03*(rh) + (rl)$) -- ($(corner1) - 1.2*(vcol) - 2*(rh) + 0.03*(rh) + (rl) + (al)$) node[color=black, midway, above, shift={(\labelxshift,\labelyshift)}] {$\omega_4$};    
    \filldraw[mybox](corner2) rectangle ($(corner2) + (rh) + (rl)$) node[color=black, midway, above, shift={(\labelxshifttwo,\labelyshifttwo)}] {\parbox{3cm}{\centering (b) \\ Degenerate \\ four-wave mixing}} node[color=black, midway] {\parbox{3cm}{\centering THG \\ (UC)}};
    \draw[myarrow] ($(corner2) + 0.97*(rh) - (al)$) -- ($(corner2) + 0.97*(rh)$) node[color=black, midway, above, shift={(\labelxshift,\labelyshift)}] {$\omega$};
    \draw[myarrow] ($(corner2) + 0.5*(rh) - (al)$) -- ($(corner2) + 0.5*(rh)$) node[color=black, midway, above, shift={(\labelxshift,\labelyshift)}] {$\omega$};
    \draw[myarrow] ($(corner2) + 0.03*(rh) - (al)$) -- ($(corner2) + 0.03*(rh)$) node[color=black, midway, above, shift={(\labelxshift,\labelyshift)}] {$\omega$};
    \draw[myarrow] ($(corner2) + 0.5*(rh) + (rl)$) -- ($(corner2) + 0.5*(rh) + (rl) + (al)$) node[color=black, midway, above, shift={(\labelxshift,\labelyshift)}] {$3\omega$};
    \filldraw[mybox] ($(corner2) - (vcol) - (rh)$) rectangle ($(corner2) - (vcol) - (rh) + (rh) + (rl)$) node[color=black, midway] {\parbox{3cm}{\centering TSHG \\ (DC)}};
    \draw[myarrow] ($(corner2) - (vcol) - (rh) + 0.5*(rh) - (al)$) -- ($(corner2) - (vcol) - (rh) + 0.5*(rh)$) node[color=black, midway, above, shift={(\labelxshift,\labelyshift)}] {$3\omega$};
    \draw[myarrow] ($(corner2) - (vcol) - (rh) + 0.97*(rh) + (rl)$) -- ($(corner2) - (vcol) - (rh) + 0.97*(rh) + (rl) + (al)$) node[color=black, midway, above, shift={(\labelxshift,\labelyshift)}] {$\omega$};
    \draw[myarrow] ($(corner2) - (vcol) - (rh) + 0.5*(rh) + (rl)$) -- ($(corner2) - (vcol) - (rh) + 0.5*(rh) + (rl) + (al)$) node[color=black, midway, above, shift={(\labelxshift,\labelyshift)}] {$\omega$};
    \draw[myarrow] ($(corner2) - (vcol) - (rh) + 0.03*(rh) + (rl)$) -- ($(corner2) - (vcol) - (rh) + 0.03*(rh) + (rl) + (al)$) node[color=black, midway, above, shift={(\labelxshift,\labelyshift)}] {$\omega$};
    \draw[ultra thick,dashed, color=cyan!60] ($(corner2) - 2.4*(vcol) + 0.5*(rl)$) -- ($(corner2) - 3*(vcol) + 0.5*(rl)$); 
     \filldraw[mybox](corner3) rectangle ($(corner3) + (rh) + (rl)$) node[color=black, midway, above, shift={(\labelxshifttwo,\labelyshifttwo)}] {\parbox{3.6cm}{\centering (c) \\ Hyper-Raman scattering, \\ type I}} node[color=black, midway] {$\rm H_IS$};
    \draw[myarrow] ($(corner3) + 0.8*(rh) - (al)$) -- ($(corner3) + 0.8*(rh)$) node[color=black, midway, above, shift={(\labelxshift,\labelyshift)}] {$\omega_L$};
    \draw[myarrow] ($(corner3) + 0.2*(rh) - (al)$) -- ($(corner3) + 0.2*(rh)$) node[color=black, midway, above, shift={(\labelxshift,\labelyshift)}] {$\omega_L$};    
    \draw[myarrow] ($(corner3) + 0.8*(rh) + (rl)$) -- ($(corner3) + 0.8*(rh) + (rl) + (al)$) node[color=black, midway, above, shift={(\labelxshift,\labelyshift)}] {$\omega_S$};
    \draw[mysquigglyarrow] ($(corner3) + 0.2*(rh) + (rl)$) -- ($(corner3) - 0.2*(rh) + (rl) + (al)$) node[color=black, midway, above, shift={(0,0.04)}] {$\omega_V$};
    \filldraw[mybox] ($(corner3) - (vcol) - (rh)$) rectangle ($(corner3) - (vcol) - (rh) + (rh) + (rl)$) node[color=black, midway] {$\rm H_IAS$};
    \draw[myarrow] ($(corner3) - (vcol) - (rh) + 0.9*(rh) - (al)$) -- ($(corner3) - (vcol) - (rh) + 0.9*(rh)$) node[color=black, midway, above, shift={(\labelxshift,\labelyshift)}] {$\omega_L$};
    \draw[myarrow] ($(corner3) - (vcol) - (rh) + 0.4*(rh) - (al)$) -- ($(corner3) - (vcol) - (rh) + 0.4*(rh)$) node[color=black, midway, above, shift={(\labelxshift,\labelyshift)}] {$\omega_L$};    
    \draw[mysquigglyarrow] ($(corner3) - (vcol) - (rh) - 0.37*(rh) - (al)$) -- ($(corner3) - (vcol) - (rh) + 0.03*(rh)$) node[color=black, midway, above, shift={(0,0.04)}] {$\omega_V$};
    \draw[myarrow] ($(corner3) - (vcol) - (rh) + 0.5*(rh) + (rl)$) -- ($(corner3) - (vcol) - (rh) + 0.5*(rh) + (rl) + (al)$) node[color=black, midway, above, shift={(\labelxshift,\labelyshift)}] {$\omega_A$};  
    \filldraw[mybox](corner4) rectangle ($(corner4) + (rh) + (rl)$) node[color=black, midway, above, shift={(\labelxshifttwo,\labelyshifttwo)}] {\parbox{3.6cm}{\centering (d) \\ Hyper-Raman scattering, \\ type II}} node[color=black, midway] {$\rm H_{II}S$};
    \draw[myarrow] ($(corner4) + 0.5*(rh) - (al)$) -- ($(corner4) + 0.5*(rh)$) node[color=black, midway, above, shift={(\labelxshift,\labelyshift)}] {$\omega_L$};           
    \draw[myarrow] ($(corner4) + 0.97*(rh) + (rl)$) -- ($(corner4) + 0.97*(rh) + (rl) + (al)$) node[color=black, midway, above, shift={(\labelxshift,\labelyshift)}] {$\omega_S$};   
    \draw[mysquigglyarrow] ($(corner4) + 0.5*(rh) + (rl)$) -- ($(corner4) + 0.1*(rh) + (rl) + (al)$) node[color=black, midway, above, shift={(0,0.04)}] {$\omega_V$};
    \draw[mysquigglyarrow] ($(corner4) + 0.03*(rh) + (rl)$) -- ($(corner4) - 0.37*(rh) + (rl) + (al)$) node[color=black, midway, above, shift={(0,0.04)}] {$\omega_V$};    
    \filldraw[mybox]($(corner4) - (vcol) - (rh)$) rectangle ($(corner4) - (vcol) - (rh) + (rh) + (rl)$) node[color=black, midway] {$\rm H_{II}AS$};
    \draw[myarrow] ($(corner4) - (vcol) - (rh) + 0.97*(rh) - (al)$) -- ($(corner4) - (vcol) - (rh) + 0.97*(rh)$) node[color=black, midway, above, shift={(\labelxshift,\labelyshift)}] {$\omega_L$};
    \draw[mysquigglyarrow] ($(corner4) - (vcol) - (rh) + 0.1*(rh) - (al)$) -- ($(corner4)  - (vcol) - (rh) + 0.5*(rh)$) node[color=black, midway, above, shift={(0,0.04)}] {$\omega_V$};
    \draw[mysquigglyarrow] ($(corner4) - (vcol) - (rh) - 0.37*(rh) - (al)$) -- ($(corner4)  - (vcol) - (rh) + 0.03*(rh)$) node[color=black, midway, above, shift={(0,0.04)}] {$\omega_V$};                 
    \draw[myarrow] ($(corner4) - (vcol) - (rh) + 0.5*(rh) + (rl)$) -- ($(corner4) - (vcol) - (rh) + 0.5*(rh) + (rl) + (al)$) node[color=black, midway, above, shift={(\labelxshift,\labelyshift)}] {$\omega_A$};   
    \end{tikzpicture}
  }
}
\caption{Schematic representations (Feynman-like diagrams) of four-wave-mixing processes. (a) Four-wave mixing can be divided into three general categories: type~I, with two incoming and two outgoing signals (above), type~II, with three incoming signals and one outgoing (middle), and type~III, with one incoming signal and three outgoing ones (below). (b) When three of the frequencies are degenerate, we have either third-harmonic generation (THG, or upconversion, above) or third-subharmonic generation (TSHG, or downconversion, below). When two of the frequencies are degenerate, four processes are possible (not pictured here, but shown in \appref{app:4WaveMixing}). (c) When a phonon is involved, the process is called hyper-Raman scattering of type~I. The only change to Stokes ($\rm H_IS$, above) and anti-Stokes ($\rm H_IAS$, below) Raman scattering from the three-wave-mixing case [see \figref{fig:3WaveMixing}(c)] is that there are two (degenerate) incoming photons instead of one. (d) With two degenerate phonons, the process is called hyper-Raman scattering of type~II. The two phonons replace the single one in the Stokes ($\rm H_{II}S$, above) and anti-Stokes ($\rm H_{II}AS$, below) versions of ordinary Raman scattering from \figref{fig:3WaveMixing}(c). \label{fig:4WaveMixing}}
\end{figure*}

In this section, we treat four-wave mixing, starting as in \secref{sec:3WaveMixing} with a general description and then treating special cases, such as degenerate four-wave mixing and hyper-Raman scattering. An overview of these processes is given in \figref{fig:4WaveMixing}. We again provide deterministic analogues for each case. Since there are many similarities to the material presented in \secref{sec:3WaveMixing}, the treatment here will be a little more concise. However, compared to \secref{sec:3WaveMixing} there are more processes to cover and longer paths of virtual transitions to consider in calculating the effective coupling for those processes.

\subsection{General description}
\label{sec:4WaveMixingGeneral}

\subsubsection{Nonlinear optics}

Four-wave mixing comes in three types, as illustrated in \figref{fig:4WaveMixing}(a). Type~I, with the interaction Hamiltonian
\be
\hat H_{\rm int} = g \hat a_1 \hat a_2 \hat a_3^\dag \hat a_4^\dag + g^* \hat a_1^\dag \hat a_2^\dag \hat a_3 \hat a_4,
\ee
has two incoming and two outgoing signals. Processes with three incoming signals and one outgoing are here called type~II, and processes with one incoming signal and three outgoing ones are here referred to as type~III. The interaction Hamiltonian for both types~II and III can be written as
\be
\hat H_{\rm int} = g \hat a_1 \hat a_2 \hat a_3 \hat  a_4^\dag + g^* \hat a_1^\dag \hat a_2^\dag \hat a_3^\dag \hat a_4.
\ee

\subsubsection{Analogous processes}
\label{sec:4WaveMixingGeneralDeterministic}

There are, just as for three-wave mixing, several possible setups that allow deterministic analogues of the four-wave mixing processes. The clearest analogy is probably four resonators all coupled to a single qubit. Adjusting the resonator frequencies to satisfy the condition $\omega_a + \omega_b \approx \omega_c + \omega_d$, the states $\ket{1,1,0,0,g}$ and $\ket{0,0,1,1,g}$ become resonant and the transitions between these states will constitute type-I four-wave mixing. Similarly, if $\omega_a + \omega_b + \omega_c \approx \omega_d$, the transition $\ket{1,1,1,0,g} \to \ket{0,1,1,1,g}$ corresponds to type-II mixing and the reverse process $\ket{0,1,1,1,g} \to \ket{1,0,0,0,g}$ will be type-III mixing.

If at least one of the excitations in the four-wave mixing can be hosted in a qubit, additional setups are possible. With three resonators coupled to a single qubit, $\ket{1,1,0,g} \leftrightarrow \ket{0,0,1,e}$ corresponds to type-I mixing and the processes $\ket{1,1,1,g} \leftrightarrow \ket{0,0,0,e}$ corresponds to type-II ($\to$) and type-III ($\leftarrow$) mixing, respectively. In the same way, with two resonators coupled to two qubits, $\ket{1,1,g,g} \leftrightarrow \ket{0,0,e,e}$ are analogues of type-I mixing and the processes $\ket{0,1,e,e} \leftrightarrow \ket{1,0,g,g}$ are some possible analogues for type-II ($\to$) and type-III ($\leftarrow$) mixing, respectively. Finally, with a single resonator coupled to three qubits, $\ket{1,e,g,g} \leftrightarrow \ket{0,g,e,e}$ corresponds to type-I mixing and the processes $\ket{0,e,e,e} \leftrightarrow \ket{1,g,g,g}$ corresponds to type-II ($\to$) and type-III ($\leftarrow$) mixing, respectively. Hosting at least one excitation in a qubit may be preferable, since such setups, in general, will require one less intermediate virtual transition than the setup with four resonators and a qubit. Barring destructive interference between the various virtual transition paths, this implies that the effective coupling will be weaker in the latter setup.

All these processes can occur due to intermediate virtual transitions as before. However, in contrast to three-wave mixing, the four-wave mixing analogues do not require the generalized Rabi interaction Hamiltonian from \eqref{eq:HintGenRabi}. The standard quantum Rabi model in \eqref{eq:HintRabi} is sufficient, since the parity of the number of excitation is conserved in four-wave mixing. In fact, for type-I processes, which do not change the number of excitations, the interaction terms from the JC model in \eqref{eq:HintJC} are sufficient to mediate the required virtual transitions. However, terms from the full quantum Rabi model can still give a significant contribution to the effective coupling between the initial and final states of such processes.

\subsection{Degenerate four-wave mixing: Third-harmonic and third-subharmonic generation}
\label{sec:Degenerate4WaveMixing}

In this subsection, we limit our analysis to the cases where three of the signals involved are degenerate. The cases with two degenerate signals are reviewed briefly in \appref{app:4WaveMixing}.

\subsubsection{Nonlinear optics}

Let us analyze a degenerate case of four-wave mixing assuming $\hat a_1 = \hat a_2 = \hat a_3 \equiv \hat a$, $\hat a_4 \equiv \hat a_+$, and $\omega_+ = 3 \omega$. The creation and annihilation of a photon in the Fock basis can be given as $\ket{n, n_+} \to \ket{n-3, n_+ +1}$ for third-harmonic generation (upconversion) and $\ket{n, n_+} \to \ket{n+3, n_+ -1}$ for third-subharmonic generation (downconversion); see also \figref{fig:4WaveMixing}(b). The interaction Hamiltonian for both processes reads as
\be
\hat H_{\rm int} = g \hat a^3  \hat a_+^\dag + g^* \hat a^{\dag 3} \hat a_+.
\label{eq:4WaveMixingThirdHarmonicHint}
\ee
The initial pure state for third-subharmonic generation is usually chosen as $\ket{\psi(t_0)} = \sum_{n=0}^{\infty} c_n \ket{n,0}$, while that for third-harmonic generation can read as $\ket{\psi(t_0)} = \sum_{n_+=0}^{\infty} c_{n_+} \ket{0, n_+}$, where $c_n$ and $c_{n_+}$ are arbitrary complex amplitudes like in \secref{sec:Degenerate3WaveMixingNonlinear}.

\subsubsection{Analogous processes}
\label{sec:Degenerate4WaveMixingDeterministic}

\begin{figure*}
\centerline{
  \resizebox{\linewidth}{!}{
    \begin{tikzpicture}[
      scale=1,
      level/.style={thick},
      virtual/.style={thick,dashed},
      ztrans/.style={thick,->,shorten >=0.1cm,shorten <=0.1cm,>=stealth,densely dashed,color=red},
      nrtrans/.style={thick,->,shorten >=0.1cm,shorten <=0.1cm,>=stealth,densely dashed,color=blue},
      rtrans/.style={thick,->,shorten >=0.1cm,shorten <=0.1cm,>=stealth,color=blue},
      strans/.style={rtrans, snake=snake, line after snake=1.5mm},
      rtrans2/.style={thick,->,shorten >=0.1cm,shorten <=0.1cm,>=stealth,color=green!35!black},
      strans2/.style={rtrans2, snake=snake, line after snake=1.5mm},
      classical/.style={thin,double,<->,shorten >=4pt,shorten <=4pt,>=stealth},
      mybox/.style={color=cyan!70, fill=cyan!20, very thick},
      myarrow/.style={color=green!35!black, thick, ->, shorten >=0.05cm, shorten <=0.05cm, >=latex},
      mysquigglyarrow/.style={color=green!35!black, thick, ->, shorten >=0.05cm, shorten <=0.05cm, >=latex, snake=snake, line after snake=2mm}
    ]
    \newcommand\labelyshift{-0.05}; 
    \newcommand\labelxshift{0}; 
    \newcommand\labelyshifttwo{0.8}; 
    \newcommand\labelxshifttwo{0}; 
    \coordinate (corner) at (0,0); 
    \coordinate (rh) at (0,1); 
    \coordinate (rl) at (5cm, 0cm); 
    \coordinate (ccenter) at ($(corner) + 0.5*(rl) + (0cm, 3cm) + 1.5*(rh)$); 
    \coordinate (ccenter2) at ($(corner) + 0.5*(rl) - (0cm, 2cm) - 0.5*(rh)$); 
    \coordinate (al) at (1.5cm, 0cm); 
    \filldraw[mybox](corner) rectangle ($(corner) + (rh) + (rl)$) node[color=black, midway] {\parbox{4cm}{\centering TSHG \\ (DC)}};
    \draw[color=cyan!70, thick, dashed] (ccenter) circle (3cm);
    \draw[color=cyan!70, thick, densely dashed] ($(corner) + (rh) + 0.3*(rl)$) --  ($(corner) + (rh) + (0.5,1.2)$);
    \draw[color=cyan!70, thick, densely dashed] ($(corner) + (rh) + 0.7*(rl)$) --  ($(corner) + (rh) + (rl) + (-0.5,1.2)$);
    \draw[color=cyan!70, thick, dashed] ($(ccenter) - (6.5,0)$) circle (3cm);
    \draw[color=cyan!70, thick, densely dashed] ($(corner) + (rh)$) --  ($(corner) + (rh) + (-4.3,0.5)$);
    \draw[color=cyan!70, thick, densely dashed] ($(corner) + (rh) + 0.05*(rl)$) --  ($(corner) + (rh) + (-0.95,3.6)$);
    \draw[color=cyan!70, thick, dashed] ($(ccenter) + (6.5,0)$) circle (3cm);
    \draw[color=cyan!70, thick, densely dashed] ($(corner) + (rh) + (rl)$) --  ($(corner) + (rh) + (rl) + (4.3,0.5)$);
    \draw[color=cyan!70, thick, densely dashed] ($(corner) + (rh) + 0.95*(rl)$) --  ($(corner) + (rh) + (rl) + (0.95,3.6)$);
    \draw[color=cyan!70, thick, dashed] (ccenter2) circle (2cm);
    \draw[color=cyan!70, thick, densely dashed] ($(corner) + 0.3*(rl)$) --  ($(corner) + (0.85,-1.3)$);
    \draw[color=cyan!70, thick, densely dashed] ($(corner) + 0.7*(rl)$) --  ($(corner) + (rl) + (-0.85,-1.3)$);
    \draw[myarrow] ($(corner) + 0.5*(rh) - (al)$) -- ($(corner) + 0.5*(rh)$) node[color=black, midway, above, shift={(\labelxshift,\labelyshift)}] {$3\omega$};
    \draw[myarrow] ($(corner) + 0.8*(rh) + (rl)$) -- ($(corner) + 0.8*(rh) + (rl) + (al)$) node[color=black, midway, above, shift={(\labelxshift,\labelyshift)}] {$\omega$};
    \draw[myarrow] ($(corner) + 0.42*(rh) + (rl)$) -- ($(corner) + 0.42*(rh) + (rl) + (al)$) node[color=black, midway, above, shift={(\labelxshift,\labelyshift)}] {$\omega$};
    \draw[myarrow] ($(corner) + 0.04*(rh) + (rl)$) -- ($(corner) + 0.04*(rh) + (rl) + (al)$) node[color=black, midway, above, shift={(\labelxshift,\labelyshift)}] {$\omega$};
    \coordinate (v) at (0cm, 0.55cm); 
    \coordinate (l) at (1.0cm, 0cm); 
    \coordinate (h) at (0.1cm, 0cm); 
    \coordinate (10g) at ($(ccenter) + (-9.2,-0.7)$); 
    \coordinate (11e) at ($(10g) + (l) + (h) + 3*(v)$);
    \coordinate (00e) at ($(10g) + (l) + (h) - 1*(v)$);
    \coordinate (12g) at ($(10g) + 2*(l) + 2*(h) + 2*(v)$);
    \coordinate (01g) at ($(10g) + 2*(l) + 2*(h) - 2*(v)$);
    \coordinate (13e) at ($(10g) + 3*(l) + 3*(h) + 5*(v)$);
    \coordinate (02e) at ($(10g) + 3*(l) + 3*(h) + 1*(v)$);
    \coordinate (03g) at ($(10g) + 4*(l) + 4*(h)$);
    \draw[level] (10g) -- ($(10g) + (l)$) node[midway,below,xshift=-0.05cm] {\footnotesize{$\ket{1,0,g}$}};
    \draw[level] (11e) -- ($(11e) + (l)$) node[midway,above] {\footnotesize{$\ket{1,1,e}$}};
    \draw[level] (00e) -- ($(00e) + (l)$) node[midway,below] {\footnotesize{$\ket{0,0,e}$}};
    \draw[level] (12g) -- ($(12g) + (l)$) node[midway,above] {\footnotesize{$\ket{1,2,g}$}};
    \draw[level] (01g) -- ($(01g) + (l)$) node[midway,below] {\footnotesize{$\ket{0,1,g}$}};
    \draw[level] (13e) -- ($(13e) + (l)$) node[midway,above] {\footnotesize{$\ket{1,3,e}$}};
    \draw[level] (02e) -- ($(02e) + (l)$) node[midway,above] {\footnotesize{$\ket{0,2,e}$}};
    \draw[level] (03g) -- ($(03g) + (l)$) node[midway,below,xshift= 0cm] {\footnotesize{$\ket{0,3,g}$}};
    \draw[nrtrans] ($(10g) + (0.85cm,0cm)$) -- ($(11e) + (0.3cm,0cm)$);
    \draw[rtrans] ($(10g) + (0.85cm,0cm)$) -- ($(00e) + (0.3cm,0cm)$);
    \draw[rtrans] ($(11e) + (0.95cm,0cm)$) -- ($(12g) + (0.05cm,0cm)$);
    \draw[nrtrans] ($(11e) + (0.9cm,0cm)$) -- ($(01g) + (0.2cm,0cm)$);
    \draw[rtrans] ($(00e) + (0.85cm,0cm)$) -- ($(01g) + (0.15cm,0cm)$);
    \draw[nrtrans] ($(12g) + (0.9cm,0cm)$) -- ($(13e) + (0.3cm,0cm)$);
    \draw[rtrans] ($(12g) + (0.9cm,0cm)$) -- ($(02e) + (0.05cm,0cm)$);
    \draw[nrtrans] ($(01g) + (0.8cm,0cm)$) -- ($(02e) + (0.3cm,0cm)$);
    \draw[nrtrans] ($(13e) + (0.8cm,0cm)$) -- ($(03g) + (0.4cm,0cm)$);
    \draw[rtrans] ($(02e) + (0.8cm,0cm)$) -- ($(03g) + (0.3cm,0cm)$);
    %
    \coordinate (v) at (0cm, 0.4cm); 
    \coordinate (l) at (1.2cm, 0cm); 
    \coordinate (h) at (0.2cm, 0cm); 
    \coordinate (0e) at ($(ccenter) + (-2.7,0)$); 
    \coordinate (1g) at ($(0e) + (l) + (h) - 3*(v)$);
    \coordinate (2e) at ($(0e) + 2*(l) + 2*(h) + 3*(v)$);
    \coordinate (3g) at ($(0e) + 3*(l) + 3*(h)$);
    \draw[level] (0e) -- ($(0e) + (l)$) node[midway,below,xshift=-0.1cm] {\footnotesize{$\ket{0,e}$}};
    \draw[level] (1g) -- ($(1g) + (l)$) node[midway,below] {\footnotesize{$\ket{1,g}$}};
    \draw[level] (2e) -- ($(2e) + (l)$) node[midway,above] {\footnotesize{$\ket{2,e}$}};
    \draw[level] (3g) -- ($(3g) + (l)$) node[midway,below,xshift=0.1cm] {\footnotesize{$\ket{3,g}$}};
    \draw[rtrans] ($(0e) + (0.9cm,0cm)$) -- ($(1g) + (0.3cm,0cm)$);
    \draw[nrtrans] ($(1g) + (0.9cm,0cm)$) -- ($(2e) + (0.3cm,0cm)$);
    \draw[rtrans] ($(2e) + (0.9cm,0cm)$) -- ($(3g) + (0.3cm,0cm)$);
    %
    \coordinate (v) at (0cm, 0.8cm); 
    \coordinate (l) at (1.2cm, 0cm); 
    \coordinate (h) at (0.2cm, 0cm); 
    \coordinate (1ggg) at ($(ccenter) + (3.8,0) - 0.5*(v)$); 
    \coordinate (2egg) at ($(1ggg) + (l) + (h) + 2*(v)$);
    \coordinate (0egg) at ($(1ggg) + (l) + (h) - 1*(v)$);
    \coordinate (1eeg) at ($(1ggg) + 2*(l) + 2*(h) + 1*(v)$);
    \coordinate (0eee) at ($(1ggg) + 3*(l) + 3*(h)$);
    \draw[level] (1ggg) -- ($(1ggg) + (l)$) node[midway,below,xshift= -0.05cm] {\footnotesize{$\ket{1,g,g,g}$}};
    \draw[level] (2egg) -- ($(2egg) + (l)$) node[midway,above,xshift=0.6cm,yshift=-0.1cm] {\footnotesize{$\frac{1}{\sqrt{3}}\left(\ket{2,e,g,g} + \ket{2,g,e,g} + \ket{2,g,g,e} \right)$}};
    \draw[level] (0egg) -- ($(0egg) + (l)$) node[midway,below,xshift=0.65cm,yshift=0.1cm] {\footnotesize{$\frac{1}{\sqrt{3}}\left(\ket{0,e,g,g} + \ket{0,g,e,g} + \ket{0,g,g,e} \right)$}};
    \draw[level] (1eeg) -- ($(1eeg) + (l)$) node[midway,below,xshift=1.2cm,yshift=0.75cm] {\parbox{1.5cm}{\centering \footnotesize{$\frac{1}{\sqrt{3}}(\ket{1,e,e,g}$ \\ $\quad + \ket{1,e,g,e} $ \\ $\quad + \ket{1,g,e,e})$}}};
    \draw[level] (0eee) -- ($(0eee) + (l)$) node[midway,below,xshift=0cm] {\footnotesize{$\ket{0,e,e,e}$}};
    \draw[nrtrans] ($(1ggg) + (1.05cm,0cm)$) -- ($(2egg) + (0.3cm,0cm)$);
    \draw[rtrans] ($(1ggg) + (1.05cm,0cm)$) -- ($(0egg) + (0.3cm,0cm)$);
    \draw[rtrans] ($(2egg) + (0.9cm,0cm)$) -- ($(1eeg) + (0.3cm,0cm)$);
    \draw[nrtrans] ($(0egg) + (0.9cm,0cm)$) -- ($(1eeg) + (0.3cm,0cm)$);
    \draw[rtrans] ($(1eeg) + (0.8cm,0cm)$) -- ($(0eee) + (0.1cm,0cm)$);
    %
    \coordinate (ground) at ($(ccenter2) + (-1cm,-1.5cm)$); 
    \coordinate (htot) at (0cm, 3cm); 
    \coordinate (ltot) at (2cm, 0cm); 
    \coordinate (lphon) at (1cm, 0cm); 
    \draw[level] (ground) -- ($(ground) + (ltot)$);
    \draw[virtual] ($(ground) + (htot)$) -- ($(ground) + (htot) + (ltot)$);
    \draw[virtual] ($(ground) + 0.667*(htot) + (ltot) - (lphon)$) -- ($(ground) + 0.667*(htot) + (ltot)$);
    \draw[virtual] ($(ground) + 0.333*(htot) + (ltot) - (lphon)$) -- ($(ground) + 0.333*(htot) + (ltot)$);
    \draw[rtrans2] ($(ground) + (0.5cm,0cm)$) -- ($(ground) + (htot) + (0.5cm,0cm)$) node[color=black, midway, left] {$3\omega$};
    \draw[rtrans2] ($(ground) + (htot) + (1.5cm,0cm)$) -- ($(ground) + 0.667*(htot) + (1.5cm,0cm)$) node[color=black, midway, right] {$\omega$};
    \draw[rtrans2] ($(ground) + 0.667*(htot) + (1.5cm,0cm)$) -- ($(ground) + 0.333*(htot) + (1.5cm,0cm)$) node[color=black, midway, right] {$\omega$};
    \draw[rtrans2] ($(ground) + 0.333*(htot) + (1.5cm,0cm)$) -- ($(ground) + (1.5cm,0cm)$) node[color=black, midway, right] {$\omega$};
    \end{tikzpicture}
  }
}
\caption{Realizations of down- and upconversion with four-wave mixing. The upper left panel shows all virtual transitions that contribute to the downconversion process (third-subharmonic generation) $\ket{1,0,g} \to \ket{0,3,g}$ to lowest order. Similarly, the upper middle panel shows all virtual transitions that contribute to the downconversion process $\ket{0,e} \to \ket{3,g}$ to lowest order, and the upper right panel shows all virtual transitions that contribute to the downconversion process $\ket{1,g,g,g} \to \ket{0,e,e,e}$ to lowest order. The lower panel show the generic level diagram for the process in nonlinear optics. The same arrow and level styles as in \figref{fig:SecondSubharmonicDetailed} are used. In the upper left panel, $\omega_q = 3\omega_a$; in the upper middle panel, we have set $\omega_a = 3\omega_b$ and $\omega_q = 2\omega_b$; in the upper right panel, $\omega_a = 3\omega_q$. If the directions of all arrows in the entire figure are reversed, upconversion (third-harmonic generation) is shown instead. \label{fig:ThirdSubharmonicThreeSchemes}}
\end{figure*}
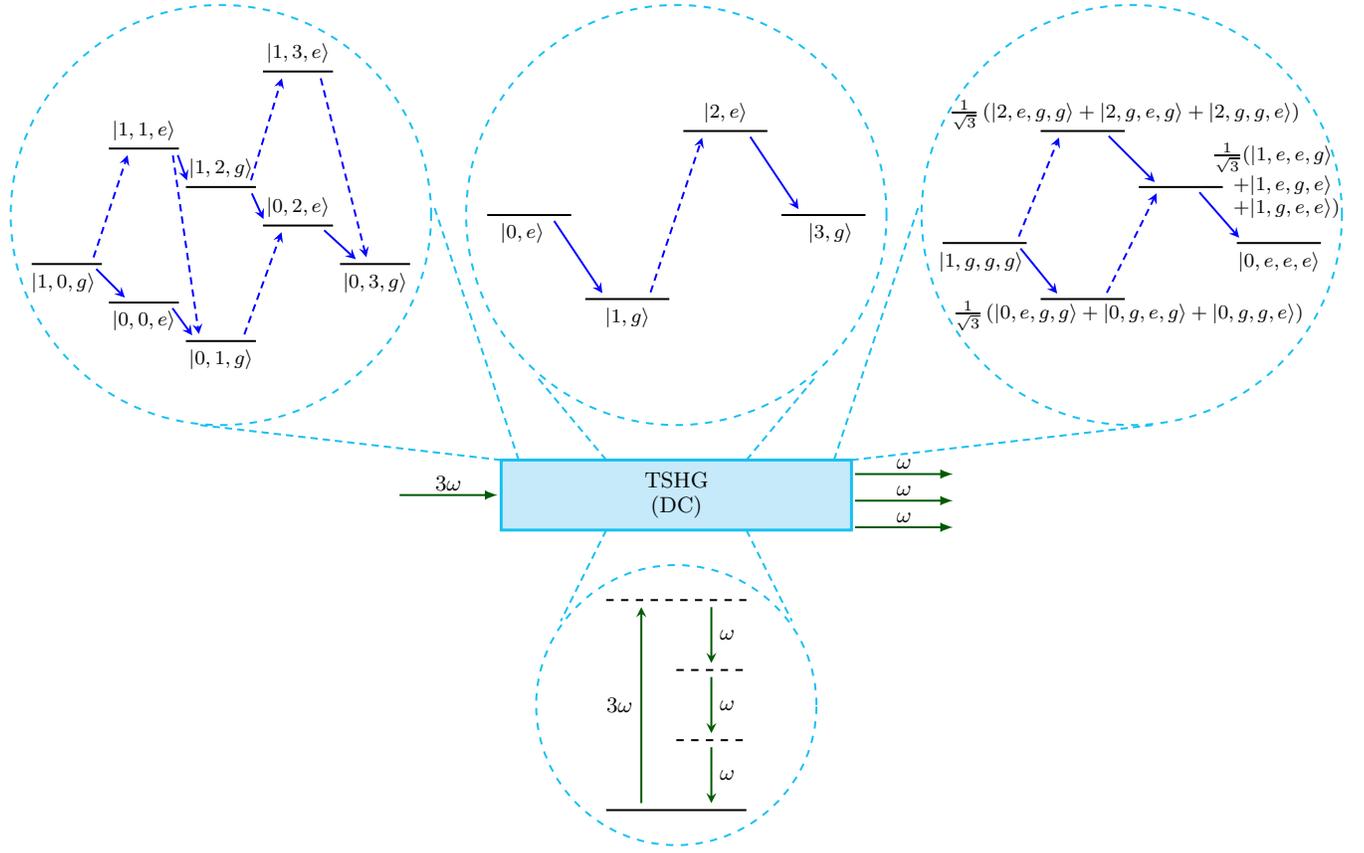

Also in this case, there are various possible deterministic setups, extensions of those discussed in \secref{sec:Degenerate3WaveMixingDeterministic}. The three most straightforward setups are illustrated in \figref{fig:ThirdSubharmonicThreeSchemes}. We note from the figure that although these setups in general require one more intermediate step than in the three-wave-mixing case, the calculations of the effective coupling are simplified by the fact that we only need to use transitions mediated by the quantum Rabi Hamiltonian (blue arrows in the figure), and not the $\sz$ terms of the generalized Rabi Hamiltonian (red arrows in \figref{fig:SecondSubharmonicDetailed}), since the excitation-number parity is conserved.

\paragraph{Two resonators}
\label{sec:Degenerate4WaveMixingDeterministicTwoResonators}

The first analogue, shown in the upper left panel of \figref{fig:ThirdSubharmonicThreeSchemes}, utilizes two resonators, with frequencies $\omega_a \approx 3\omega_b$, coupled to a single qubit such that virtual intermediate transitions enable the process $\ket{1,0,g} \leftrightarrow \ket{0,3,g}$, which realizes both up- and downconversion. The full Hamiltonian for this system is given by \eqref{eq:H2res1qb} and 
\be
\hat H_{\rm int} = \left[g_a \left( \hat a + \hat a^\dag \right) + g_b \left( \hat b + \hat b^\dag \right) \right] \sx.
\label{eq:Hint2res1qbRabi}
\ee

We can derive, in the same way as before, an effective Hamiltonian
\be
\hat H_{\rm int}^{\rm eff} = g_{\rm eff} \ketbra{1,0,g}{0,3,g} +  {\rm H.c.}
\ee
The effective coupling requires fourth-order perturbation theory to calculate. Summing the four contributing paths using \eqref{eq:PerturbationFormula} with $\ket{i} = \ket{0,3,g}$ and $\ket{f} = \ket{1,0,g}$ gives
\bea
g_{\rm eff} &=& \sqrt{6} g_a g_b^3 \bigg( - \frac{1}{\left( \Omega_{aq} - 2\omega_b \right) \Delta_{ab} \Omega_{aq}} \nn\\
&&+ \frac{1}{\left( \Omega_{aq} - 2\omega_b \right) \Delta_{ab} \Delta_{bq}} - \frac{1}{ 2\omega_b \left( \Omega_{aq} - 2\omega_b \right) \Delta_{bq}} \nn\\
&&+ \frac{1}{ 2\omega_b \left( 3\omega_b + \omega_q \right) \Delta_{bq}} \bigg).
\eea
Applying the resonance condition $\omega_a = 3\omega_b$ simplifies this result to
\bea
g_{\rm eff} &=& \frac{\sqrt{6} g_a g_b^3}{2\omega_b} \left( \frac{1}{\Delta_{bq} \left(3 \omega_b - \omega_q \right)} - \frac{1}{\Omega_{bq} \left(3 \omega_b + \omega_q \right)} \right) \nn\\
&=& \frac{4 \sqrt{6} g_a g_b^3 \omega_q}{9 \omega_b^4 - 10 \omega_b^2 \omega_q^2 + \omega_q^4},
\eea
which scales as $(g_j/\omega)^3$, $j=a,b$, as expected for a fourth-order process.

\paragraph{Multiphoton Rabi oscillations}

The second option, shown in the upper middle panel of \figref{fig:ThirdSubharmonicThreeSchemes}, is multiphoton Rabi oscillations between the states $\ket{0,e}$ and $\ket{3,g}$ with a single resonator coupled to a single qubit ($\omega_q \approx 3\omega_a$), a process studied in Refs.~\cite{Ma2015, Garziano2015}. The Hamiltonian for the system is given by Eqs.~(\ref{eq:H1res1qb}) and (\ref{eq:HintRabi}). 

The effective interaction Hamiltonian for this process is
\be
\hat H_{\rm int}^{\rm eff} = g_{\rm eff} \ketbra{0,e}{3,g} +  {\rm H.c.}
\ee
The effective coupling for three-photon Rabi oscillations follows immediately from third-order perturbation theory as there is only a single path contributing. With $\ket{i} = \ket{3,g}$ and $\ket{f} = \ket{0,e}$, \eqref{eq:PerturbationFormula} gives
\be
g_{\rm eff} = \frac{\sqrt{6} g^3}{2\omega_a \Delta_{aq}} = - \frac{9 \sqrt{6} g^3}{4 \omega_q^2},
\ee
where we used the resonance condition $\omega_q = \omega_a/3$ in the last step. This result was also derived in Ref.~\cite{Ma2015} using adiabatic elimination.

\paragraph{Three identical qubits}
\label{sec:Degenerate4WaveMixingDeterministicThreeQubits}

A third possibility, shown in the upper right panel of \figref{fig:ThirdSubharmonicThreeSchemes}, is coupling a single resonator to three identical qubits ($\omega_a = 3\omega_q$), such that the process $\ket{1,g,g,g} \leftrightarrow \ket{0,e,e,e}$ is implemented, as discussed in Ref.~\cite{Garziano2016}. In this case, the Hamiltonian for the system is
\bea
\hat H &=& \omega_a \hat a^\dag \hat a + \sum_{j=1}^3 \omega_{q}\frac{\sz^{(j)}}{2} + \hat H_{\rm int}, \\
\hat H_{\rm int} &=& g \left( \hat a + \hat a^\dag \right) \sum_{j=1}^3 \sx^{(j)},
\eea
and the effective interaction Hamiltonian of interest is 
\be
\hat H_{\rm int}^{\rm eff} = g_{\rm eff} \ketbra{1,g,g,g}{0,e,e,e} +  {\rm H.c.}
\ee

The effective coupling can be calculated with fourth-order perturbation theory. Following \eqref{eq:PerturbationFormula}, adding up the contributions from the two paths with $\ket{i} = \ket{0,e,e,e}$ and $\ket{f} = \ket{1,g,g,g}$, leads to
\be
g_{\rm eff} = g^3 \left( \frac{3}{\Delta_{qa}^2} + \frac{6}{2\omega_q \Delta_{qa}} \right) =  - \frac{3 g^3 \left(\omega_a - 3\omega_q \right)}{\omega_q \Delta_{qa}^2},
\ee
which goes to zero on resonance ($\omega_a = 3\omega_q$); the two paths interfere destructively then. However, as shown numerically in the appendix of Ref.~\cite{Garziano2016}, a coupling between the states $\ket{1,g,g,g}$ and $\ket{0,e,e,e}$ nevertheless exists close to that resonance. This is partly due to the fact that the energy levels are shifted from their bare-state values to the dressed states induced by the ultrastrong interaction and partly due to the influence of higher-order processes.

Comparing the three analogues given here, we note that the multiphoton Rabi oscillations and the single photon exciting three qubits both require one less intermediate step than the setup with two resonators and one qubit. However, in the three-qubit case this does not necessarily translate into a stronger effective coupling due to destructive interference between the virtual transitions. The possibility of such destructive interference diminishing the effective coupling needs to be kept in mind when designing analogues of nonlinear optics in these setups. We will see one more example of this phenomenon below.

\subsection{Hyper-Raman scattering, type I: Two-photon processes}
\label{sec:4WaveMixingHyperRamanScatteringTypeI}

\subsubsection{Nonlinear optics}

Hyper-Raman scattering is a generalization of Raman scattering (see \secref{sec:3WaveMixingRaman}) to include either multiple incoming photons or multiple phonons. Here, we first analyze hyper-Raman scattering based on two-photon processes (we refer to this as type~I hyper-Raman scattering), as described by the following Hamiltonians for Stokes frequency,
\be
\hat{H}^{(S)}_{\rm int} = g_S \hat a^2_L \hat a_S^\dagger \hat a_V^\dag + {\rm H.c.},
\ee
and anti-Stokes frequency (which could also be called sideband hypercooling of type I),
\be
\hat{H}^{(A)}_{\rm int} = g_A^* \hat a^2_L \hat a^\dag_A \hat a_V + {\rm H.c.}
\ee
These processes are sketched in \figref{fig:4WaveMixing}(c). We note that here $\omega_S > \omega_L$, contrary to the standard Raman scattering case. For simplicity, we have omitted multiphonon versions analogous to \eqref{eq:StokesRamanSum}.

\subsubsection{Analogous processes}
\label{sec:4WaveMixingHyperRamanScatteringTypeIDeterministic}

Just as in \secref{sec:3WaveMixingRamanDeterministic}, we consider setups where qubit excitations play the role of phonons in the deterministic analogues of hyper-Raman scattering. For the type~I process, two resonators (one corresponding to the $L$ mode, one corresponding to the $S$ or $A$ mode) are coupled to a single qubit. This setup is studied further in our forthcoming work Ref.~\cite{Kockum2017} as a means to implement deterministic up- and downconversions controlled by a qubit.

\paragraph{Stokes Hyper-Raman scattering, type~I}

\begin{figure}[ht!]
\centerline{
  \resizebox{0.9\columnwidth}{!}{
    \begin{tikzpicture}[
      scale=1,
      level/.style={thick},
      virtual/.style={thick,dashed},
      ztrans/.style={thick,->,shorten >=0.1cm,shorten <=0.1cm,>=stealth,densely dashed,color=red},
      nrtrans/.style={thick,->,shorten >=0.1cm,shorten <=0.1cm,>=stealth,densely dashed,color=blue},
      rtrans/.style={thick,->,shorten >=0.1cm,shorten <=0.1cm,>=stealth,color=blue},
      strans/.style={rtrans, shorten >=0.04cm, snake=snake, line after snake=1.7mm},
      rtrans2/.style={thick,->,shorten >=0.1cm,shorten <=0.1cm,>=stealth,color=green!35!black},
      strans2/.style={rtrans2, snake=snake, line after snake=1.7mm},
      classical/.style={thin,double,<->,shorten >=4pt,shorten <=4pt,>=stealth},
      mybox/.style={color=cyan!70, fill=cyan!20, very thick},
      myarrow/.style={color=green!35!black, thick, ->, shorten >=0.05cm, shorten <=0.05cm, >=latex},
      mysquigglyarrow/.style={color=green!35!black, thick, ->, shorten >=0.05cm, shorten <=0.05cm, >=latex, snake=snake, line after snake=2mm}
    ]
    \newcommand\labelyshift{-0.05}; 
    \newcommand\labelxshift{0}; 
    \newcommand\labelyshifttwo{0.8}; 
    \newcommand\labelxshifttwo{0}; 
    \coordinate (corner) at (0cm, 0cm); 
    \coordinate (rh) at (0cm, 1cm); 
    \coordinate (rl) at (2cm, 0cm); 
    \coordinate (ccenter) at ($(corner) + 0.5*(rl) + (0cm, 3cm) + 1.5*(rh)$); 
    \coordinate (ccenter2) at ($(corner) + 0.5*(rl) - (0cm, 2cm) - 0.5*(rh)$); 
    \coordinate (al) at (1.5cm, 0cm); 
    \filldraw[mybox](corner) rectangle ($(corner) + (rh) + (rl)$) node[color=black, midway] {$\rm H_{I}S$};
    \draw[color=cyan!70, thick, dashed] (ccenter) circle (3cm);
    \draw[color=cyan!70, thick, densely dashed] ($(corner) + (rh)$) --  ($(corner) + (rh) + (-1.2cm,1.4cm)$);
    \draw[color=cyan!70, thick, densely dashed] ($(corner) + (rh) + (rl)$) --  ($(corner) + (rh) + (rl) + (1.2cm,1.4cm)$);
    \draw[color=cyan!70, thick, dashed] (ccenter2) circle (2cm);
    \draw[color=cyan!70, thick, densely dashed] (corner) --  ($(corner) + (-0.5cm,-1.2cm)$);
    \draw[color=cyan!70, thick, densely dashed] ($(corner) + (rl)$) --  ($(corner) + (rl) + (0.5cm,-1.2cm)$);
    \draw[myarrow] ($(corner) + 0.75*(rh) - (al)$) -- ($(corner) + 0.75*(rh)$) node[color=black, midway, above, shift={(\labelxshift,\labelyshift)}] {$\omega_L$};
    \draw[myarrow] ($(corner) + 0.25*(rh) - (al)$) -- ($(corner) + 0.25*(rh)$) node[color=black, midway, above, shift={(\labelxshift,\labelyshift)}] {$\omega_L$};    
    \draw[myarrow] ($(corner) + 0.75*(rh) + (rl)$) -- ($(corner) + 0.75*(rh) + (rl) + (al)$) node[color=black, midway, above, shift={(\labelxshift,\labelyshift)}] {$\omega_S$};
    \draw[mysquigglyarrow] ($(corner) + 0.25*(rh) + (rl)$) -- ($(corner) - 0.25*(rh) + (rl) + (al)$) node[color=black, midway, above, shift={(\labelxshift,0.04)}] {$\omega_V$};
    \coordinate (v) at (0cm, 0.55cm); 
    \coordinate (l) at (1.2cm, 0cm); 
    \coordinate (h) at (0.2cm, 0cm); 
    \coordinate (02g) at ($(ccenter) + (-2.7cm,0cm) + 0*(v) $); 
    \coordinate (12e) at ($(02g) + (l) + (h) + 4*(v)$);
    \coordinate (01e) at ($(02g) + (l) + (h) - 1*(v)$);
    \coordinate (11g) at ($(02g) + 2*(l) + 2*(h) +1*(v)$);
    \coordinate (00g) at ($(02g) + 2*(l) + 2*(h) - 4*(v)$);
    \coordinate (10e) at ($(02g) + 3*(l) + 3*(h)$);
    \draw[level] (02g) -- ($(02g) + (l)$) node[midway,below,xshift=-0.1cm] {\footnotesize{$\ket{0,2,g}$}};
    \draw[level] (12e) -- ($(12e) + (l)$) node[midway,above] {\footnotesize{$\ket{1,2,e}$}};
    \draw[level] (01e) -- ($(01e) + (l)$) node[midway,below] {\footnotesize{$\ket{0,1,e}$}};
    \draw[level] (11g) -- ($(11g) + (l)$) node[midway,above] {\footnotesize{$\ket{1,1,g}$}};
    \draw[level] (00g) -- ($(00g) + (l)$) node[midway,below] {\footnotesize{$\ket{0,0,g}$}};
    \draw[level] (10e) -- ($(10e) + (l)$) node[midway,below,xshift=-0.1cm] {\footnotesize{$\quad\:\:\ket{1,0,e}$}};
    \draw[nrtrans] ($(02g) + (0.95cm,0cm)$) -- ($(12e) + (0.3cm,0cm)$);
    \draw[rtrans] ($(02g) + (0.95cm,0cm)$) -- ($(01e) + (0.3cm,0cm)$);
    \draw[nrtrans] ($(12e) + (1cm,0cm)$) -- ($(11g) + (0.1cm,0cm)$);
    \draw[rtrans] ($(01e) + (1cm,0cm)$) -- ($(11g) + (0.3cm,0cm)$);    
    \draw[nrtrans] ($(01e) + (1cm,0cm)$) -- ($(00g) + (0.3cm,0cm)$);   
    \draw[rtrans] ($(11g) + (1cm,0cm)$) -- ($(10e) + (0.2cm,0cm)$);
    \draw[nrtrans] ($(00g) + (1cm,0cm)$) -- ($(10e) + (0.2cm,0cm)$);       
    \coordinate (ground) at ($(ccenter2) + (-1cm,-1.5cm)$); 
    \coordinate (htot) at (0cm, 3cm); 
    \coordinate (ltot) at (2cm, 0cm); 
    \coordinate (ls) at (1cm, 0cm); 
    \draw[level] (ground) -- ($(ground) + (ltot)$);
    \draw[virtual] ($(ground) + 0.5*(htot) $) -- ($(ground) + 0.5*(htot) + (ls)$);    
    \draw[virtual] ($(ground) + (htot)$) -- ($(ground) + (htot) + (ltot)$);
    \draw[level] ($(ground) + 0.25*(htot) + (ls)$) -- ($(ground) + 0.25*(htot) + (ltot)$);
    \draw[rtrans2] ($(ground) + (0.5cm,0cm)$) -- ($(ground) + 0.5*(htot) + (0.5cm,0cm)$) node[color=black, midway, left] {$\omega_L$};
    \draw[rtrans2] ($(ground) + (0.5cm,0cm) + 0.5*(htot)$) -- ($(ground) + (htot) + (0.5cm,0cm)$) node[color=black, midway, left] {$\omega_L$};
    \draw[rtrans2] ($(ground) + (htot) + (1.5cm,0cm)$) -- ($(ground) + 0.25*(htot) + (1.5cm,0cm)$) node[color=black, midway, right, shift={(0.03,0)}] {$\omega_S$};
    \draw[strans2] ($(ground) + 0.25*(htot) + (1.5cm,0cm)$) -- ($(ground) + (1.5cm,0cm)$) node[color=black, midway, right, shift={(0.03,0)}] {$\omega_V$};
    \end{tikzpicture}
  }
}
\caption{Stokes hyper-Raman scattering of type~I and its deterministic analogue. The upper panel shows all virtual transitions that contribute to lowest order to the transition $\ket{0,2,g} \to \ket{1,0,e}$. The lower panel shows the generic level diagram for the process in nonlinear optics. The connection becomes clear with the identifications $a = S$, $b = L$, and $q = V$. The same arrow and level styles as in \figref{fig:SecondSubharmonicDetailed} are used; we have set $\omega_a = 3 \omega_q$ and $\omega_b = 2 \omega_q$. \label{fig:HyperRaman1StokesDetails}}
\end{figure}
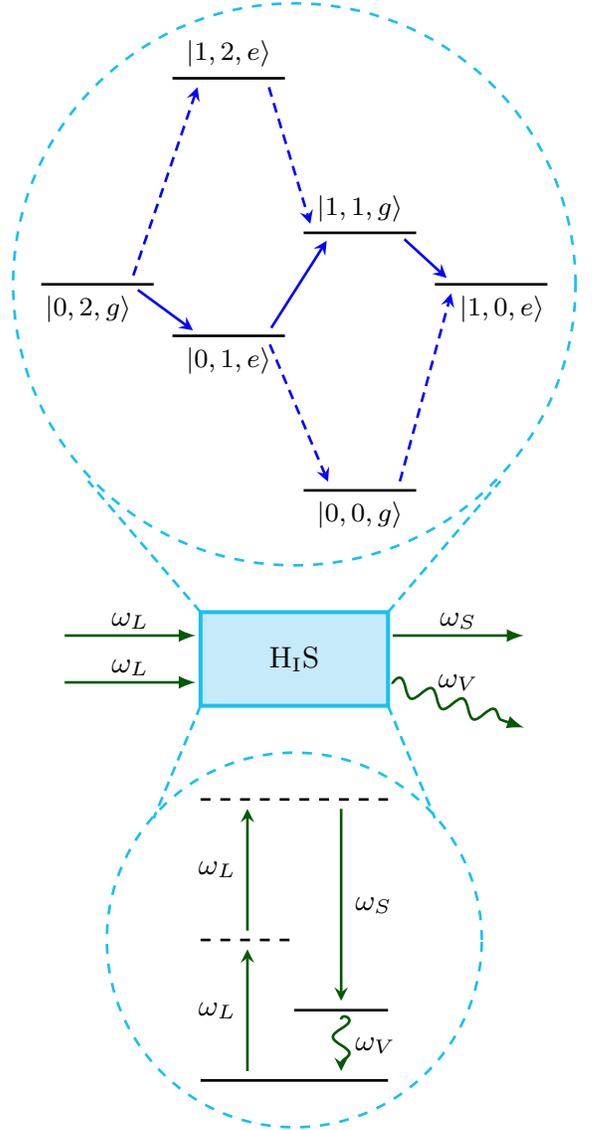

Setting $\omega_a + \omega_q \approx 2 \omega_b$ and making the connections $a = S$, $b = L$, and $q = V$, we see that the process $\ket{0,2,g} \to \ket{1,0,e}$ corresponds to Stokes hyper-Raman scattering of type~I. In the upper panel of \figref{fig:HyperRaman1StokesDetails}, we show the virtual transitions contributing to this process. From the full system Hamiltonian, given by Eqs.~(\ref{eq:H2res1qb}) and (\ref{eq:Hint2res1qbRabi}) just as for the three-photon frequency conversion in \secref{sec:Degenerate4WaveMixingDeterministicTwoResonators}, we can derive the effective Hamiltonian
\be
\hat H_{\rm int,H_I S}^{\rm eff} = g_{\rm eff} \ketbra{0,2,g}{1,0,e} +  {\rm H.c.}
\ee
Third-order perturbation theory following \eqref{eq:PerturbationFormula} gives
\bea
g_{\rm eff} &=& \sqrt{2} g_a g_b^2 \left( \frac{1}{-2 \omega_b \Delta_{qb}} + \frac{1}{\Delta_{ab} \Delta_{qb}} + \frac{1}{\Delta_{ab} \Omega_{aq}} \right) \nn\\
&=& \frac{\sqrt{2} g_a g_b^2 \left(\omega_a - 2\omega_b \right)}{\omega_b \Delta_{ab}^2},
\label{eq:GeffHyperRaman1Stokes}
\eea
where we used the resonance condition $\omega_q = 2\omega_b - \omega_a$ in the last step.

We note that one of the paths contributing to the coupling [the second term in \eqref{eq:GeffHyperRaman1Stokes}] only requires interactions given by the JC version of the interaction Hamiltonian,
\be
\hat H_{\rm int} = g_a \left( \hat a \sp + \hat a^\dag \sm \right) + g_b \left( \hat b \sp + \hat b^\dag \sm \right).
\ee
Omitting the other terms from \eqref{eq:GeffHyperRaman1Stokes} and setting $\omega_q = 2\omega_b - \omega_a$, the result is
\be
g_{\rm eff} = - \frac{\sqrt{2} g_a g_b^2}{\Delta_{ab}^2}.
\ee
These effective couplings are also calculated in Ref.~\cite{Kockum2017} using adiabatic elimination. In that case, the results are a little more complicated since some higher-order contributions are included, but to lowest order the results coincide with those given here.

\paragraph{Anti-Stokes Hyper-Raman scattering, type~I}

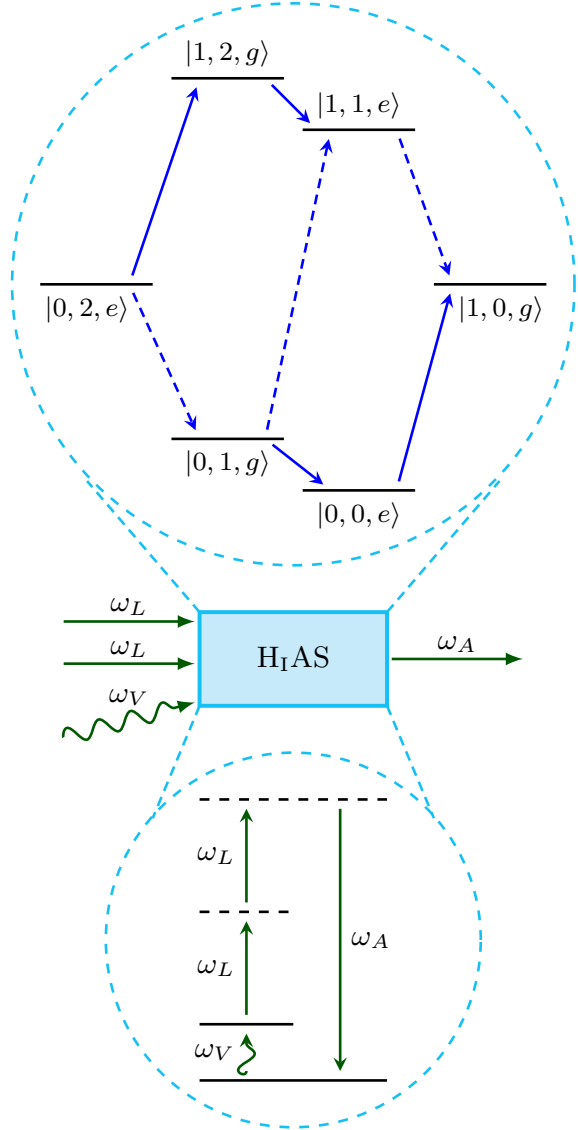
\begin{figure}
\centerline{
  \resizebox{0.9\columnwidth}{!}{
    \begin{tikzpicture}[
      scale=1,
      level/.style={thick},
      virtual/.style={thick,dashed},
      ztrans/.style={thick,->,shorten >=0.1cm,shorten <=0.1cm,>=stealth,densely dashed,color=red},
      nrtrans/.style={thick,->,shorten >=0.1cm,shorten <=0.1cm,>=stealth,densely dashed,color=blue},
      rtrans/.style={thick,->,shorten >=0.1cm,shorten <=0.1cm,>=stealth,color=blue},
      strans/.style={rtrans, shorten >=0.04cm, snake=snake, line after snake=0.05cm},
      rtrans2/.style={thick,->,shorten >=0.1cm,shorten <=0.1cm,>=stealth,color=green!35!black},
      strans2/.style={rtrans2, snake=snake, line after snake=0.05cm},
      classical/.style={thin,double,<->,shorten >=4pt,shorten <=4pt,>=stealth},
      mybox/.style={color=cyan!70, fill=cyan!20, very thick},
      myarrow/.style={color=green!35!black, thick, ->, shorten >=0.05cm, shorten <=0.05cm, >=latex},
      mysquigglyarrow/.style={color=green!35!black, thick, ->, shorten >=0.05cm, shorten <=0.05cm, >=latex, snake=snake, line after snake=2mm}
    ]
    \newcommand\labelyshift{-0.05}; 
    \newcommand\labelxshift{0}; 
    \newcommand\labelyshifttwo{0.8}; 
    \newcommand\labelxshifttwo{0}; 
    \coordinate (corner) at (0cm, 0cm); 
    \coordinate (rh) at (0cm, 1cm); 
    \coordinate (rl) at (2cm, 0cm); 
    \coordinate (ccenter) at ($(corner) + 0.5*(rl) + (0cm, 3cm) + 1.5*(rh)$); 
    \coordinate (ccenter2) at ($(corner) + 0.5*(rl) - (0cm, 2cm) - 0.5*(rh)$); 
    \coordinate (al) at (1.5cm, 0cm); 
    \filldraw[mybox](corner) rectangle ($(corner) + (rh) + (rl)$) node[color=black, midway] {$\rm H_{I}AS$};
    \draw[color=cyan!70, thick, dashed] (ccenter) circle (3cm);
    \draw[color=cyan!70, thick, densely dashed] ($(corner) + (rh)$) --  ($(corner) + (rh) + (-1.2cm,1.4cm)$);
    \draw[color=cyan!70, thick, densely dashed] ($(corner) + (rh) + (rl)$) --  ($(corner) + (rh) + (rl) + (1.2cm,1.4cm)$);
    \draw[color=cyan!70, thick, dashed] (ccenter2) circle (2cm);
    \draw[color=cyan!70, thick, densely dashed] (corner) --  ($(corner) + (-0.5cm,-1.2cm)$);
    \draw[color=cyan!70, thick, densely dashed] ($(corner) + (rl)$) --  ($(corner) + (rl) + (0.5cm,-1.2cm)$);
    \draw[myarrow] ($(corner) + 0.9*(rh) - (al)$) -- ($(corner) + 0.9*(rh)$) node[color=black, midway, above, shift={(\labelxshift,\labelyshift)}] {$\omega_L$};
    \draw[myarrow] ($(corner) + 0.45*(rh) - (al)$) -- ($(corner) + 0.45*(rh)$) node[color=black, midway, above, shift={(\labelxshift,\labelyshift)}] {$\omega_L$};    
    \draw[mysquigglyarrow] ($(corner) - 0.35*(rh) - (al)$) -- ($(corner) + 0.05*(rh)$) node[color=black, midway, above, shift={(\labelxshift,0.03)}] {$\omega_V$};
    \draw[myarrow] ($(corner) + 0.5*(rh) + (rl)$) -- ($(corner) + 0.5*(rh) + (rl) + (al)$) node[color=black, midway, above, shift={(\labelxshift,\labelyshift)}] {$\omega_A$};
    \coordinate (v) at (0cm, 0.55cm); 
    \coordinate (l) at (1.2cm, 0cm); 
    \coordinate (h) at (0.2cm, 0cm); 
    \coordinate (02e) at ($(ccenter) + (-2.7cm,0cm) + 0*(v) $); 
    \coordinate (12g) at ($(02e) + (l) + (h) + 4*(v)$);
    \coordinate (01g) at ($(02e) + (l) + (h) - 3*(v)$);
    \coordinate (11e) at ($(02e) + 2*(l) + 2*(h) + 3*(v)$);
    \coordinate (00e) at ($(02e) + 2*(l) + 2*(h) - 4*(v)$);
    \coordinate (10g) at ($(02e) + 3*(l) + 3*(h)$);
    \draw[level] (02e) -- ($(02e) + (l)$) node[midway,below,xshift=-0.1cm] {\footnotesize{$\ket{0,2,e}$}};
    \draw[level] (12g) -- ($(12g) + (l)$) node[midway,above] {\footnotesize{$\ket{1,2,g}$}};
    \draw[level] (01g) -- ($(01g) + (l)$) node[midway,below] {\footnotesize{$\ket{0,1,g}$}};
    \draw[level] (11e) -- ($(11e) + (l)$) node[midway,above] {\footnotesize{$\ket{1,1,e}$}};
    \draw[level] (00e) -- ($(00e) + (l)$) node[midway,below] {\footnotesize{$\ket{0,0,e}$}};
    \draw[level] (10g) -- ($(10g) + (l)$) node[midway,below,xshift=-0.1cm] {\footnotesize{$\quad\:\:\ket{1,0,g}$}};
    \draw[rtrans] ($(02e) + (0.95cm,0cm)$) -- ($(12g) + (0.3cm,0cm)$);
    \draw[nrtrans] ($(02e) + (0.95cm,0cm)$) -- ($(01g) + (0.3cm,0cm)$);
    \draw[rtrans] ($(12g) + (1cm,0cm)$) -- ($(11e) + (0.15cm,0cm)$);
    \draw[nrtrans] ($(01g) + (1cm,0cm)$) -- ($(11e) + (0.3cm,0cm)$);    
    \draw[rtrans] ($(01g) + (1cm,0cm)$) -- ($(00e) + (0.3cm,0cm)$);   
    \draw[nrtrans] ($(11e) + (1cm,0cm)$) -- ($(10g) + (0.2cm,0cm)$);
    \draw[rtrans] ($(00e) + (1cm,0cm)$) -- ($(10g) + (0.2cm,0cm)$);       
    \coordinate (ground) at ($(ccenter2) + (-1cm,-1.5cm)$); 
    \coordinate (htot) at (0cm, 3cm); 
    \coordinate (ltot) at (2cm, 0cm); 
    \coordinate (ls) at (1cm, 0cm); 
    \draw[level] (ground) -- ($(ground) + (ltot)$);
    \draw[level] ($(ground) + 0.2*(htot) $) -- ($(ground) + 0.2*(htot) + (ls)$);
    \draw[virtual] ($(ground) + 0.6*(htot) $) -- ($(ground) + 0.6*(htot) + (ls)$);    
    \draw[virtual] ($(ground) + (htot)$) -- ($(ground) + (htot) + (ltot)$);
    \draw[strans2] ($(ground) + (0.5cm,0cm)$) -- ($(ground) + 0.2*(htot) + (0.5cm,0cm)$) node[color=black, midway, left] {$\omega_V$};
    \draw[rtrans2] ($(ground) + (0.5cm,0cm) + 0.2*(htot)$) -- ($(ground) + 0.6*(htot) + (0.5cm,0cm)$) node[color=black, midway, left] {$\omega_L$};
    \draw[rtrans2] ($(ground) + (0.5cm,0cm) + 0.6*(htot)$) -- ($(ground) + 1.0*(htot) + (0.5cm,0cm)$) node[color=black, midway, left] {$\omega_L$};
    \draw[rtrans2] ($(ground) + (htot) + (1.5cm,0cm)$) -- ($(ground) + (1.5cm,0cm)$) node[color=black, midway, right] {$\omega_A$};
    \end{tikzpicture}
  }
}
\caption{Anti-Stokes hyper-Raman scattering of type~I (sideband hypercooling) and its deterministic analogue. The upper panel shows all virtual transitions that contribute to lowest order to the transition $\ket{0,2,e} \to \ket{1,0,g}$. The lower panel shows the generic level diagram for the process in nonlinear optics. The connection becomes clear with the identifications $a = A$, $b = L$, and $q = V$. The same arrow and level styles as in \figref{fig:SecondSubharmonicDetailed} are used; we have set $\omega_a = 5 \omega_q$ and $\omega_b = 2 \omega_q$. \label{fig:HyperRaman1AntiStokesDetails}}
\end{figure}

If we instead set $\omega_a \approx 2\omega_b + \omega_q$ in the same setup as for Stokes hyper-Raman scattering of type~I (and make the connections $a = A$, $b = L$, and $q = V$), the transition $\ket{0,2,e} \to \ket{1,0,g}$ corresponds to anti-Stokes hyper-Raman scattering of type~I. The effective interaction Hamiltonian becomes
\be
\hat H_{\rm int,H_I AS}^{\rm eff} = g_{\rm eff} \ketbra{0,2,e}{1,0,g} +  {\rm H.c.}
\ee
The virtual transitions contributing to the process $\ket{0,2,e} \to \ket{1,0,g}$ are shown in \figref{fig:HyperRaman1AntiStokesDetails}. Third-order perturbation theory following \eqref{eq:PerturbationFormula} gives
\bea
g_{\rm eff} &=& \sqrt{2} g_a g_b^2 \left( \frac{1}{2 \omega_b \Omega_{qb}} - \frac{1}{\Delta_{ab} \Omega_{qb}} + \frac{1}{\Delta_{ab} \Delta_{aq}} \right) \nn\\
&=& \frac{\sqrt{2} g_a g_b^2 \left(\omega_a - 2\omega_b \right)}{\omega_b \Delta_{ab}^2},
\label{eq:GeffHyperRaman1AntiStokes}
\eea
where we used the resonance condition $\omega_q = \omega_a - 2\omega_b$ in the last step. We note that the expression is the same as the one obtained for type-I Stokes hyper-Raman scattering in \eqref{eq:GeffHyperRaman1Stokes}, despite the resonance condition being different.

This effective coupling is also calculated in Ref.~\cite{Kockum2017} using adiabatic elimination. Again, in that case, the result is a little more complicated since some higher-order contributions are included, but to lowest order it coincides with \eqref{eq:GeffHyperRaman1AntiStokes}.

\subsection{Hyper-Raman scattering, type II: Two-phonon processes}
\label{sec:4WaveMixingHyperRamanScatteringTypeII}

\subsubsection{Nonlinear optics}

Hyper-Raman scattering can also be based on two-phonon processes (we refer to this as type~II hyper-Raman scattering), as described by the following interaction Hamiltonians with Stokes frequency,
\be
\hat{H}^{(S)}_{\rm int} = g_S \hat a_L \hat a_S^\dagger \hat a_{V1}^\dag \hat a_{V2}^\dag + {\rm H.c.},
\ee
and with anti-Stokes frequency (which could also be called sideband hypercooling of type II),
\be
\hat{H}^{(A)}_{\rm int} = g_A^* \hat a_L \hat a^\dag_A \hat a_{V1} \hat a_{V2} + {\rm H.c.}
\ee
These processes are sketched in \figref{fig:4WaveMixing}(d).

\subsubsection{Analogous processes}

The closest analogue here is a setup with two resonators both coupled to two identical qubits. The full system Hamiltonian is given by
\bea
\hat H &=& \omega_a \hat a^\dag \hat a + \omega_b \hat b^\dag \hat b + \sum_{j=1}^2 \omega_q \frac{\sz^{(j)}}{2} + \hat H_{\rm int}, \\
\hat H_{\rm int} &=& \left[ g_a \left( \hat a + \hat a^\dag \right) + g_b \left( \hat b + \hat b^\dag \right) \right] \sum_{j=1}^2 \sx^{(j)}.
\eea
If the frequencies satisfy the resonance condition $\omega_a \approx \omega_b + 2\omega_q$, the process $\ket{1,0,g,g} \to \ket{0,1,e,e}$, whose virtual transitions are shown in \figref{fig:HyperRaman2Details}, corresponds to Stokes hyper-Raman scattering of type~II, given that we make the connections $a = L$, $b = S$, and $q = V$. The reverse process corresponds to anti-Stokes hyper-Raman scattering of type~II, if we instead identify $a = A$ and $b = L$.

\begin{figure}
\centerline{
  \resizebox{0.9\columnwidth}{!}{
    \begin{tikzpicture}[
      scale=1,
      level/.style={thick},
      virtual/.style={thick,dashed},
      ztrans/.style={thick,->,shorten >=0.1cm,shorten <=0.1cm,>=stealth,densely dashed,color=red},
      nrtrans/.style={thick,->,shorten >=0.1cm,shorten <=0.1cm,>=stealth,densely dashed,color=blue},
      rtrans/.style={thick,->,shorten >=0.1cm,shorten <=0.1cm,>=stealth,color=blue},
      strans/.style={rtrans, shorten >=0.04cm, snake=snake, line after snake=1.7mm},
      rtrans2/.style={thick,->,shorten >=0.1cm,shorten <=0.1cm,>=stealth,color=green!35!black},
      strans2/.style={rtrans2, snake=snake, line after snake=1.7mm},
      classical/.style={thin,double,<->,shorten >=4pt,shorten <=4pt,>=stealth},
      mybox/.style={color=cyan!70, fill=cyan!20, very thick},
      myarrow/.style={color=green!35!black, thick, ->, shorten >=0.05cm, shorten <=0.05cm, >=latex},
      mysquigglyarrow/.style={color=green!35!black, thick, ->, shorten >=0.05cm, shorten <=0.05cm, >=latex, snake=snake, line after snake=2mm}
    ]
    \newcommand\labelyshift{-0.05}; 
    \newcommand\labelxshift{0}; 
    \newcommand\labelyshifttwo{0.8}; 
    \newcommand\labelxshifttwo{0}; 
    \coordinate (corner) at (0cm, 0cm); 
    \coordinate (rh) at (0cm, 1cm); 
    \coordinate (rl) at (2cm, 0cm); 
    \coordinate (ccenter) at ($(corner) + 0.5*(rl) + (0cm, 3cm) + 1.5*(rh)$); 
    \coordinate (ccenter2) at ($(corner) + 0.5*(rl) - (0cm, 2cm) - 0.5*(rh)$); 
    \coordinate (al) at (1.5cm, 0cm); 
    \filldraw[mybox](corner) rectangle ($(corner) + (rh) + (rl)$) node[color=black, midway] {$\rm H_{II}S$};
    \draw[color=cyan!70, thick, dashed] (ccenter) circle (3cm);
    \draw[color=cyan!70, thick, densely dashed] ($(corner) + (rh)$) --  ($(corner) + (rh) + (-1.2cm,1.4cm)$);
    \draw[color=cyan!70, thick, densely dashed] ($(corner) + (rh) + (rl)$) --  ($(corner) + (rh) + (rl) + (1.2cm,1.4cm)$);
    \draw[color=cyan!70, thick, dashed] (ccenter2) circle (2cm);
    \draw[color=cyan!70, thick, densely dashed] (corner) --  ($(corner) + (-0.5cm,-1.2cm)$);
    \draw[color=cyan!70, thick, densely dashed] ($(corner) + (rl)$) --  ($(corner) + (rl) + (0.5cm,-1.2cm)$);
    \draw[myarrow] ($(corner) + 0.5*(rh) - (al)$) -- ($(corner) + 0.5*(rh)$) node[color=black, midway, above, shift={(\labelxshift,\labelyshift)}] {$\omega_L$};           
    \draw[myarrow] ($(corner) + 0.95*(rh) + (rl)$) -- ($(corner) + 0.95*(rh) + (rl) + (al)$) node[color=black, midway, above, shift={(\labelxshift,\labelyshift)}] {$\omega_S$};   
    \draw[mysquigglyarrow] ($(corner) + 0.5*(rh) + (rl)$) -- ($(corner) + 0.1*(rh) + (rl) + (al)$) node[color=black, midway, above, shift={(0,0.02)}] {$\omega_V$};
    \draw[mysquigglyarrow] ($(corner) + 0.03*(rh) + (rl)$) -- ($(corner) - 0.37*(rh) + (rl) + (al)$) node[color=black, midway, above, shift={(0,0.02)}] {$\omega_V$};
    \coordinate (v) at (0cm, 1.2cm); 
    \coordinate (l) at (1.4cm, 0cm); 
    \coordinate (h) at (0.2cm, 0cm); 
    \coordinate (10gg) at ($(ccenter) + (-2.3cm,0cm)$); 
    \coordinate (11eg) at ($(10gg) + (l) + (h) + (v)$);
    \coordinate (00eg) at ($(10gg) + (l) + (h) - (v)$);
    \coordinate (01ee) at ($(10gg) + 2*(l) + 2*(h)$);
    \draw[level] (10gg) -- ($(10gg) + (l)$) node[midway,below,xshift=-0.1cm] {\footnotesize{$\ket{1,0,g,g}$}};
    \draw[level] (11eg) -- ($(11eg) + (l)$) node[midway,above] {\footnotesize{$\frac{1}{\sqrt{2}} \left( \ket{1,1,e,g} + \ket{1,1,g,e} \right)$}};
    \draw[level] (00eg) -- ($(00eg) + (l)$) node[midway,below] {\footnotesize{$\frac{1}{\sqrt{2}} \left( \ket{0,0,e,g} + \ket{0,0,g,e} \right)$}};
    \draw[level] (01ee) -- ($(01ee) + (l)$) node[midway,below,xshift=-0.1cm] {\footnotesize{$\quad\:\:\ket{0,1,e,e}$}};
    \draw[nrtrans] ($(10gg) + (1.2cm,0cm)$) -- ($(11eg) + (0.3cm,0cm)$);
    \draw[rtrans] ($(10gg) + (1.2cm,0cm)$) -- ($(00eg) + (0.3cm,0cm)$); 
    \draw[rtrans] ($(11eg) + (1.2cm,0cm)$) -- ($(01ee) + (0.2cm,0cm)$);
    \draw[nrtrans] ($(00eg) + (1.2cm,0cm)$) -- ($(01ee) + (0.2cm,0cm)$);       
    \coordinate (ground) at ($(ccenter2) + (-1cm,-1.5cm)$); 
    \coordinate (htot) at (0cm, 3cm); 
    \coordinate (ltot) at (2cm, 0cm); 
    \coordinate (ls) at (1cm, 0cm); 
    \draw[level] (ground) -- ($(ground) + (ltot)$);
    \draw[virtual] ($(ground) + (htot)$) -- ($(ground) + (htot) + (ltot)$);
    \draw[level] ($(ground) + 0.5*(htot) + (ls)$) -- ($(ground) + 0.5*(htot) + (ltot)$);
    \draw[level] ($(ground) + 0.25*(htot) + (ls)$) -- ($(ground) + 0.25*(htot) + (ltot)$);
    \draw[rtrans2] ($(ground) + (0.5cm,0cm)$) -- ($(ground) + (htot) + (0.5cm,0cm)$) node[color=black, midway, left] {$\omega_L$};
    \draw[rtrans2] ($(ground) + (htot) + (1.5cm,0cm)$) -- ($(ground) + 0.5*(htot) + (1.5cm,0cm)$) node[color=black, midway, right, shift={(0.03,0)}] {$\omega_S$};
    \draw[strans2] ($(ground) + 0.5*(htot) + (1.5cm,0cm)$) -- ($(ground) + 0.25*(htot) + (1.5cm,0cm)$) node[color=black, midway, right, shift={(0.03,0)}] {$\omega_V$};
    \draw[strans2] ($(ground) + 0.25*(htot) + (1.5cm,0cm)$) -- ($(ground) + (1.5cm,0cm)$) node[color=black, midway, right, shift={(0.03,0)}] {$\omega_V$};
    \end{tikzpicture}
  }
}
\caption{Stokes hyper-Raman scattering of type~II and its deterministic analogue. The upper panel shows all virtual transitions that contribute to lowest order to the transition $\ket{1,0,g,g} \to \ket{0,1,e,e}$. The lower panel shows the generic level diagram for the process in nonlinear optics. The connection becomes clear with the identifications $a = L$, $b = S$, and $q = V$. The same arrow and level styles as in \figref{fig:SecondSubharmonicDetailed} are used. If the directions of all arrows in the entire figure are reversed, and the labels are changed such that $S \to L$ and $L \to A$, anti-Stokes hyper-Raman scattering of type~II is shown instead. \label{fig:HyperRaman2Details}}
\end{figure}

The virtual transitions give rise to the effective Hamiltonian
\bea
\hat H_{\rm int}^{\rm eff} &=& g_{\rm eff} \ketbra{1,0,g,g}{0,1,e,e} +  {\rm H.c.},
\eea
which describes both processes. Adding up the two paths in \figref{fig:HyperRaman2Details} using second-order perturbation theory following \eqref{eq:PerturbationFormula}, we obtain
\be
g_{\rm eff} = 2 g_a g_b \left(\frac{1}{\Delta_{qa}} + \frac{1}{\Omega_{bq}} \right) =  - \frac{2 g_a g_b \left( \Delta_{ba} + 2 \omega_q \right)}{\Delta_{aq} \Omega_{bq}},
\ee
which goes to zero on resonance ($\omega_a = \omega_b + 2\omega_q$). However, a finite coupling should result from the fact that bare energy levels are shifted to dressed ones and higher-order processes can contribute, analogous to the situation for the single photon exciting three qubits discussed in \secref{sec:Degenerate4WaveMixingDeterministicThreeQubits}.

\section{Other nonlinear processes}
\label{sec:OtherProcesses}

While three- and four-wave mixing have been the main focus of this article, there are several other nonlinear-optics processes for which analogues can be found. In this section, we treat a few of these.

\subsection{Higher-harmonic and -subharmonic generation}

A plethora of processes are possible when considering wave-mixing involving five or more frequencies. To shed light on the relevant considerations for the deterministic analogues of these processes, it is sufficient to consider higher-harmonic and -subharmonic generation as a simple representative example.

\subsubsection{Nonlinear optics}

We consider the degenerate case of $m$-wave mixing assuming $\hat a_1 = \hat a_2 = \ldots = \hat a_{m-1} \equiv \hat a$, $\hat a_m \equiv \hat a_+$, and $\omega_+ = (m-1)\omega$. The creation and annihilation of a photon in the Fock basis can then be given as $\ket{n, n_+} \to \ket{n - m + 1, n_+ + 1}$ for $(m-1)$th-harmonic generation (upconversion) and $\ket{n, n_+} \to \ket{n + m - 1, n_+ -1}$ for $(m-1)$th-subharmonic generation (downconversion). The interaction Hamiltonian for both processes can be written as
\be
\hat H_{\rm int} = g \hat a^{m-1} \hat a^\dag_+ + g^* \hat a^{\dag (m-1)} \hat a_+,
\ee
generalizing \eqref{eq:UpDownConversionHint} for three-wave mixing and \eqref{eq:4WaveMixingThirdHarmonicHint} for four-wave mixing.

\subsubsection{Analogous processes}
\label{sec:OtherProcessesHigherHarmonicDeterministic}

It is straightforward to extend the three approaches discussed in \secref{sec:Degenerate3WaveMixingDeterministic} for three-wave mixing and in \secref{sec:Degenerate4WaveMixingDeterministic} for four-wave mixing. One can use two resonators, with frequencies $\omega_a = (m-1)\omega_b$, coupled to a single qubit such that the process $\ket{1, 0, g} \leftrightarrow \ket{0, m-1, g}$ is enabled by virtual intermediate transitions, realizing both up- and downconversion. The other approaches are to use multiphoton Rabi oscillations between $\ket{0, e}$ and $\ket{m-1, g}$ using a single resonator coupled to a single qubit with $\omega_q = (m-1)\omega_a$~\cite{Garziano2015}; or to couple a single resonator to $m-1$ identical qubits [$\omega_a = (m-1)\omega_q$] such that the process $\ket{1, g, \ldots, g} \leftrightarrow \ket{0, e, \ldots, e}$ is realized~\cite{Garziano2016}. 

What these three approaches, and all other analogues of $m$-wave mixing, have in common are that they require an increasing number of intermediate virtual transitions as $m$ increases. In general, the effective coupling $g_{\rm eff}$, determining the transition rate, will be proportional to $(g/\omega)^{n-1}$ if $n$ steps of intermediate virtual transitions are required to go between the initial and final states. Here, $g$ and $\omega$ are the coupling and the relevant system frequencies, respectively, in the quantum Rabi model discussed in \secref{sec:QuantumRabiModel}. Considering this, the fact that the multiphoton Rabi oscillations require one less intermediate step than the other two approaches described above (see Secs.~\ref{sec:Degenerate3WaveMixingDeterministic} and \ref{sec:Degenerate4WaveMixingDeterministic}) makes them the most suited to implement an analogue of higher-harmonic and -subharmonic generation. 

We also note that the standard quantum Rabi model, \eqref{eq:HintRabi}, is sufficient to mediate the virtual transitions needed for $m$-wave mixing when $m$ is even. If $m$ is odd, the interaction terms from the generalized quantum Rabi model, \eqref{eq:HintGenRabi}, are necessary to realize the analogues discussed here.

\subsection{Multiphoton absorption}

\subsubsection{Nonlinear optics}

Simultaneous absorption of multiple photons in a system is a nonlinear process, first predicted by G\"{o}ppert-Mayer~\cite{Goppert-Mayer1931}. Unlike most of the wave-mixing processes discussed above (except Raman scattering), this process changes the net energy of the system.

\subsubsection{Analogous processes}

A clear analogy of multiphoton absorption is provided by the multiphoton Rabi oscillations~\cite{Garziano2015} already discussed in the context of harmonic and subharmonic generation in Secs.~\ref{sec:Degenerate3WaveMixingDeterministic}, \ref{sec:Degenerate4WaveMixingDeterministic}, and \ref{sec:OtherProcessesHigherHarmonicDeterministic}. During a multiphoton Rabi oscillation, a single qubit absorbs $n$ photons from the resonator it is coupled to; this is the process $\ket{n,g} \rightarrow \ket{0,e}$. We also note that circuit-QED experiments with flux qubits have demonstrated multiphoton absorption in a driven qubit-resonator system~\cite{Deppe2008, Chen2016}. 

\subsection{Parametric processes}

\subsubsection{Nonlinear optics}

Many of the processes discussed above can be analyzed for the case where one of the fields is a strong drive that can be approximated as classical. As an example, consider the general three-wave-mixing processes described by Eqs.~(\ref{TWMsum}) and (\ref{TWMdif}) in \secref{sec:3WaveMixingGeneralNonlinear}. We denote the frequencies by $\omega_p \equiv \omega_1$ for the pump (drive) mode, $\omega_s \equiv \omega_2$ for the signal mode, and $\omega_i \equiv \omega_{\pm}$ for the idler mode. We then apply the parametric approximation $\hat a_p (t) \approx \mean{a_p (t)} \approx \alpha_p (t) \equiv \abs{\alpha_p} \exp \left[-i (\omega t + \phi _p) \right]$, which is usually valid if $\mean{n_p (t)} \approx \mean{n_p (t_0)} \gg \max\{1, \mean{n_s (t)}, \mean{n_i (t)}\}$, where $\hat n_x$ is the number of photons in mode $x$. For the case $\omega_p = \omega_s + \omega_i$, \eqref{TWMdif} then becomes
\bea
\hat H^{(\text{amp})} &=& g^* \alpha_p^* \hat a_i \hat a_s + g \alpha_p \hat a_i^\dag \hat a_s^\dag \nn \\
&=& \kappa \left[ \hat a_i \hat a_s e^{i (\omega t + \phi)} + \hat a_i^\dag \hat a_s^\dag e^{-i (\omega t + \phi)} \right],
\label{eq:ParAmp}
\eea
where $\omega \equiv \omega_p$, the coupling constant $g$ is rescaled as $\kappa = \abs{g \alpha_p}$, and $\phi = \phi_p - {\rm arg} (g)$. This equation describes parametric amplification (downconversion). For the special case $\omega_s = \omega_i$, it corresponds to degenerate parametric downconversion. Similarly, under the parametric approximation, with $\hat a_1 \equiv \hat a_p \approx \alpha_p$, \eqref{TWMsum} describes parametric frequency conversion.

\subsubsection{Analogous processes}
\label{sec:OtherProcessesParametricDeterministic}

As discussed in \secref{sec:Degenerate3WaveMixingDeterministicTwoResonators}, Ref.~\cite{Moon2005} showed that a setup with two resonator modes coupled to a qubit with the interaction of \eqref{eq:HintGenRabi} can give an effective interaction of the form
\be
\hat H_{\rm int}^{\rm eff} = \zeta \left( \hat a^{\dag 2} \hat b + \hat a^2 \hat b^\dag \right) \sz,
\label{eq:ParDownConversionDeterministic}
\ee
where 
\be
\zeta = \frac{g_a^2 g_b^2 \sin \theta \sin (2\theta)}{\omega_a \left(\omega_q - \omega_b \right)}.
\ee
This interaction results in degenerate parametric downconversion and squeezing. Note that the qubit state will affect the process since the interaction is proportional to $\sz$, which is absent in \eqref{eq:ParAmp}. However, if we assume that the qubit remains in its ground (or excited) state during the system evolution, we can recover \eqref{eq:ParAmp} from \eqref{eq:ParDownConversionDeterministic}. Such qubit ``freezing'' can be achieved, e.g., by Zeno-type effects.

In general, similar effective interaction Hamiltonians should be possible to derive for all setups considered in this article where the initial and final states for the system have the same qubit state. Thus, most parametric processes from nonlinear optics have deterministic analogues involving virtual photons.  

\subsection{Kerr, cross-Kerr, and Pockels effects}

\subsubsection{Nonlinear optics}

The Kerr, cross-Kerr, and Pockels effects differ from the other nonlinear-optics phenomena discussed so far in that they do not involve any change in the number of excitations in some mode. Instead, the frequency of a mode $a$ is modified, either through self-interaction (Kerr effect) or through interaction with a second mode $b$ (Pockels effect when the change is proportional to the amplitude of the field; cross-Kerr effect when the change is proportional to the square of said amplitude). These effects can be described by the following Hamiltonians:
\bea
\hat H_{\rm K} &=& \chi_{\rm K} \left( \hat a^\dag \hat a \right)^2, \label{eq:HKerr}\\
\hat H_{\rm cK} &=& \chi_{\rm cK} \hat a^\dag \hat a \hat b^\dag \hat b, \\
\hat H_{\rm P} &=& \chi_{\rm P} \hat a^\dag \hat a \left( \hat b + \hat b^\dag \right),
\eea
where $\chi_x$ gives the strength of the nonlinear interaction.

\subsubsection{Analogous processes}

The Kerr effect can be realized with a single qubit coupled to a resonator with only the JC interaction of \eqref{eq:HintJC}. In the dispersive regime, where $g \ll \abs{\omega_a - \omega_q}$, a perturbation expansion in the small parameter $g/(\omega_a - \omega_q)$ yields a term~\cite{Boissonneault2009}
\be
\hat H_{\rm K}^{\rm disp} = \chi_{\rm K} \left( \hat a^\dag \hat a \right)^2 \sz,
\label{eq:DispKerr}
\ee
where
\be
\chi_{\rm K} = - \frac{g^4}{(\omega_a - \omega_q)^3}
\ee
This Hamiltonian reduces to the standard Kerr Hamiltonian, given in \eqref{eq:HKerr}, if we can assume that the qubit remains in one and the same state during the system evolution, as discussed above in \secref{sec:OtherProcessesParametricDeterministic}. More general derivations for multiple resonator modes and a multi-level atom in the dispersive regime have shown how both Kerr and cross-Kerr effects can be realized~\cite{Nigg2012, Zhu2013}. In particular, Ref.~\cite{Zhu2013} demonstrates clearly how an atom coupled to two resonators in the dispersive regime via a general coupling like \eqref{eq:HintGenRabi} gives rise to the Kerr and cross-Kerr effects due to fourth-order processes involving virtual photons in the same way as all other analogues of nonlinear optics discussed previously in this article. We note that these Kerr and cross-Kerr terms, just like in Eqs.~(\ref{eq:ParDownConversionDeterministic}) and (\ref{eq:DispKerr}), involve sums over the diagonal qubit operators $\ketbra{g}{g}$ and $\ketbra{e}{e}$.

Based on the above theory, experiments in circuit QED have recently demonstrated both the single-photon Kerr~\cite{Kirchmair2013} and cross-Kerr effects~\cite{Holland2015}. A large cross-Kerr effect for propagating photons interacting with a three-level artificial atom in circuit QED has also been studied theoretically~\cite{Fan2013} and experimentally~\cite{Hoi2013a}. However, to the best of our knowledge, no such experimental demonstrations exists for the Pockels effect and we have been unable to find a mechanism for engineering it in the setups we consider.

\section{Summary and outlook}
\label{sec:SummaryOutlook}

We have shown how analogues of nonlinear optics can be realized in systems where one or more qubits are coupled to one or more resonator modes. These analogous processes are all based on the light-matter interaction between a qubit and a photonic mode described by the quantum Rabi Hamiltonian or some generalized version thereof. This interaction allows the number of excitations in the system to change, which makes possible the creation and annihilation of virtual photons and qubit excitations. In this way, initial and final states of the nonlinear-optics processes can be connected via a number of virtual transitions, creating an effective deterministic coupling between the states. The effective coupling decreases when the number of intermediate transition steps increases. However, with the recent experimental demonstrations of USC in a variety of systems, circuit QED in particular, it should now be possible to observe many of these nonlinear-optics phenomena in new settings.

For the case of three-wave mixing, we have shown how analogues can be constructed for sum- and difference-frequency generation, including the special cases of second-harmonic generation (upconversion) and second-subharmonic generation (downconversion) as well as Stokes and anti-Stokes spontaneous and stimulated Raman scattering. A summary of all the three-wave-mixing processes and their analogues is given in \tabref{tab:Summary3Wave}.

Similarly, for the case of four-wave mixing, we have shown how analogues can be realized for all types of non-degenerate and degenerate mixing, including third-harmonic and third-subharmonic generation as well as all forms of hyper-Raman scattering. We provide a summary of all the four-wave-mixing processes and their analogues in \tabref{tab:Summary4Wave}. Finally, we have also shown that analogues working according to the same principle are available for higher-harmonic and -subharmonic generation, multiphoton absorption, parametric processes, and the Kerr and cross-Kerr effects. 

It is noteworthy that some of the setups we consider, especially the relatively simple setups of a single qubit coupled to one or two resonators, can be used to realize many analogues of nonlinear-optics phenomena in one universal system. While some processes that we discuss here have been investigated in previous and forthcoming publications, we have now provided a unified and clear picture of how and why nonlinear-optics analogues can be constructed in these setups.

There are many directions for future work following this article. They include deriving effective Hamiltonians for more parametric processes based on the setups discussed here and finding an analogue of the Pockels effect. An interesting way to take the ideas of the current work one step further is to consider analogues of nonlinear-optics processes where the excitations are exchanged only between atoms and any resonators in the setups are only excited virtually, which will be treated in a forthcoming publication~\cite{Stassi2017}. We also see a great potential for using the processes described here to create various superposition states with applications in quantum information. Considering an experimental implementation, we believe that most of the processes discussed here can be realized with currently available technology in circuit QED.

\section*{Acknowledgements}

A.F.K. acknowledges support from a JSPS Postdoctoral Fellowship for Overseas Researchers. A.M. and F.N. acknowledge the support of a grant from the John Templeton Foundation. F.N. was also partially supported by the RIKEN iTHES Project, the MURI Center for Dynamic Magneto-Optics via the AFOSR award number FA9550-14-1-0040, the IMPACT program of JST, CREST, and a Grant-in-Aid for Scientific Research (A).

\clearpage
\begin{turnpage}

\begin{table}
\centering
\caption{A summary of three-wave-mixing processes in nonlinear optics and their deterministic analogues with single atoms and virtual photons. In the case of nondegenerate three-wave mixing, with the exception of Raman scattering, the given frequencies and transitions are just some of the possibilities. \label{tab:Summary3Wave}}
\renewcommand{\arraystretch}{1.2}
\renewcommand{\tabcolsep}{0.1cm}
\begin{tabular}{|c|c|c|c|c|c|c|c|} \hline
\multicolumn{3}{|c|}{\textbf{Nonlinear-optics process}} & \textbf{Analogous setup} & \textbf{Frequencies} & \textbf{Transition} & \textbf{Hamiltonian} & \textbf{Reference}  \\ \hline
\multirow{6}{*}{\parbox{2cm}{\textbf{Degenerate three-wave mixing}}} & \multicolumn{2}{|c|}{\multirow{3}{*}{\parbox{4cm}{Second-harmonic generation \\ (upconversion)}}} & 1 resonator, 1 qubit & $\omega_q = 2 \omega_a$ & $\ket{2, g} \to \ket{0, e}$ & Gen.~Rabi & \secref{sec:Degenerate3WaveMixing},~\cite{Garziano2015} \\ \cline{4-8}
 & \multicolumn{2}{|c|}{} & 2 resonators, 1 qubit & $\omega_a = 2 \omega_b$ & $\ket{0, 2, g} \to \ket{1, 0, g}$ & Gen.~Rabi & \secref{sec:Degenerate3WaveMixing},~\cite{Moon2005} \\ \cline{4-8}
 & \multicolumn{2}{|c|}{} & 1 resonator, 2 qubits & $\omega_a = 2 \omega_q$ & $\ket{0, e, e} \to \ket{1, g, g}$ & Gen.~Rabi & \secref{sec:Degenerate3WaveMixing},~\cite{Garziano2016} \\ \cline{2-8}
 & \multicolumn{2}{|c|}{\multirow{3}{*}{\parbox{4.5cm}{Second-subharmonic generation \\ (downconversion)}}} & 1 resonator, 1 qubit & $\omega_q = 2 \omega_a$ & $\ket{0, e} \to \ket{2, g}$ & Gen.~Rabi & \secref{sec:Degenerate3WaveMixing},~\cite{Garziano2015} \\ \cline{4-8}
 & \multicolumn{2}{|c|}{} & 2 resonators, 1 qubit & $\omega_a = 2 \omega_b$ & $\ket{1, 0, g} \to \ket{0, 2, g}$ & Gen.~Rabi & \secref{sec:Degenerate3WaveMixing},~\cite{Moon2005} \\ \cline{4-8}
 & \multicolumn{2}{|c|}{} & 1 resonator, 2 qubits & $\omega_a = 2 \omega_q$ & $\ket{1, g, g} \to \ket{0, e, e}$ & Gen.~Rabi & \secref{sec:Degenerate3WaveMixing},~\cite{Garziano2016} \\ \hline
 \multirow{10}{*}{\parbox{2.1cm}{\textbf{Non- \\ degenerate three-wave mixing}}} & \multirow{2}{*}{\parbox{2.6cm}{Spontaneous \\ Raman scattering}} & Stokes & \multirow{4}{*}{2 resonators, 1 qubit} & \multirow{4}{*}{$\omega_a = \omega_b + \omega_q$} & $\ket{1, 0, g} \to \ket{0, 1, e}$ & Gen.~Rabi & \secref{sec:3WaveMixingRaman},~\cite{Kockum2017} \\ \cline{3-3} \cline{6-8}
 & & anti-Stokes & & & $\ket{0, 1, e} \to \ket{1, 0, g}$ & Gen.~Rabi & \secref{sec:3WaveMixingRaman},~\cite{Kockum2017} \\ \cline{2-3} \cline{6-8}
 & \multirow{2}{*}{\parbox{2.6cm}{Stimulated \\ Raman scattering}} & Stokes & & & $\ket{1, n, g} \to \ket{0, n+1, e}$ & Gen.~Rabi & \secref{sec:3WaveMixingRaman} \\ \cline{3-3} \cline{6-8}
 & & anti-Stokes & & & $\ket{n, 1, e} \to \ket{n+1, 0, g}$ & Gen.~Rabi & \secref{sec:3WaveMixingRaman} \\ \cline{2-8}
 & \multicolumn{2}{|c|}{\multirow{3}{*}{Sum-frequency generation}} & 1 resonator, 2 qubits & $\omega_a = \omega_{q1} + \omega_{q2}$ & $\ket{0, e, e} \to \ket{1, g, g}$ & Gen.~Rabi & \secref{sec:3WaveMixingGeneral},~\cite{Garziano2016} \\ \cline{4-8}
 & \multicolumn{2}{|c|}{} & 2 resonators, 1 qubit & $\omega_a + \omega_b = \omega_q$ & $\ket{1, 1, g} \to \ket{0, 0, e}$ & Gen.~Rabi & \secref{sec:3WaveMixingGeneral} \\ \cline{4-8}
 & \multicolumn{2}{|c|}{} & 3 resonators, 1 qubit & $\omega_a + \omega_b = \omega_c$ & $\ket{1, 1, 0, g} \to \ket{0, 0, 1, g}$ & Gen.~Rabi & \secref{sec:3WaveMixingGeneral} \\ \cline{2-8}
 & \multicolumn{2}{|c|}{\multirow{3}{*}{Difference-frequency generation}} & 1 resonator, 2 qubits & $\omega_a = \omega_{q1} + \omega_{q2}$ & $\ket{1, g, g} \to \ket{0, e, e}$ & Gen.~Rabi & \secref{sec:3WaveMixingGeneral},~\cite{Garziano2016}  \\ \cline{4-8}
 & \multicolumn{2}{|c|}{} & 2 resonators, 1 qubit & $\omega_a + \omega_b = \omega_q$ & $\ket{0, 0, e} \to \ket{1, 1, g}$ & Gen.~Rabi & \secref{sec:3WaveMixingGeneral} \\ \cline{4-8}
 & \multicolumn{2}{|c|}{} & 3 resonators, 1 qubit & $\omega_a + \omega_b = \omega_c$ & $\ket{0, 0, 1, g} \to \ket{1, 1, 0, g}$ & Gen.~Rabi & \secref{sec:3WaveMixingGeneral} \\ \hline
\end{tabular}
\end{table}

\end{turnpage}

\pdfpageattr\expandafter{\the\pdfpageattr/Rotate 90}

\clearpage

\pdfpageattr\expandafter{\the\pdfpageattr/Rotate 0}

\clearpage
\begin{turnpage}

\begin{table}
\centering
\caption{A summary of four-wave-mixing processes in nonlinear optics and their deterministic analogues with single atoms and virtual photons. In the case of nondegenerate four-wave mixing, the given frequencies and transitions are just some of the possibilities. For degenerate four-wave mixing with two degenerate signals, see \appref{app:4WaveMixing}. \label{tab:Summary4Wave}}
\renewcommand{\arraystretch}{1.2}
\renewcommand{\tabcolsep}{0.1cm}
\begin{tabular}{|c|c|c|c|c|c|c|c|} \hline
\multicolumn{3}{|c|}{\textbf{Nonlinear-optics process}} & \textbf{Analogous setup} & \textbf{Frequencies} & \textbf{Transition} & \textbf{Hamiltonian} & \textbf{Reference}  \\ \hline
 \multirow{10}{*}{\parbox{2cm}{\textbf{Degenerate four-wave mixing}}} & \multicolumn{2}{|c|}{\multirow{3}{*}{\parbox{4cm}{Third-harmonic generation \\ (upconversion)}}} & 1 resonator, 1 qubit & $\omega_q = 3 \omega_a$ & $\ket{3, g} \to \ket{0, e}$ & Rabi & \secref{sec:Degenerate4WaveMixing},~\cite{Ma2015, Garziano2015} \\ \cline{4-8}
  & \multicolumn{2}{|c|}{} & 2 resonators, 1 qubit & $\omega_a = 3 \omega_b$ & $\ket{0, 3, g} \to \ket{1, 0, g}$ & Rabi & \secref{sec:Degenerate4WaveMixing} \\ \cline{4-8}
 & \multicolumn{2}{|c|}{} & 1 resonator, 3 qubits & $\omega_a = 3 \omega_q$ & $\ket{0, e, e, e} \to \ket{1, g, g, g}$ & Rabi & \secref{sec:Degenerate4WaveMixing},~\cite{Garziano2016} \\ \cline{2-8}
 & \multicolumn{2}{|c|}{\multirow{3}{*}{\parbox{4.5cm}{Third-subharmonic generation \\ (downconversion)}}} & 1 resonator, 1 qubit & $\omega_q = 3 \omega_a$ & $\ket{0, e} \to \ket{3, g}$ & Rabi & \secref{sec:Degenerate4WaveMixing},~\cite{Ma2015, Garziano2015} \\ \cline{4-8}
  & \multicolumn{2}{|c|}{} & 2 resonators, 1 qubit & $\omega_a = 3 \omega_b$ & $\ket{1, 0, g} \to \ket{0, 3, g}$ & Rabi & \secref{sec:Degenerate4WaveMixing} \\ \cline{4-8}
 & \multicolumn{2}{|c|}{} & 1 resonator, 3 qubits & $\omega_a = 3 \omega_q$ & $\ket{1, g, g, g} \to \ket{0, e, e, e}$ & Rabi & \secref{sec:Degenerate4WaveMixing},~\cite{Garziano2016} \\ \cline{2-8}
 & \multirow{2}{*}{\parbox{3.6cm}{Hyper-Raman scattering,\\ type I}} & Stokes & \multirow{2}{*}{2 resonators, 1 qubit} & $\omega_a + \omega_q = 2 \omega_b$ & $\ket{0, 2, g} \to \ket{1, 0, e}$ & JC & \secref{sec:4WaveMixingHyperRamanScatteringTypeI},~\cite{Kockum2017} \\ \cline{3-3} \cline{5-8}
 & & anti-Stokes & & $\omega_a = 2 \omega_b + \omega_q$  & $\ket{0, 2, e} \to \ket{1, 0, g}$ & Rabi & \secref{sec:4WaveMixingHyperRamanScatteringTypeI},~\cite{Kockum2017} \\ \cline{2-8}
 & \multirow{2}{*}{\parbox{3.6cm}{Hyper-Raman scattering, \\ type II}} & Stokes & \multirow{2}{*}{2 resonators, 2 qubits} & \multirow{2}{*}{$\omega_a = \omega_b + 2 \omega_q$} & $\ket{1, 0, g, g} \to \ket{0, 1, e, e}$ & Rabi & \secref{sec:4WaveMixingHyperRamanScatteringTypeII} \\ \cline{3-3} \cline{6-8}
 & & anti-Stokes & & & $\ket{0, 1, e, e} \to \ket{1, 0, g, g}$ & Rabi & \secref{sec:4WaveMixingHyperRamanScatteringTypeII} \\ \hline
 \multirow{12}{*}{\parbox{2.1cm}{\textbf{Non- \\ degenerate four-wave mixing}}} & \multicolumn{2}{|c|}{\multirow{4}{*}{\parbox{5.1cm}{Type I \\ (2 inputs, 2 outputs)}}} & 3 resonators, 1 qubit & $\omega_a + \omega_b = \omega_c + \omega_q$ & $\ket{1, 1, 0, g} \to \ket{0, 0, 1, e}$ & JC & \secref{sec:4WaveMixingGeneral} \\ \cline{4-8}
  & \multicolumn{2}{|c|}{} & 4 resonators, 1 qubit & $\omega_a + \omega_b = \omega_c + \omega_d$ & $\ket{1, 1, 0, 0, g} \to \ket{0, 0, 1, 1, g}$ & JC & \secref{sec:4WaveMixingGeneral} \\ \cline{4-8}
  & \multicolumn{2}{|c|}{} & 2 resonators, 2 qubits & $\omega_a + \omega_b = \omega_{q1} + \omega_{q2}$ & $\ket{1, 1, g, g} \to \ket{0, 0, e, e}$ & JC & \secref{sec:4WaveMixingGeneral} \\ \cline{4-8}
  & \multicolumn{2}{|c|}{} & 1 resonator, 3 qubits & $\omega_a + \omega_{q1} = \omega_{q2} + \omega_{q3}$ & $\ket{1, e, g, g} \to \ket{0, g, e, e}$ & JC & \secref{sec:4WaveMixingGeneral} \\ \cline{2-8}
 & \multicolumn{2}{|c|}{\multirow{4}{*}{\parbox{5.1cm}{Type II \\ (3 inputs, 1 output)}}} & 3 resonators, 1 qubit & $\omega_a + \omega_b + \omega_c = \omega_q$ & $\ket{1, 1, 1, g} \to \ket{0, 0, 0, e}$ & Rabi & \secref{sec:4WaveMixingGeneral} \\ \cline{4-8}
  & \multicolumn{2}{|c|}{} & 4 resonators, 1 qubit & $\omega_a + \omega_b + \omega_c = \omega_d$ & $\ket{1, 1, 1, 0, g} \to \ket{0, 0, 0, 1, g}$ & Rabi & \secref{sec:4WaveMixingGeneral} \\ \cline{4-8}
  & \multicolumn{2}{|c|}{} & 2 resonators, 2 qubits & $\omega_a = \omega_b + \omega_{q1} + \omega_{q2}$ & $\ket{0, 1, e, e} \to \ket{1, 0, g, g}$ & Rabi & \secref{sec:4WaveMixingGeneral} \\ \cline{4-8}
  & \multicolumn{2}{|c|}{} & 1 resonator, 3 qubits & $\omega_a = \omega_{q1} + \omega_{q2} + \omega_{q3}$ & $\ket{0, e, e, e} \to \ket{1, g, g, g}$ & Rabi & \secref{sec:4WaveMixingGeneral} \\ \cline{2-8}
 & \multicolumn{2}{|c|}{\multirow{4}{*}{\parbox{5.1cm}{Type III \\ (1 input, 3 outputs)}}} & 3 resonators, 1 qubit & $\omega_a + \omega_b + \omega_c = \omega_q$ & $\ket{0, 0, 0, e} \to \ket{1, 1, 1, g}$ & Rabi & \secref{sec:4WaveMixingGeneral} \\ \cline{4-8}
  & \multicolumn{2}{|c|}{} & 4 resonators, 1 qubit & $\omega_a = \omega_b + \omega_c + \omega_d$ & $\ket{1, 0, 0, 0, g} \to \ket{0, 1, 1, 1, g}$ & Rabi & \secref{sec:4WaveMixingGeneral} \\ \cline{4-8}
  & \multicolumn{2}{|c|}{} & 2 resonators, 2 qubits & $\omega_a = \omega_b + \omega_{q1} + \omega_{q2}$ & $\ket{1, 0, g, g} \to \ket{0, 1, e, e}$ & Rabi & \secref{sec:4WaveMixingGeneral} \\ \cline{4-8}
  & \multicolumn{2}{|c|}{} & 1 resonator, 3 qubits & $\omega_a = \omega_{q1} + \omega_{q2} + \omega_{q3}$ & $\ket{1, g, g, g} \to \ket{0, e, e, e}$ & Rabi & \secref{sec:4WaveMixingGeneral} \\ \hline
\end{tabular}
\end{table}

\end{turnpage}

\pdfpageattr\expandafter{\the\pdfpageattr/Rotate 90}

\clearpage

\pdfpageattr\expandafter{\the\pdfpageattr/Rotate 0}

\appendix

\section{Classical description of nonlinear optical phenomena}
\label{app:NonlinearSusceptibility}

Here we give a few examples showing how mixing of classical waves can be explained classically by applying the principal relation of nonlinear optics,
\bea
\mathbf{P} &=& \epsilon_0 \left( \chi ^{(1)} \mathbf{E} + \chi^{(2)} \mathbf{E}^2 + \chi^{(3)} \mathbf{E}^3 + \ldots \right) \nn\\ 
&=& \mathbf{P}^{(1)} + \mathbf{P}^{(2)} + \mathbf{P}^{(3)} + \ldots. 
\label{nonlinearP}
\eea
In this pedagogical introduction to nonlinear optics, based on Ref.~\cite{Lindsay1975}, we give the classical explanations of a few standard wave-mixing processes by applying the lowest-order required nonlinear polarization $\mathbf{P}^{(n)}$ and the corresponding nonlinear susceptibility $\chi^{(n)}$. Our examples include the linear (Pockels) and quadratic (Kerr) electro-optical phenomena. Second-harmonic generation in a $\chi^{(2)}$ medium was already treated in \secref{sec:MechanismsNonlinearOptics}.

\subsection{Wave mixing in a $\chi^{(2)}$ medium and the Pockels effect}

Assume that two monochromatic scalar electric waves, $E_1(t) = E_{10} \cos (\omega_1 t)$ and $E_2(t) = E_{20} \cos (\omega_2 t)$, are applied to a medium described by the second-order frequency-independent susceptibility $\chi^{(2)}$. Then, the induced second-order polarization $P^{(2)}$ is given by
\bea
P^{(2)} &=& \epsilon_0 \chi^{(2)} E^2 = \epsilon_0 \chi^{(2)} \left\{ E_{10} \cos (\omega_1 t) + E_{20} \cos (\omega _2 t) \right\}^2 \nn \\
&=& \epsilon_0 \chi^{(2)} \Big\{ E^2_{10} \cos^2 (\omega_1 t) + E^2_{20} \cos^2 (\omega_2 t) \nn \\ 
&&+ 2 E_{10} E_{20} \cos (\omega_1 t) \cos (\omega_2 t) \Big\} \nn \\
&=& \frac{1}{2} \epsilon_0 \chi^{(2)} \Big\{ E^2_{10} [1 + \cos (2 \omega_1 t)] + E^2_{20} [1 + \cos (2 \omega_2 t)] \nn \\
&&+ 2 E_{10} E_{20} \left( \cos [ (\omega_1 - \omega_2) t ] + \cos [ (\omega_1 + \omega_2) t ] \right) \Big\} \nn \\
&=& \frac{1}{2} \epsilon_0 \chi^{(2)} \Big\{ \left( E_{10}^2 + E_{20}^2 \right) + E_{10}^2 \cos (2 \omega_1 t) \nn \\ 
&&+ E_{20}^2 \cos (2 \omega_2 t) + 2 E_{10} E_{20} \cos [ (\omega_1 - \omega_2) t ] \nn \\
&&+ 2 E_{10} E_{20} \cos [ (\omega_1 + \omega_2) t ] \Big\} \nn \\
&\equiv& P_0^{(2)} + P_{2 \omega _1}^{(2)} + P_{2 \omega_2}^{(2)} + P_{\omega_1 - \omega_2}^{(2)} + P_{\omega_1 + \omega_2}^{(2)}, 
\label{chi2b}
\eea
where the induced second-order nonlinear polarization $P_{\omega_x}^{(2)}$, oscillating with frequency $\omega_x = 0, 2 \omega_1, \ldots$, is defined by the corresponding $\omega_x$-dependent term in the second-last equation in \eqref{chi2b}.

We see that this process can be interpreted as mixing of two waves with frequencies $\omega_1$ and $\omega_2$. Alternatively, in a general case, this effect can be interpreted as six-wave mixing if we include also the four output (mixed) frequencies $2 \omega_1$, $2 \omega_2$, $\abs{\omega_1 - \omega_2}$, and $\omega_1 + \omega_2$. In a quantum description, the latter interpretation is conventionally applied.

In a special case, let us assume that $\omega_2=0$; then $E_2 = E_{20} = {\rm const}$, and 
\be
P_{\omega _1 - \omega_2}^{(2)} + P_{\omega _1 + \omega_2}^{(2)} = 2 P_{\omega_1}^{(2)} = \epsilon_0 \left( 2 \chi^{(2)} E_{20} \right) E_1(t).
\ee
We see that the effective first-order-like susceptibility $\chi_{\rm eff}^{(1)} \equiv 2 \chi^{(2)} E_{20}$ is proportional to the amplitude of the constant electric field. This phenomenon is usually referred to as the (linear) Pockels effect or linear electro-optical effect.

A few comments can be made on the momentum (and energy) conservation when fields of frequencies $\omega_1$ and $\omega_2$ are mixed to generate fields with sum ($\omega_+ = \omega_1 + \omega_2$) and difference ($\omega_- = \abs{\omega_1 - \omega_2}$) frequencies. These new fields can be amplified depending on which momentum condition $\mathbf{k}_+ = \mathbf{k}_1 + \mathbf{k}_2$ or $\mathbf{k}_- = \mathbf{k}_1 - \mathbf{k}_2$ is satisfied for the corresponding wave vectors $\mathbf{k}_j$. Usually only one of these conditions is satisfied. If both conditions are fulfilled, then the wave mixing has a local character. For example, if $\omega_1 = \omega_2 \equiv \omega$ and $\mathbf{k}_1 = \mathbf{k}_2 \equiv \mathbf{k}$, then $\omega_+ = 2 \omega$, $\omega_- = 0$, $k_+ = 2 k$, and $k_- = 0$.

\subsection{Third-harmonic generation in a $\chi^{(3)}$ medium}

Assume that a monochromatic electric wave $E(t) = E_0 \cos (\omega t)$ is applied to a medium described solely by a third-order susceptibility $\chi^{(3)}$. Then we observe
\bea
P^{(3)} &=& \epsilon_0 \chi^{(3)} E^3 = \epsilon_0 \chi^{(3)} E_0^3 \cos^3 (\omega t) \nn \\
&=& \epsilon_0 \chi^{(3)} E_0^3 \left[ \frac{3 \cos (\omega t) + \cos (3 \omega t)}{4} \right] \nn \\
&=& \frac{3}{4} \epsilon_0 \chi^{(3)} E_0^3\cos (\omega t) + \frac{1}{4} \epsilon_0 \chi^{(3)} E_0^3 \cos (3 \omega t) \nn \\
&=& P_{\omega}^{(3)} + P_{3 \omega}^{(3)},
\label{chi3a}
\eea
where the term $P_{3 \omega}^{(3)}$ describes the induced polarization, oscillating with triple the frequency of the input field, which can be interpreted as third-harmonic generation.

\subsection{Wave mixing in a $\chi^{(3)}$ medium and the Kerr effect}

Assume that two monochromatic electric beams, $E_1 (t) = E_{10} \cos (\omega_1 t)$ and $E_2(t) = E_{20} \cos (\omega_2 t)$, are applied to a medium described by a third-order susceptibility $\chi^{(3)}$, and that $\chi^{(3)}$ is frequency independent. Then the third-order induced polarization $P^{(3)}$ of the medium is given by
\bea
P^{(3)} &=& \epsilon _0 \chi^{(3)} E^3 = \epsilon_0 \chi^{(3)} \left[ E_{10} \cos (\omega_1 t) + E_{20} \cos (\omega_2 t) \right]^3 \nn \\
&=& P_{\omega_1}^{(3)} + P_{\omega_2}^{(3)} + P_{3 \omega_1}^{(3)} + P_{3 \omega_2}^{(3)} + P_{2 \omega_1 - \omega_2}^{(3)} \nn \\
&&+ P_{2 \omega_1 + \omega_2}^{(3)} + P_{2 \omega_2 - \omega_1}^{(3)} + P_{2 \omega_2 + \omega_1}^{(3)},
\label{chi3b}
\eea
where we do not give (except two terms) an explicit form of these induced third-order nonlinear polarizations $P_{\omega_x}^{(3)}$, but only indicate their frequencies $\omega_x$.

Analogously to wave mixing in a $\chi^{(2)}$ medium, one can interpret this process as mixing of two waves with frequencies $\omega_1$ and $\omega_2$. Alternatively, this effect, in a general case, can be interpreted as mixing of eight waves (including the output waves) with frequencies $\omega_1$, $\omega_2$, $3 \omega_1$, $3 \omega_2$, $\abs{2 \omega_1 \pm \omega_2}$, and $\abs{2 \omega_2 \pm \omega_1}$. In a quantum description, the latter convention is usually applied.

In a special case, we have
\be
P_{2 \omega_2 \pm \omega_1}^{(3)} = \frac{3}{4} \epsilon_0 \chi^{(3)} E_{20}^2 E_{10} \cos [ (2 \omega_2 \pm \omega_1) t ].
\label{chi3c}
\ee
If we assume that $\omega_2 = 0$, we obtain
\be
P_{2 \omega_2 \pm \omega_1}^{(3)} = \frac{3}{4} \epsilon_0 \chi^{(3)} E_{20}^2 E_{10} \cos (\omega_1 t) \equiv \epsilon_0 \chi^{(1)}_{\rm eff} E_1(t).
\label{chi3d}
\ee
Thus, the effective first-order-like susceptibility $\chi^{(1)}_{\rm eff} \equiv \frac{3}{4} \chi ^{(3)} E^{2}_{20}$ is proportional to the square of the constant electric field $E_2(t) = E_{20}$. This is a standard classical explanation of the Kerr effect, which is also referred to as the quadratic electro-optical effect.

\section{Perturbation theory}
\label{app:PerturbationTheory}

In this appendix, we show how to derive the expression for the effective coupling given in \eqref{eq:PerturbationFormula}. In all processes we considered in Secs.~\ref{sec:3WaveMixing} and \ref{sec:4WaveMixing}, there is an initial state $\ket{i}$ and a final state $\ket{f}$ connected by the effective coupling in an effective interaction Hamiltonian
\be
\hat H_{\rm int}^{\rm eff} = g_{\rm eff} \ketbra{f}{i} +  {\rm H.c.}
\label{eq:HeffApp}
\ee
As stated in \secref{sec:QuantumRabiModel}, if the shortest path between $\ket{i}$ and $\ket{f}$ is an $n$th-order process, the effective coupling $g_{\rm eff}$ is given to lowest order by
\be
g_{\rm eff} = \sum_{j_1, j_2, \ldots, j_{n-1}} \frac{V_{f j_{n-1}} \ldots V_{j_2 j_1} V_{j_1 i}}{\left( E_i - E_{j_1} \right) \left( E_i - E_{j_2} \right) \ldots \left( E_i - E_{j_{n-1}} \right) },
\label{eq:PerturbationFormulaApp}
\ee
where the sum goes over all virtual transitions forming $n$-step paths between $\ket{i}$ and $\ket{f}$. The formula in \eqref{eq:PerturbationFormulaApp} can be derived by considering the Dyson series of the time evolution operator in the interaction picture,
\bea
&&\hat U_I (t,t_0) = 1 - i \int_{t_0}^t \id t' \hat H_{\rm int}(t') \nn\\
&&+ (- i)^2 \int_{t_0}^t \id t' \int_{t_0}^{t'} \id t'' \hat H_{\rm int}(t') \hat H_{\rm int}(t'') + \ldots, \quad
\label{eq:U}
\eea
when the interaction Hamiltonian $\hat H_{\rm int}$ is time-independent. Assuming the system starts in the eigenstate $\ket{i}$ of the noninteracting Hamiltonian at time $t_0$, the probability of the transition $\ket{i} \to \ket{f}$ is given to lowest ($n$th) order by the $n$th-order term in \eqref{eq:U}, $U_I^{(n)} (t,t_0)$, through
\bea
&&P \left( \ket{i} \to \ket{f} \right) = \abssq{\brakket{f}{\hat U_I^{(n)} (t,t_0)}{i}} = \frac{\left( 1 - e^{i (E_f - E_i)} \right)^2}{\left( E_f - E_i \right)^2} \nn\\
&&\times \abssq{\sum_{j_1, j_2, \ldots, j_{n-1}} \frac{V_{fj_{n-1}} \ldots V_{j_2j_1} V_{j_1i}}{\left( E_i - E_{j_1} \right) \left( E_i - E_{j_2} \right) \ldots \left( E_i - E_{j_{n-1}} \right) }}, \nn\\
\:
\eea
which in the limit $t \to \infty$ gives the transition rate
\bea
&&W_{(\ket{i} \to \ket{f})} = 2\pi \delta \left( E_f - E_i \right) \nn\\
&& \times \abssq{\sum_{j_1, j_2, \ldots, j_{n-1}} \frac{V_{fj_{n-1}} \ldots V_{j_2j_1} V_{j_1i}}{\left( E_i - E_{j_1} \right) \left( E_i - E_{j_2} \right) \ldots \left( E_i - E_{j_{n-1}} \right) }}. \nn\\
\:
\eea
This is just Fermi's golden rule, showing that the effective Hamiltonian in \eqref{eq:HeffApp} with the coupling strength $g_{\rm eff}$ given by \eqref{eq:PerturbationFormula} gives the correct coupling matrix element between $\ket{i}$ and $\ket{f}$.

\section{Four-wave mixing with two degenerate frequencies}
\label{app:4WaveMixing}

\begin{figure}
\centerline{
  \resizebox{\linewidth}{!}{
    \begin{tikzpicture}[
      scale=1,
      level/.style={thick},
      virtual/.style={thick,dashed},
      ztrans/.style={thick,->,shorten >=0.1cm,shorten <=0.1cm,>=stealth,densely dashed,color=red},
      nrtrans/.style={thick,->,shorten >=0.1cm,shorten <=0.1cm,>=stealth,densely dashed,color=blue},
      rtrans/.style={thick,->,shorten >=0.1cm,shorten <=0.1cm,>=stealth,color=blue},
      strans/.style={rtrans, snake=snake, line after snake=1.5mm},
      classical/.style={thin,double,<->,shorten >=4pt,shorten <=4pt,>=stealth},
      mybox/.style={color=cyan!70, fill=cyan!20, very thick},
      myarrow/.style={color=green!35!black, thick, ->, shorten >=0.05cm, shorten <=0.05cm, >=latex},
      mysquigglyarrow/.style={color=green!35!black, thick, ->, shorten >=0.05cm, shorten <=0.05cm, >=latex, snake=snake, line after snake=2mm}
    ]
    \coordinate (corner1) at (0cm, 0cm); 
    \coordinate (rh) at (0, 1); 
    \coordinate (rl) at (1.2, 0); 
    \coordinate (vcol) at (0, 1); 
    \coordinate (hrow) at (0.5, 0); 
    \coordinate (corner2) at ($(corner1) - 0.6*(vcol) - (rh)$); 
    \coordinate (al) at ($(rl) + (0,0)$); 
    \coordinate (corner3) at ($(corner1) + (rl) + (al) + (hrow) + (al)$); 
    \coordinate (corner4) at ($(corner3) - 0.6*(vcol) - (rh)$); 
    \newcommand\labelyshift{-0.06}; 
    \newcommand\labelxshift{0}; 
    \newcommand\labelyshifttwo{0.8}; 
    \newcommand\labelxshifttwo{0}; 
    \filldraw[mybox](corner1) rectangle ($(corner1) + (rh) + (rl)$) node[color=black, midway] {I};
    \draw[myarrow] ($(corner1) + 0.8*(rh) - (al)$) -- ($(corner1) + 0.8*(rh)$) node[color=black, midway, above, shift={(\labelxshift,\labelyshift)}] {$\omega_1$};
    \draw[myarrow] ($(corner1) + 0.2*(rh) - (al)$) -- ($(corner1) + 0.2*(rh)$) node[color=black, midway, above, shift={(\labelxshift,\labelyshift)}] {$\omega_1$};
    \draw[myarrow] ($(corner1) + 0.8*(rh) + (rl)$) -- ($(corner1) + 0.8*(rh) + (rl) + (al)$) node[color=black, midway, above, shift={(\labelxshift,\labelyshift)}] {$\omega_2$};
    \draw[myarrow] ($(corner1) + 0.2*(rh) + (rl)$) -- ($(corner1) + 0.2*(rh) + (rl) + (al)$) node[color=black, midway, above, shift={(\labelxshift,\labelyshift)}] {$\omega_3$};
    \filldraw[mybox](corner2) rectangle ($(corner2) + (rh) + (rl)$) node[color=black, midway] {I};
    \draw[myarrow] ($(corner2) + 0.8*(rh) - (al)$) -- ($(corner2) + 0.8*(rh)$) node[color=black, midway, above, shift={(\labelxshift,\labelyshift)}] {$\omega_1$};
    \draw[myarrow] ($(corner2) + 0.2*(rh) - (al)$) -- ($(corner2) + 0.2*(rh)$) node[color=black, midway, above, shift={(\labelxshift,\labelyshift)}] {$\omega_2$};
    \draw[myarrow] ($(corner2) + 0.8*(rh) + (rl)$) -- ($(corner2) + 0.8*(rh) + (rl) + (al)$) node[color=black, midway, above, shift={(\labelxshift,\labelyshift)}] {$\omega_3$};
    \draw[myarrow] ($(corner2) + 0.2*(rh) + (rl)$) -- ($(corner2) + 0.2*(rh) + (rl) + (al)$) node[color=black, midway, above, shift={(\labelxshift,\labelyshift)}] {$\omega_3$};       
    \filldraw[mybox](corner3) rectangle ($(corner3) + (rh) + (rl)$) node[color=black, midway] {II};
    \draw[myarrow] ($(corner3) + 0.95*(rh) - (al)$) -- ($(corner3) + 0.95*(rh)$) node[color=black, midway, above, shift={(\labelxshift,\labelyshift)}] {$\omega_1$};
    \draw[myarrow] ($(corner3) + 0.5*(rh) - (al)$) -- ($(corner3) + 0.5*(rh)$) node[color=black, midway, above, shift={(\labelxshift,\labelyshift)}] {$\omega_1$};
    \draw[myarrow] ($(corner3) + 0.05*(rh) - (al)$) -- ($(corner3) + 0.05*(rh)$) node[color=black, midway, above, shift={(\labelxshift,\labelyshift)}] {$\omega_2$};
    \draw[myarrow] ($(corner3) + 0.5*(rh) + (rl)$) -- ($(corner3) + 0.5*(rh) + (rl) + (al)$) node[color=black, midway, above, shift={(\labelxshift,\labelyshift)}] {$\omega_3$};
    \filldraw[mybox](corner4) rectangle ($(corner4) + (rh) + (rl)$) node[color=black, midway] {III};
    \draw[myarrow] ($(corner4) + 0.5*(rh) - (al)$) -- ($(corner4) + 0.5*(rh)$) node[color=black, midway, above, shift={(\labelxshift,\labelyshift)}] {$\omega_1$};
    \draw[myarrow] ($(corner4) + 0.95*(rh) + (rl)$) -- ($(corner4) + 0.95*(rh) + (rl) + (al)$) node[color=black, midway, above, shift={(\labelxshift,\labelyshift)}] {$\omega_2$};
    \draw[myarrow] ($(corner4) + 0.5*(rh) + (rl)$) -- ($(corner4) + 0.5*(rh) + (rl) + (al)$) node[color=black, midway, above, shift={(\labelxshift,\labelyshift)}] {$\omega_2$};        
    \draw[myarrow] ($(corner4) + 0.05*(rh) + (rl)$) -- ($(corner4) + 0.05*(rh) + (rl) + (al)$) node[color=black, midway, above, shift={(\labelxshift,\labelyshift)}] {$\omega_3$};    
    \end{tikzpicture}
  }
}
\caption{Schematic representations (Feynman-like diagrams) of the four-wave-mixing processes with two degenerate frequencies. Going clockwise from the upper left corner, they are: type-I four-wave mixing with the frequencies of the two incoming signals degenerate ($2\omega_1 = \omega_2 + \omega_3$), type-II four-wave mixing with the frequencies of two of the incoming signals degenerate ($2\omega_1 + \omega_2 = \omega_3$), type-III four-wave mixing with the frequencies of two of the outgoing signals degenerate ($\omega_1 = 2\omega_2 + \omega_3$), and type-I four-wave mixing with the frequencies of the two outgoing signals degenerate ($\omega_1 + \omega_2 = 2 \omega_3$). \label{fig:4WaveMixingTwoDegenerateFreqs}}
\end{figure}
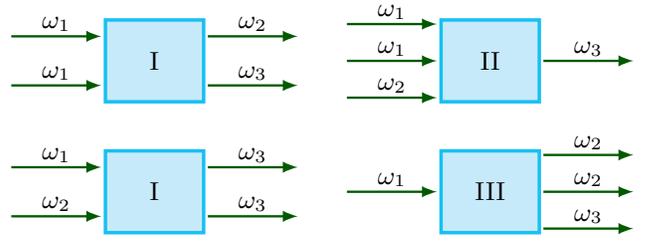

For completeness, we here show, in \figref{fig:4WaveMixingTwoDegenerateFreqs}, schematic representations of the four degenerate four-wave-mixing processes where two frequencies are degenerate (omitted from \figref{fig:4WaveMixing}). Analogues for these processes can be constructed in the same way as for the other four-wave mixing processes treated in \secref{sec:4WaveMixing} and listed in \tabref{tab:Summary4Wave}. The most obvious setup is three resonators all coupled to a single qubit. In that case, the process $\ket{2,0,0,g} \leftrightarrow \ket{0,1,1,g}$ corresponds to the type-I mixing shown in the figure. Similarly, $\ket{2,1,0,g} \leftrightarrow \ket{0,0,1,g}$ realizes analogues of the pictured type-II ($\to$) and type-III ($\leftarrow$) processes, respectively.

Just as for the nondegenerate mixing processes discussed in Secs.~\ref{sec:3WaveMixingGeneralDeterministic} and \ref{sec:4WaveMixingGeneralDeterministic}, additional setups become possible if we allow at least one of the excitations to be hosted in a qubit. With two resonators coupled to a single qubit, the two type-I mixing processes shown in \figref{fig:4WaveMixingTwoDegenerateFreqs} could be emulated by $\ket{2,0,g} \leftrightarrow \ket{0,1,e}$, which we recognise as the analogue of type-I hyper-Raman scattering already treated in \secref{sec:4WaveMixingHyperRamanScatteringTypeIDeterministic}. The pictured type-II and type-III mixing could similarly be emulated by, e.g., the process $\ket{2,1,g} \leftrightarrow \ket{0,0,g}$. In the same way, a setup with a single resonator coupled to two qubits could realize analogues of the pictured type-I processes through the transition $\ket{2,g,g} \leftrightarrow \ket{0,e,e}$, and of the pictured type-II and type-III processes through the transition $\ket{2,e,g} \leftrightarrow \ket{0,g,e}$.

\bibliography{NonlinearOpticsRefs}

\end{document}